\documentstyle[12pt,epsfig]{article}
\newcommand{\beq}{\begin{equation}}
\newcommand{\eeq}{\end{equation}}
\newcommand{\bea}{\begin{eqnarray}}
\newcommand{\eea}{\end{eqnarray}}

\newcommand{\ra}{\rightarrow}

\newcommand{\gsim}{\lower.7ex\hbox{$
\;\stackrel{\textstyle>}{\sim}\;$}}
\newcommand{\lsim}{\lower.7ex\hbox{$
\;\stackrel{\textstyle<}{\sim}\;$}}

\def\lsim{\mathrel{\rlap{\lower3pt\hbox{\hskip0pt$\sim$}}
    \raise1pt\hbox{$<$}}}         
\def\gsim{\mathrel{\rlap{\lower4pt\hbox{\hskip1pt$\sim$}}
    \raise1pt\hbox{$>$}}}         

\renewcommand{\Im}{{\rm Im}\,}

\newcommand{\bibit}[1]{\bibitem{#1}}

\newcommand{\aver}[1]{\langle #1\rangle}

\newcommand{\La}{\overline{\Lambda}}
\newcommand{\Lam}{\Lambda_{\rm QCD}}
\newcommand{\mhad}{\mu_{\rm hadr}}

\newcommand{\al}{\alpha}
\newcommand{\be}{\beta}
\newcommand{\as}{\alpha_s}
\newcommand{\GeV}{\,\mbox{GeV}}
\newcommand{\MeV}{\,\mbox{MeV}}
\newcommand{\matel}[3]{\langle #1|#2|#3\rangle}
\newcommand{\state}[1]{|#1\rangle}

\begin{document}

\begin{titlepage}
\renewcommand{\thefootnote}{\fnsymbol{footnote}}

\begin{flushright}
UND-HEP-00-BIG\hspace*{.03em}07\\
hep-ph/0010328\\
\end{flushright}
\vspace*{1.3cm}

\begin{center} \Large
{\bf Topics in the Heavy Quark Expansion}
\end{center}
\vspace*{1.5cm}
\begin{center} {\Large
Nikolai Uraltsev} \\
\vspace{1.2cm}
{\normalsize
{\it Dept. of Physics, Univ. of Notre Dame du Lac, Notre Dame, IN 46556,
U.S.A.}\\
{\small \rm and} \\
{\it St.\,Petersburg Nuclear Physics Institute,
Gatchina, St.\,Petersburg 188350, Russia
}\\
}

\vspace*{1.7cm}

{\large{\bf Abstract}}\\
\end{center}
\vspace*{.2cm}
Achievements in the heavy quark theory over the last decade are
reviewed, with the main emphasis put on dynamical methods which
quantify nonperturbative effects via application of the Operator
Product Expansion. These include the total weak decay rates of heavy
flavor hadrons and nonperturbative corrections to heavy quark sum
rules. Two new exact superconvergent sum rules are derived; they
differ from the known ones in that they are finite and normalization
point independent in perturbation theory. A new hadronic parameter
$\overline\Sigma$ is introduced which is a spin-nonsinglet analogue of 
$\La\!=\!M_B\!-\!m_b$; it is expected to be about $0.25\GeV$. The first sum
rule implies the bound $\rho^2\!>\! 3/4$ for the slope of the Isgur-Wise
function.
The heavy quark potential is discussed and its connection to the
infrared contributions in the heavy quark mass. Among applications
extraction of $|V_{cb}|$ from the total semileptonic and from the
$B\!\to\! D^*$ zero recoil rates is addressed, as well as extracting
$|V_{ub}|$ from $\Gamma_{\rm sl}(b\!\to\! u)$. Practical aspects of local
quark-hadron duality are briefly discussed.
\vfill
\noindent
\end{titlepage}

\newpage
\thispagestyle{empty}
\tableofcontents

\newpage

\section{Introduction}

Quark-gluon dynamics are governed by the QCD Lagrangian
\begin{equation}
{\cal L} =-\frac{1}{4} G_{\mu\nu}^a G_{\mu\nu}^a
+\sum_{\psi}\bar \psi (i\!\not\!\!{D} \!-\! m_\psi) \psi
\label{n1}
\eeq
where $G_{\mu\nu}^a$ is the gluon field strength tensor, and $\psi$ are
quark fields.
Masses of $u$ and $d$ quarks are
only a few $\MeV$, much smaller than $\Lam$, and in most applications
of hadron physics $u$ and $d$
can be considered as massless.  The strange
quark mass is about $150\MeV$, only a little smaller
than $\Lam$. Nevertheless, there is ample evidence that treating
$s$ quark as light is justified, and corrections to the (light) $SU(3)$
symmetry are reasonably small. Thus, $m_s$ lies essentially below the
actual typical hadronic QCD scale $\mhad \sim 500\mbox{ to } 700 \MeV$.

Those quarks $Q$ for which $m_Q \gg \mhad \sim (2\,$--$\,3) \Lam$ are
heavy quarks. The sixth $t$ quark is the heaviest, $m_t\approx
170\GeV$. However, it is too heavy. Its width due to the `semiweak'
decay $t\ra b+W^+$ is $\Gamma_t \approx 1\GeV$. It decays too fast for
the $t$-hadrons, the bound states with light quarks to be formed.

The best candidates for application of the Heavy Quark Expansion (HQE)
are beauty hadrons. $b$ quark is heavy enough to confidently use the
expansion in $1/m_b$, yet the leading corrections to the heavy quark
limit are not negligible.

The charmed quark $c$ can be called heavy only with some
reservations.
While in some cases this yields a reasonable approximation,
$m_c$ often appears manifestly too low for a quantitative treatment of
charmed physics in the $1/m_Q$ expansion. There is no universal answer
here, and considerable care must be exercised in
every particular case.

Usually considered heavy flavor hadrons are
composed from
one heavy quark $Q$, and a light cloud: a light antiquark $\bar q$, or
diquark $qq$,
together with the gluon medium and light
quark-antiquark pairs. The role of the gluon medium is to
keep all `valence' constituents together, in a
colorless bound state which will be generically denoted by $H_Q$.
Therefore, we have the following simplified picture of a heavy flavor
hadron.  The heavy quark has a small
size $\sim 1/m_Q$, and is surrounded by a static Coulomb-like color
field $A_0$ at small distances. Non-Abelian selfinteraction slightly
modifies the potential, but the nonlinearity is driven by the coupling
$\as\left(r^{-1}\right)/\pi$ and is not significant. At larger
distances the selfinteraction strengthens, at $R \gsim \Lam^{-1}$ it
is completely nonperturbative: the soft modes of the light fields are
strongly coupled and strongly fluctuate.

In weak decays
the standard situation is that the external (to QCD) forces like
$W$ bosons interact with the heavy quark, say, instantaneously replace
the $b$ quark by $c$ quark generally changing its velocity.
Such an event excites first the typical modes of the heavy
quark, hard gluons with $\vec{k}_{\rm typ} \sim m_Q$. Since
$$
\as \left(k_{\rm typ} \right)\sim  \as(m_Q) \ll 1
$$
{\it perturbation theory} is adequate there.

An actual decay process, nevertheless, eventually runs into the
strong-interaction nonperturbative domain of $\vec{k} \sim \omega \sim
\Lam$, where $\omega$ denotes the characteristic frequencies. The final
hadronization dynamics shaping the hadrons observed in experiment, is
a result of soft nonperturbative physics which is responsible for
confinement. Since the complicated final state dynamics
involve $\omega \ll m_Q$, to deal with nonperturbative
effects one can use the {\it nonrelativistic expansion} for
the heavy quark.

The main subject of the HQE is nonperturbative physics.
First, the perturbative corrections are conceptually simple, even if
the actual
computations are often cumbersome and, beyond the first-order effects,
require sophisticated state-of-the-art technique.

Second, the perturbative corrections are
calculated in the full QCD rather than in the effective low-energy
theory, since they come just from the gluon
momenta $k\sim m_Q$ where the nonrelativistic approximation is not
applicable. Still, the interplay of the
perturbative and nonperturbative effects is quite nontrivial and often
involves theoretical subtleties.

The treatment of the nonperturbative effects is a nontrivial problem, and
different methods of QCD are used here. The basic tool for all of them
in heavy quarks is the Wilson operator product expansion (OPE)
\cite{wilson}.

\subsection{Nonrelativistic expansion}

The main simplification
of a nonrelativistic treatment is that the number of heavy
quarks $n_Q$ and antiquarks $\bar n_Q$ are separately conserved.
Propagation with usual relativistic
Green function for the heavy quark contains also the process
of the $Q\bar Q$ pair creation if the time ordering of the
vertices along the line is reversed somewhere.  In the nonrelativistic
kinematics, however, such
configurations yield a power-suppressed contribution, since the
virtuality of the intermediate state (the energy denominator, in
the language of noncovariant perturbation theory of quantum
mechanics) is of the order of $2m_Q$. In order to observe such
processes as real ones, it would be necessary to supply energy at least
as large as $\omega \approx 2m_Q$.

Heavy quarks can appear also in closed loops, for example, in the
processes of the virtual gluon conversion. For heavy quarks
such effects are also suppressed, $\sim
\vec{k}^2/m_Q^2$ if the gluon momentum $k$ is much smaller than $m_Q$.

As a result, the field-theoretic, or second-quantized
description of the heavy  quark  becomes redundant, and
it is sufficient to resort to its usual quantum-mecha\-nical (QM)
treatment.
For example, the wavefunction of a heavy flavor hadron takes the form
\beq
\Psi_\al\left[\vec{x}_Q,\, \left\{x_{\rm light} \right\} \right]\;\;,
\label{4}
\eeq
where $\vec{x}_Q$ is the heavy quark coordinate, $x_{\rm light}$
generically represents an infinite number of light degrees of freedom;
index $\al$ describes the heavy quark spin.

A relativistic $S=\frac{1}{2}$ particle has four components, i.e.
$\al=1...4\,$. The nonrelativistic spinor $\Psi(x)$ has only two of
them, $\al=1,2$ describing two spin states. The remaining $\al=\{3,4\}$
components describe antiparticles
which decouple in the nonrelativistic theory:
\beq
Q(x) \;=\;
\left(
\begin{array}{cc} \Psi(x) \\ \chi(x) \end{array}
\right) \qquad \qquad
\begin{array}{ll}
\Psi(x) \sim {\cal O}(1) \\
\chi(x) \sim \frac{\vec{p}}{m_Q} \ra 0
\end{array}
\label{6}
\eeq
The nonrelativistic Hamiltonian of the spin-$\frac{1}{2}$
particle is the well known
Pauli Hamiltonian:\footnote{We use the convention where the coupling
$g_s$ is absorbed in the gauge fields; this is convenient for
nonperturbative analysis.}
\beq
{\cal H}_{\rm Pauli}\;=\;
- A_0 \;+ \;
\frac{\left(i\vec\partial\!-\! \vec{A}\right)^2}{2m} \;+\;
\frac{\vec \sigma \vec B}{2m}\;.
\label{8}
\eeq
The last operator is written with the coefficient appearing 
for an `elementary' pointlike
particle. Presence of the interaction generally renormalizes its
strength, the chromomagnetic moment of the heavy quark. In the heavy
quark limit $m_Q\ra \infty$ only the first term survives while the last
two terms describing space propagation and interaction of spin with the
chromomagnetic field $\vec{B}$, disappear. An infinitely heavy quark is
static and interacts only with the color Coulomb
potential.\footnote{Speaking of the Coulomb interaction in QCD we mean
the interaction with the timelike component $A_0$ of the color gauge
potential. It differs from the simple $1/R$ electrodynamic potential.} 
In turn, the heavy quark is the source of a static color field independent
of the heavy quark spin. The actual dynamics of the heavy quark spin
reveal itself in full only when the quark interacts with quanta having
large momentum $|\vec{q}\,| \gsim m_Q$.

It is easy to illustrate the $b$ quark spin-independence of the strong
forces in the following way. One can imagine the QCD world where,
instead of the actual $b$ quark there exists a scalar spinless $\tilde
b$ quark with the same mass. The usual spin-$0$ $B$ meson, $b\bar{q}$
would become a spin-$\frac{1}{2}$ particle $\tilde B \sim \tilde b
\bar{q}$. The $\Lambda_b$ baryon $bud$ having spin-$\frac{1}{2}$ would,
in turn become a scalar spinless $\tilde bud$ state; $\Sigma_b$
would have spin $1$. Nevertheless, at
$m_b \ra \infty$ the properties of $B$ and $\tilde B$ or $\Lambda_b$
and $\tilde \Lambda_b$ would be identical. The two degenerate spin
states of $\tilde B$  are counterparts of $B$ and $B^*$ of the actual
QCD.

An immediate question comes to one's mind at this point: what about
the relation between spin and statistics? In such a gedanken operation
we replace spin--$\frac{1}{2}$ hadrons  by $S\!=\!0$ ones, and {\it vice
versa}, {\it i.e.} interchange fermions and bosons. It is clear,
however, that for the states or processes with a single heavy quark the
statistics symmetry properties do not play a role.

Such an independence of the strong dynamics of the heavy quark spin is
called the heavy quark {\it Spin Symmetry}.

The rest mass $m_Q$ of the heavy quark also enters in a trivial way:
the Hamiltonian simply contains $(n_Q\!+\!\bar n_Q) m_Q$. Since both are
fixed, it is an overall additive constant. For a moving quark this
constant is $E=\sqrt{m_Q^2+\vec{p}^{\,2}} = m_Q\sqrt{1+\vec{v}^{\,2}}$.
Therefore, the actual dynamics is not affected by the concrete value of
the mass $m_Q$. This is the  {\it Heavy Flavor Symmetry} which states, for
example, the equal properties of charmed and beauty hadrons to the
extent they both can be considered heavy enough.

The heavy flavor symmetry leads also to certain scaling behavior of the
transition amplitudes with heavy flavor hadrons: the amplitudes
depend on
their velocities rather than on the absolute values of momenta:
\beq
A\left(
P_Q^{in}, p_i^{in};\: P_{Q'}^{out}, p_l^{out}
\right) \;=\;
{\cal A} \left(
\frac{P_Q^{in}}{m_Q}, p_i^{in};\: \frac{P_{Q'}^{out}}{m_{Q'}}, p_l^{out}
\right) \;.
\label{10}
\eeq
Here $P$ denote the momenta of the heavy flavor hadrons and $p$ refer
to other participating particles. 
Such a scaling behavior is valid only with respect to the soft part of
the interaction, and no hard gluons with $\vec{k} \sim m_Q$
are involved.

The one-particle (or QM) description of the heavy quark degrees of
freedom is the key simplification of the nonrelativistic expansion. The light
cloud, however, still requires a full-fledged field-theoretic
treatment. Even considering the static limit $m_Q \ra \infty$ where
only interaction with the Coulomb field $A_0$ remains,
the latter strongly fluctuates, in contrast
to simple potential QM models where the corresponding potential
$V(x)$ is a $c$-number function of coordinates.

For this reason even the quantum mechanics of heavy quarks is highly
nontrivial. Exploiting the symmetry properties of the heavy quark
interactions does not require understanding of these complicated
strong interaction dynamics. This was the main field of applications at
an early stage of theoretical development of heavy quark physics.
The recent progress is mainly related to a better treatment of basic
properties of this complicated strongly interacting system
via application of dynamic QCD methods based on the short distance
expansion.

\subsection{Operator Product Expansion}

The basic theoretical tool of the heavy quark theory is the Wilson
operator product expansion  \cite{wilson}.
The idea of the OPE was formulated by K.\,\,Wilson in the late 60's,
originally in the context of the statistical problems which are closely
related to the field theories in Euclidean space. This idea, in
general, is a separation of effects originating at large and small
distances. The application to real physical processes in
Minkowski space is often less transparent and technically more
complicated, however it is always based on the same underlying
concept.

The original QCD Lagrangian
\beq
{\cal L} =-\frac{1}{4} G_{\mu\nu}^a G_{\mu\nu}^a
+\sum_{q}\bar q (i\!\not\!\!{D}\!-\!m_q) q
 +\sum_{Q}\bar Q (i\!\not\!\!{D} \!-\! m_Q) Q =
{\cal L}_{\rm light}
+ \sum_{Q}\bar Q (i\!\not\!\!{D} \!-\! m_Q) Q
\label{12}
\eeq
is formulated at very short
distances, or, equivalently, at a very high normalization point
$\mu \!=\! M_0$, where $M_0$ is the mass of an ultraviolet
regulator. In other words, the
normalization point is assumed to be much
higher
than all mass scales in the theory, in particular, $\mu\gg m_Q$.
An
effective theory for  describing the
low energy
properties
is obtained by evolving the
Lagrangian from the high scale $M_0$ down to a lower
normalization point
$\mu$.
This means that we  integrate out, step by step,
all high frequency modes in the theory thus calculating the
Lagrangian ${\cal L}(\mu )$. The latter is a full-fledged
Lagrangian  with respect to  the soft modes
with characteristic frequencies less than $\mu$.  The hard
(high frequency) modes determine the coefficient  functions
in ${\cal L}(\mu )$, while the contribution of the soft modes
is hidden in the matrix elements of (an infinite set of)
operators appearing in ${\cal L}(\mu )$.
The value of this approach,  outlined by Wilson long ago
\cite{wilson}, has become
widely recognized and exploited in countless applications.
The peculiarity
of the heavy quark theory lies in the fact that the {\em in} and
{\em out} states contain heavy quarks. Although we
integrate out the  field fluctuations with the frequencies down to
$\mu \ll m_Q$, the heavy quark fields themselves are not integrated
out since we consider the sector with nonzero heavy flavor
charge. The effective Lagrangian ${\cal L}(\mu )$ acts in this
sector. Since the heavy quarks are neither produced nor annihilated,
any sector with the given $Q$-number is treated separately
from all others.

\section{Basics of the Heavy Quark Theory}

\subsection{Effective Hamiltonian}

The strategy for integrating out virtual degrees of freedom to
pass on to an effective theory of heavy particles is described in the
textbooks. The nonrelativistic fermion field $Q(x)$ has four degrees
of freedom. Two of them, $\Psi(x)$ in Eq.~(\ref{6}) are nearly on-shell
and two $\chi(x)$ are highly virtual describing excitation of
antiquarks. One needs to integrate out first the antiparticle fields
$\chi(x)$. For simplicity, we consider this in the rest frame.

Integrating out the antiparticle fields can be done straightforwardly
since the QCD Lagrangian is bilinear in the quark fields. In this way
one would obtain the (tree level version of the) so-called Lagrangian
of HQET ${\cal L}_{\rm HQET}$.
It is a correct nonrelativistic Lagrangian up to the first order in
$1/m_Q$.

This does not complete the program, however: one yet has a full
`particle' field $\Psi(x)$ which includes all frequencies from $0$ to
$\infty$. Such a problem does not show up in
the potential models where modes with $k \gg \Lam$ are not excited.
However, in actual QCD all radiative corrections would diverge due to
large momentum gluons.

Therefore, besides the antiparticle fields one needs to integrate out
hard gluons and the high frequency components of the heavy quark field
$\Psi(x)$ itself, those for which $|\vec{k}\,|, \omega > \mu$.
The scale $\mu$ is the normalization point of the effective theory.
Since the configurations we integrate out depend on $\mu$, the
remaining effective low energy theory is also
$\mu$-dependent.

In practice, we want to have $\mu\ll m_b$ and, actually, as low as
possible. On the other hand, $\mu$ must still belong to the perturbative
domain. In practice this means that the best choice routinely adopted
for applications is $\mu\sim \mbox{ a few } \times \Lam$, from $0.7
\mbox{ to } 1\GeV$. All coefficients in the effective Lagrangian
obtained in this way are functions of $\mu$.

To summarize, in treating heavy quarks we separate all strong
interaction effects into `hard' and `soft' introducing a
normalization scale $\mu$. To calculate the effect of the
short distance (perturbative) physics we use the original QCD
Lagrangian Eq.\,(\ref{12}). The soft physics is treated by the
nonrelativistic Lagrangian where the heavy quarks are represented by
the corresponding nonrelativistic fields $\varphi_Q$:
$$
{\cal L}_{\rm eff}=
-\frac{1}{4} G_{\mu\nu}^2
+\sum_{q}\bar q (i\!\not\!\!{D}\!-\!m_q) q
+\sum_{Q} \left\{
-m_Q \varphi_Q^+ \varphi_Q + \varphi_Q^+ iD_0 \varphi_Q  \;-
\right.
$$
\beq
\left.
-\;\frac{1}{2m_Q} \varphi_Q^+ \left(\vec\sigma i\vec{D}\right)^2
\varphi_Q -
\frac{1}{8 m_Q^2}\,\varphi_Q^+ \, \left[
-(\vec D\vec E)+\vec\sigma\!\cdot\! 
\{\vec E\!\times\!\vec\pi\!-\!\vec\pi\!\times\!\vec E\}
\right]\, \varphi_Q \right\}
\label{21}
\eeq
where
\beq
\vec{\pi} \equiv i\vec{D} = \vec{p} - \vec{A}\;,
\qquad
\left(\vec\sigma i\vec{D}\right)^2 =
\left(\vec\sigma \vec{\pi}\right)^2=
\vec\pi^{\,2}+\vec\sigma \vec{B}\;;
\label{22}
\eeq
for simplicity
only the tree level $\mu$-independent coefficients are given.  By the
standard rules one constructs from the Lagrangian (\ref{21}) the
corresponding Hamiltonian.
The heavy quark part takes the
form~\footnote{The leading term $A_0$ is often omitted here. It is then
implied that the time evolution operator is
$\pi_0=i\frac{\partial}{\partial t} + A_0$ rather than
$i\frac{\partial}{\partial t}$ in the usual Schr\"{o}dinger equation.
This can be consistently carried out through the analysis.}
\beq
{\cal H}_Q\,=\,-A_0+\frac{1}{2m_Q}\,({\vec\pi}^2 + \vec\sigma \vec
B)\,+\, \frac{1}{8m_Q^2}\,\left[-(\vec D\vec E)+ \vec\sigma \!\cdot\!
\{\vec E\!\times\!\vec\pi\!-\!\vec\pi\!\times\!\vec E\} \right] +{\cal
O}(1/m_Q^3) \;.
\label{ham}
\eeq
The first term in the $1/m_Q^2$ part is called the  Darwin term and
the second one is the convection current (spin-orbital, or $LS$) 
interaction.

Since the external interactions (electromagnetic, weak {\it etc.}) are
given in terms of the full QCD fields $Q(x)$, one needs also the relation
between $Q(x)$ and $\varphi_Q(x)$:
\beq
\varphi_Q=\left(1+\frac{(\vec\sigma\vec\pi)^2}{8m_Q^2}+ ...
\right)\frac{1\!+\!\gamma_0}{2}\,Q\;, \qquad \;
\frac{1\!-\!\gamma_0}{2} Q = \frac{\not\!\!{\,\pi} }{2m_Q} Q\, .
\label{23}
\eeq

Let us briefly recall the textbook procedure for
obtaining the nonrelativistic Lagrangian.
One starts with
\beq
{\cal L}_{\rm heavy}^0 = \bar Q(x) (i\!\not\!\!D \!-\!m_Q)Q(x)
\label{24}
\eeq
and factors out of the $Q(x)$ field the ``mechanical'' time-dependent
factor associated with the rest energy $m_Q$:
\beq
Q(x) = {\rm e}\,^{-im_Qt}{\tilde Q} (x) \; .
\label{25}
\eeq
In an arbitrary frame moving with four-velocity $v_\mu$ it
takes the following form:
\beq
Q(x) = {\rm e}\,^{-im_Qv_\mu x_\mu}{\tilde Q} (x) \;.
\label{13a}
\eeq
Then
\beq
iD_\mu Q(x) = {\rm e}\,^{-im_Q(v x)}\, \left( m_Q v_\mu +
\pi_\mu\right) \tilde Q (x)\;, \qquad \qquad \pi_\mu
\equiv \hat P_\mu-m_Qv_\mu\;.
\label{26}
\eeq
The Dirac equation  $\left(i\!\not\!\!D \!-\!m_Q\right)Q = 0$ takes the form
(now the tilde on $Q$ is omitted)
\beq
\frac{1\!-\!\gamma_0}{2}\, Q = \frac{\not\!\!{\,\pi} }{2m_Q}\, Q\: ,
\qquad\qquad \pi_0 Q = -\frac{\pi^2 \!+\!\frac{i}{2}\sigma G}{2m_Q}\, Q\, .
\label{27}
\eeq
$$
\frac{i}{2} \sigma G =\frac{i}{2} \sigma_{\mu\nu} G^{\mu\nu}\:,
\qquad\qquad
i G_{\mu\nu}= \left[\pi_\mu, \pi_\nu\right] = \left[\hat P_\mu,
\hat P_\nu\right]
$$
which allows one to exclude the small low components
$\frac{1-\gamma_0}{2} Q(x)$ expressing them {\it via} the `large' upper
components $\frac{1+\gamma_0}{2} Q(x)$.
For example, we have the useful
identity
\beq
\bar{Q}Q = \bar{Q}\gamma_0 Q +
2\bar{Q}\mbox{$\left(\!\frac{1\!-\!\gamma_0}{2}\!\right)^{\!2\!}$} Q =
\bar{Q}\gamma_0 Q + \bar{Q}\,\frac{\pi^2 \!+\!\frac{i}{2}\sigma
G}{2m_Q^2}\,Q \:+\; \mbox{total derivative}\;\:.
\label{ident}
\eeq

A subtlety emerges on this route  that must be treated properly: at
order $1/m_Q^2$ the time derivative $\partial_0$ appears with the 
nontrivial coefficient depending, for example, on the gluon field. This
can be eliminated, and the time derivative returned to its canonical
form performing the Foldy-Wouthuysen transformation
\beq
\varphi(x)=\left(1+\frac{(\vec\sigma\vec\pi)^2}{8m_Q^2} + ...
\right)\frac{1\!+\!\gamma_0}{2}\,Q(x)\;.
\label{28}
\eeq
The illustration of the necessity for this field redefinition in the
context of the heavy quark applications can be found in
Ref.~\cite{varenna}, Sect.~2.1.

\subsection{Applications to spectroscopy of heavy flavor hadrons}

To illustrate the consequences of the heavy quark Hamiltonian, let us
consider the masses of hadrons containing a single heavy quark. This
is not a dynamic question and requires only symmetry
properties of ${\cal H}_Q$. It is described in much detail in a number
of old reviews \cite{reviews}.

The mass of a hadron $H_Q$ is given by the expectation value of the
Hamiltonian:
\beq
M_{H_Q}\;=\; \frac{1}{2M_{H_Q}} \matel{H_Q}{{\cal H}_{\rm tot}}{H_Q}\;,
\label{27a}
\eeq
and we have
$$
{\cal H}_{\rm tot}\; =\;{\cal H}_{\rm light}  +{\cal H}_Q + m_Q\;,
$$
where we can expand
\beq
{\cal H}_Q\; =\;{\cal H}_{0} + \frac{1}{m_Q} {\cal H}_1 +
\frac{1}{m_Q^2} {\cal H}_2 +\, ...
\label{29a}
\eeq
with
$$
{\cal H}_{0}\;=\; -\varphi_Q^+ A_0 \varphi_Q \;
\stackrel{\rm QM}{\longrightarrow}\; -A_0(0)\left[\left\{x_{\rm
light}\right\}\right]\;,
$$
$$
\frac{1}{m_Q} {\cal H}_1 \;=\;
\,\frac{1}{2m_Q}\,({\vec\pi}^2 +
\vec\sigma \vec B)\;,
$$
\beq
\frac{1}{m_Q^2} {\cal H}_2 \;=\;
\frac{1}{8m_Q^2}\,\left[-(\vec D\vec E)+
\vec\sigma \cdot
\{\vec E\!\times\!\vec\pi-\vec\pi\!\times\!\vec E\} \right] \;,
\label{30}
\eeq
{\it etc.} 
Therefore,
\beq
M_{H_Q} = m_Q +\bar\Lambda + \frac{1}{2m_Q}
\frac{\matel{H_Q}{{\vec\pi}^2 \!+\!\vec\sigma\vec B }{H_Q} }
{2M_{H_Q}}+ ... =
m_Q +\bar\Lambda + \frac{(\mu_\pi^2 \!-\! \mu_G^2)_{H_Q}}{2m_Q}
+ ...
\label{32}
\eeq
The mass expansion was alternatively derived in Ref.~\cite{optical} 
using the formalism based on the trace of the energy-momentum tensor.

In Eq.~(\ref{32}) we introduced the notations $\mu_\pi^2$, $\mu_G^2$
for the expectation values of two $D\!=\!5$ heavy quark operators which
will often appear in our discussion:
$$ \mu_\pi^2 =
\matel{H_Q}{{\vec\pi}^2}{H_Q}_{\rm QM} \equiv
\frac{1}{2M_{H_Q}} \matel{H_Q}{\bar{Q}{\vec\pi}^2Q(0)}{H_Q}_{\rm QFT}
$$
\beq
\mu_G^2 = - 
\matel{H_Q}{\vec\sigma\vec B }{H_Q}_{\rm QM} \equiv
\frac{1}{2M_{H_Q}} \matel{H_Q}{\bar{Q}
\frac{i}{2}\sigma_{\mu\nu}G^{\mu\nu} Q(0)}{H_Q}_{\rm QFT}\;.
\label{34}
\eeq
The physical meaning of $\mu_\pi^2$ is quite evident:
the heavy quark inside $H_Q$ experiences a {\em zitterbewegung}
due to its coupling to light cloud.
Its average spatial momentum squared is $\mu_\pi^2$. The second
expectation value measures the amount of the chromomagnetic field
produced by the light cloud at the position of the heavy quark.
In principle, the actual heavy hadron states $H_Q$ depend on $m_Q$
{\it via} the $1/m_Q$-suppressed terms of the Hamiltonian. Therefore,
the above expectation values also have such terms. Often it is
convenient to consider the asymptotic values at $m_Q \ra \infty$, and
to use, correspondingly, the eigenstates of the $m_Q \!\ra\! \infty\,$
Hamiltonian ${\cal H}_0+{\cal H}_{\rm light}$.

The  parameter $\bar\Lambda$ appearing in
Eq.~(\ref{32})
was  introduced as a constant in the Heavy Quark Effective Theory
(HQET \cite{HQET}) in \cite{luke}; it is associated with those terms in the
effective Lagrangian  ${\cal L}(\mu )$ (disregarded so far) which are
entirely due to the light degrees of freedom. Needless to say that in
the Wilsonian approach $\bar\Lambda$ is actually $\mu$ dependent,
$\bar\Lambda (\mu )$.  Wherever there is the menace of confusion the
$\mu$ dependence of $\bar\Lambda $ will be indicated explicitly.

There exists an interesting expression for this parameter through the
anomaly in the trace of the energy-momentum
tensor, derived in Ref.~\cite{optical}:
\beq
\bar\Lambda = \frac{1}{2M_{H_Q}}\langle H_Q |
\frac{\beta (\alpha_s)}{4\alpha_s}G^2 +
\sum_q m_q (1\!+\!\gamma_m)\bar{q}q
|H_Q\rangle_{m_Q\rightarrow\infty} \;.
\label{defl}
\eeq
Here $\gamma_m$ is the anomalous dimension of the light quark mass.
This equation parallels the similar relation
for the nucleon mass \cite{GAN} well known in the chiral limit when all
quark masses are neglected:
$$
M_N = \frac{1}{2M_N}\langle N |
\frac{\beta (\alpha_s)}{4\alpha_s}G^2
|N\rangle\, .
$$
It should be noted, however, that the operator $G^2$ in
Eq.~(\ref{defl}), although local and gauge invariant, is not a 
{\it local heavy-quark} operator as they are understood in the framework
of the heavy quark expansion: the latter must be of the form
$\bar{Q}...Q$ where ellipses denote a local operator at the same
space-time point as the heavy quark fields $Q$.  Quantum mechanically,
the operators in Eq.~(\ref{defl}) describe the space integral over the
whole volume occupied by the heavy hadron, rather than the value of the
gluonic and quark fields at the position of the heavy quark.

The renormalization properties of the operator $G^2$ are quite
different in the sector of QCD with the heavy quark. An accurate
consideration reveals \cite{optical} that an additional term
$-\mu \frac{{\rm d}m_Q}{{\rm d}\mu}$ must be added in the r.h.s.\ of
Eq.~(\ref{defl}), where $\mu$ stands for the normalization point. The
normalization point dependence of the heavy quark parameters will be
discussed in more detail in the subsequent sections.

The value of $\La = \lim_{m_Q \to  \infty} \left(M_{H_Q}\!-\!m_Q\right)$ 
has the scale of $\Lam$ and depends on the state of light degrees of
freedom. These states generally carry spin $j$. In the limit $m_Q\ra
\infty$ the heavy quark spin decouples since the spin-dependent parts
are present only starting ${\cal H}_1$. Thus, the heavy flavor hadrons
can be classified not only by their total spin $J$ but by the spin of
light degrees of freedom $j$. It would be just the overall spin of a
hadron in the hypothetical world with the spinless heavy quarks
discussed in Sect.~1.1. Unless $j\!=\!0$, there are two values of the total
spin $J=j\pm \frac{1}{2}$. The corresponding states form `hyperfine'
multiplets and are degenerate up to $1/m_Q$ corrections. They are, for
example
$$
j=1/2 \;\; \left\{
\begin{array}{lll} D,\; & B\;\, & J=0\\
D^*, & B^*  & J=1
\end{array}
\right.\;,
\qquad \qquad
j=0\,,\;J=1/2 \qquad \Lambda_c\,,\;\,\Lambda_b
$$
For $\Lambda_Q$-baryons all spin is carried by the heavy quark (up to
$1/m_Q$ corrections). The observed spectroscopy of these states clearly
supports this picture:
$$
M_{\Lambda_b}-M_B \simeq 350\MeV\qquad\qquad M_{B_s}-M_{B^-} \simeq
90\MeV
$$
\beq
M_{\Lambda_c}-M_D \simeq 420\MeV\qquad\qquad M_{D_s}-M_{D^0} \simeq
104\MeV
\label{36}
\eeq

These relations are easily improved including $1/m_Q$ terms,
Eq.~(\ref{32}). The operator $\bar{Q} \vec{\pi}^{\,2} Q$ is
spin-independent and its expectation value is the same for all members
of a hyperfine multiplet. It does not split masses inside the multiplet.

The chromomagnetic operator $\bar{Q} \frac{i}{2} \sigma G Q = - 2
\vec{S}_Q \vec{B}(0)$ depends on the heavy quark spin $\vec{S}_Q$ and
thus lifts degeneracy leading to the hyperfine splitting among the
members of the multiplet. Restricted to a particular
hyperfine multiplet, the chromomagnetic field $\vec{B}(0)$ 
is proportional to the spin of light degrees of
freedom:
$
\vec{B} = c \cdot \vec{j}
$.
Therefore,
\beq
\aver{\vec\sigma \vec{B}} \;=\; 2c \aver{\vec{S}_Q \vec{j}\,}\;=\;
c\left(J(J\!+\!1)\!-\!j(j\!+\!1)\!-\!\frac{3}{4} \right)\;.
\label{38}
\eeq
It is easy to see that
\beq
\sum_{H_Q} \:\matel{H_Q}{\bar{Q} \vec{\sigma}\vec{B} Q}{H_Q}
\;\equiv \;
{\rm Tr}\,\bar{Q} \vec{\sigma}\vec{B} Q \;=\;0
\label{39}
\eeq
always holds if the summation is performed over a hyperfine
multiplet. Therefore, for example,
\beq
\mu_G^2(B) + 3 \mu_G^2(B^*)\;=\;0\;.
\label{40}
\eeq
In the $\Lambda_b$ family the expectation value of $\bar{b} \frac{i}{2}
\sigma G b$ vanishes.

So far most of the practical applications refer to $B$ mesons; as a
result, usually $\mu_\pi^2$ proper denotes the expectation value of the
kinetic operator just in $B$ or $B^*$. Likewise,  $\mu_G^2\equiv
\mu_G^2(B) = -3\mu_G^2(B^*)\,$.
Experimentally,
\beq
M_{B^*}\!-\!M_B\;\simeq \; \left(\frac{1}{3}+1\right)
\frac{\mu_G^2}{2m_b}\;=\; \frac{4}{3}\frac{\mu_G^2}{2m_b} \simeq
46\MeV\;.
\label{41}
\eeq
Neglecting the difference between
$M_B+M_{B^*}$ and $2m_b$ (which is formally an effect of higher order
in $1/m_b$), one can write
\beq
\mu_G^2\;\simeq \; \frac{4}{3}
\left(M_{B^*}^2\!-\!M_B^2\right) \;\simeq\; 0.36\GeV^2\;.
\label{42}
\eeq
In charmed mesons $M_{D^*}\!-\!M_D\simeq 140\MeV$ which agrees with the
fact that this hyperfine splitting is proportional to $1/m_c$.
The hadron mass averaged over a hyperfine multiplet, {\it e.g.}
$\overline{M}_B = \frac{3M_{B^*}+M_B}{4}$ is affected at order $1/m_Q$ by
only the kinetic energy term $\mu_\pi^2/2m_b$.

The mass expansion can be extended to higher orders in $1/m_Q$. For
example, to order $1/m_Q^2$ the hadron mass takes the form \cite{optical}
$$
M_{H_Q} = m_Q
+\overline\Lambda
+\frac{1}{m_Q}\,\left\{
\frac{1}{2M_{H_Q}}\langle H_Q |\bar Q \frac{(\vec\sigma\vec\pi)^2}{2}
Q|H_Q\rangle\right\}_{m_Q=\infty}\;+
$$
\begin{equation}
+\;
\frac{1}{8m_Q^2}\left\{\frac{1}{2M_{H_Q}}\langle H_Q |\bar Q
\left(-(\vec D\vec E) +2\vec\sigma\!\cdot\!\vec E\!\times\!\vec\pi\right)
Q|H_Q\rangle\right\}_{m_Q=\infty}
-\;\frac{\rho^3}{4m_Q^2}\;
+\;{\cal O}(m_Q^{-3})\;.
\label{opt26}
\eeq
The nonlocal (positive) correlator $\rho^3$ is the second-order
iteration of the $1/m_Q$ part of the Hamiltonian:
\beq
\rho^3 = i \int d^4 x\,
\; \frac{1}{4M_{H_Q}}\langle H_Q|T\{\bar Q(\vec\sigma\vec\pi)^2Q(x),
\,\bar Q (\vec\sigma\vec\pi)^2Q(0)\}|H_Q\rangle '_{m_Q=\infty}
\label{opt24}
\eeq
(the prime indicates that the diagonal transitions must be removed
from the correlation function).
In $B$ mesons the expectation values of the two terms in $1/m_Q^2$
Hamiltonian, the Darwin and the convection current 
interactions, are denoted as
$$
\rho_D^3\;=\;\frac{1}{2M_B}\,
\langle B |\,\bar b (-\frac{1}{2}\vec D \vec E) 
b\,|B\rangle\;=\;
\frac{1}{2M_{B^*}}\,
\langle B^* |\,\bar b (- \frac{1}{2}\vec D \vec E) b\,|B^*\rangle
$$
\begin{equation}
\rho_{LS}^3\;=\;\frac{1}{2M_B}\,
\langle B |\,\bar b\; \vec\sigma \!\cdot\! \vec E \!\times \!\vec \pi\;
b\,|B\rangle
\;=\;
-\frac{3}{2M_{B^*}}\,
\langle B^* |\,\bar b\; \vec\sigma \!\cdot\! \vec E \!\times\! 
\vec \pi\;
b\,|B^*\rangle\;\;.
\label{opt27}
\eeq
The Darwin operator by virtue of QCD equations of motions equals to the
local four-fermion operator of the form 
$-\frac{g_s^2}{2} \bar{Q}\frac{\lambda^a}{2} Q
\sum_q \bar{q}\gamma_0 \frac{\lambda^a}{2} q$. 

A similar decomposition can be made for the nonlocal correlators whose
expectation values were generically denoted by $\rho^3$. In particular,
in pseudoscalar and vector ground states
$$
\rho_{\pi\pi}^3=
i \int d^4 x\,
\; \frac{1}{4M_{B}}\langle B|T\{\bar b \vec\pi^{\,2}b(x),
\;\bar b\vec\pi^{\,2}b(0)\}|B\rangle '
$$
$$
\rho_{\pi G}^3=
i \int d^4 x\,
\; \frac{1}{2M_{B}}\langle B|T\{\bar b \vec\pi^{\,2}b(x),
\;\bar b\vec\sigma \vec B b(0)\}|B\rangle '
$$
\begin{equation}
\frac{1}{3}\rho_{S}^3\delta_{ij}\delta_{kl}+\frac{1}{6}
\rho_{A}^3(\delta_{ik}\delta_{jl} \!-\!
\delta_{il}\delta_{jk})=
i \int d^4 x\,
\; \frac{1}{4M_{B}}\langle B|T\{\bar b \sigma_i B_kb(x),\;
\bar b \sigma_j B_l b(0)\}|B\rangle '\;\;.
\label{opt28}
\eeq
Then one has for $\rho^3$ in $B$ and $B^*$, respectively,
\beq
\left(\rho^3\right)_B=\rho_{\pi\pi}^3+\rho_{\pi G}^3+\rho_{S}^3+
\rho_{A}^3\;\;\;,\;\;\;
\left(\rho^3\right)_{B^*}=\rho_{\pi\pi}^3-\frac{1}{3}\rho_{\pi G}^3+
\rho_{S}^3 - \frac{1}{3}\rho_{A}^3\;\;.
\label{opt29}
\eeq
These parameters also give the $1/m_Q$ corrections to the expectation 
values of the kinetic and chromomagnetic operators \cite{optical}.

In many applications one needs to know the difference between
$m_b$ and $m_c$.
This
difference is well constrained in the heavy quark expansion. For
example,
\beq
m_b-m_c=\frac{M_B\!+\!3M_{B^*}}{4}-
\frac{M_D\!+\!3M_{D^*}}{4} +
\frac{\mu_{\pi\!}^{\!2}}{2} \left(\!\frac{1}{m_{c\!}}\!-\!\frac{1}{m_b}\!\right) +
\frac{\rho_D^3\!-\!\bar\rho^{3\!}}{4}
\!\left(\!\frac{1}{m_c^2}\!-\!\frac{1}{m_b^2}\!\right) \,+\, {\cal
O}\!\left(\!\frac{1}{m^3}\!\right)\!.
\label{3.16}
\eeq
Here $\mu_\pi^2$ is the asymptotic expectation value of the kinetic
operator
and
$\bar\rho^3 \equiv \rho_{\pi\pi}^3\!+\!\rho_S^3$ is the sum of two
positive nonlocal correlators. As will be discussed
in the subsequent sections, all quantities in Eq.~(\ref{3.16}) but the
meson masses depend on the normalization point which can be
arbitrary, except that it must be much lower than
$m_{c,b}$. Evaluating the above expansion we arrive at 
\beq
m_b\!-\!m_c \simeq 3.50\GeV \,+\, 40\MeV\,
\frac{\mu_\pi^2\!-\!0.5\, {\rm GeV}^2}{0.1\,{\rm GeV^2}} \:+ \: \Delta
M_2\;, \qquad |\Delta M_2|\lsim 0.02\GeV \: ,
\label{3.17}
\eeq
where plausible assumptions about the $D\!=\!6$ expectation values have
been made.
The values of the hadronic parameters will be discussed later.
Let us mention that the $m_b\!-\!m_c$ estimate at
$\mu_\pi^2\approx 0.5\GeV^2$ appears to be in a good agreement with the
separate determinations of $m_b$ and $m_c$ from the sum  rules in
charmonia and $\Upsilon$.

\subsection{Heavy quark symmetry for formfactors}

The amplitudes of the semileptonic weak $b\ra c$ decays are described
by the corresponding transition formfactors. Typical semileptonic $b\ra
c$ decays as they look like in Feynman diagrams are shown in Figs.~1.
The hadronic part of the weak decay Hamiltonian mediating such decays
is
\beq
{\cal H}_{\rm weak} = \int \; {\rm d}^3 x\:
{\rm e}\, ^{-i\vec{q}\vec{x}}\:
\bar{c} \gamma_\mu(1\!-\!\gamma_5) b(x) \;.
\label{47}
\eeq
It says that $b$ with a momentum $\vec{p}$ is instantaneously replaced
by the $c$ quark with the momentum $\vec{p}-\vec{q}$. The resulting
state hadronizes into eigenstates of the Hamiltonian corresponding to
$m_Q\!=\!m_c$, that is, must be projected onto such states.

\thispagestyle{plain}
\begin{figure}[hhh]
 \begin{center}
 \mbox{\epsfig{file=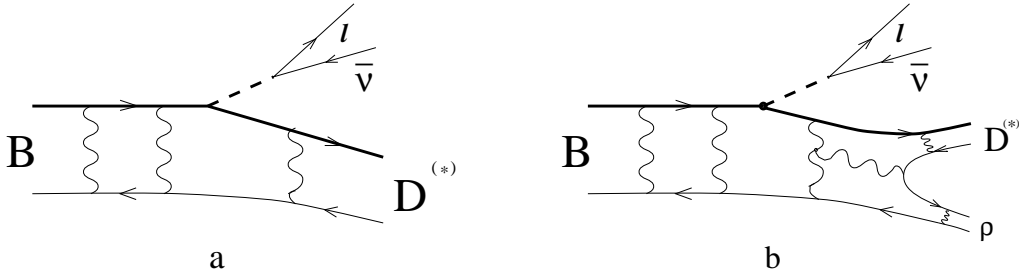,width=13.5cm}}
 \end{center}
\caption{ \small
Quark diagrams for the exclusive $B\ra D^{(*)}\:$
({\bf a}) and generic ({\bf b}) semileptonic decays.}
\end{figure}

The space-time picture of the decay is simple for heavy quarks. At 
$t\!<\!0$
the initial $b$ hadron is at rest and constitutes a coherent state of
light degrees of freedom in the static field of the heavy quark. At
$t\!=\!0$ the $b$ quark emits the lepton pair with the momentum $\vec{q}$
and transforms into a $c$ quark. The $c$ quark gets the recoil momentum
$-\vec{q}$ and starts moving with the velocity $\vec{v}=-\vec{q}/m_c$.
Such a state is not anymore an eigenstate of the Hamiltonian, and
afterwards undergoes nontrivial evolution. The light cloud can get a
coherent boost along the direction of $-\vec q$ and form again the same
ground-state, or excited hadron. Alternatively, it can crack
apart and produce a few-body final hadronic state.

The heavy quark symmetry {\it per se} cannot help calculating the
amplitudes to create such final states. However, it tells one that the
hadronization process does not depend on the heavy quark spin, or on the
concrete value of the mass $m_Q$ but rather on the velocity of the final
state heavy hadron. This velocity cannot change in the process of
hadronization if only soft gluons are exchanged between the heavy quark
and the light cloud.  This independence holds, of course, only when
$m_Q$ is very large (and the final state quark does not move too fast);
there are various $1/m_Q$ corrections at finite masses.

Let us consider, for example, the ground state transition $B\ra D$. Its
amplitude depends on the velocity $\vec{v}$ of $D$, $f(\vec{v}^{\,2})$.
The very same function would describe also decays $B\ra D^*$, or
the elastic amplitude of scattering of a photon on the $b$ quark  $B\ra
B$, $B\ra B^*$. Moreover, in the proper normalization $f(0)=1$ holds.
This fact follows from the conservation of the $b$-quark vector
current, for the amplitude at zero momentum transfer measures the total
`$b$-quark charge' of the hadron $n_b \!-\!n_{\bar{b}}$. Its origin is
simply understood: if $\vec{v}\!=\!0$, the final state is not really
disturbed by movement of the static source. A nontrivial rearrangement
of the light cloud for $B\ra D$ is associated only with
the mass-dependent terms in ${\cal H}_Q$ vanishing when $m_Q \ra
\infty$.

Consider the vector $\bar{b} \gamma_\mu b$ current in $B$ meson.
It is described by the single formfactor $f_+(q^2)$:
\beq
\matel{B(p')}{\bar{b}\gamma_\mu b(0)}{B(p)}\;=\;f_+(q^2)\,
(p+p')_\mu\;, \qquad \qquad q_\mu = p_\mu - p'_\mu
\label{49}
\eeq
(the second structure $(p\!+\!p')_\mu$ is forbidden by $T$ invariance or
current conservation). The value $f_+(0)$ measures the total `beauty
charge' of the hadron and is not renormalized by the strong
interaction, $f_+(0)=1$.\footnote{For simplicity, we use the convention
that $B$ consists of $b$ and $\bar{q}$. That is, $B^-$ is a $B$ meson
while $B^+$ is $\bar{B}$.}
Passing to the velocities, we use instead of
$q^2$ the scalar product $v_\mu v'_\mu$:
\beq
(vv')\;=\; \frac{(pp')}{M_B^2} \;=\; 1- \frac{(q^2)}{2M_B^2} \,\ge\,
1\;,
\label{50}
\eeq
and
\beq
f_+(q^2)\;=\; \xi(vv')\;, \;\; \xi(1)=1\;.
\label{51}
\eeq
$\xi(vv')$ is called the Isgur-Wise function. The heavy quark symmetry 
then states that
\beq
\matel{D(v')}{\bar{c}\gamma_\mu b(0)}{B(v)}\;=
\left(
\frac{M_B\!+\!M_D}{2\sqrt{M_BM_D}}
(p\!+\!p')_\mu\; -  \frac{M_B\!-\!M_D}{2\sqrt{M_B M_D}}
(p\!-\!p')_\mu \right) \,\xi(vv')\;.
\label{52}
\eeq
$D^*$ differs from $D$ only by the alignment of the $c$ quark spin.
Taking this into account yields
$$
\matel{D^*(v',\epsilon)}{\bar{c}\gamma_\mu  b(0)}{B(v)}\;=\;
i\epsilon_{\mu\nu\al\be}\epsilon^*_\nu v'_\al v_\be \,\xi(vv') \:
\sqrt{M_B M_D^*}
$$
\beq
\matel{D^*(v',\epsilon)}{\bar{c}\gamma_\mu \gamma_5 b(0)}{B(v)}\;= \;
\left\{\epsilon^*_\mu (vv'\!+\!1)-v'_\mu(\epsilon^* v)\right\}
\,\xi(vv')\:\sqrt{M_B M_D^*}\;.
\label{54}
\eeq
It is important that these relations are valid in the limit $m_{b,
c}\ra \infty$ and if no short-distance radiative corrections were
present.  The corrections to the symmetry limit are minimal at
$\vec{q}=0$ ($v=v'$, the so-called zero recoil point); numerically the
value of the axial formfactor was estimated to be $F_{B\ra
D^*}(0)\simeq 0.9$ \cite{vcb}.  The corrections at arbitrary $\vec{v}
\sim 1$ are generally significant.

\subsection{Feynman rules at $m_Q\ra \infty$}

The Feynman rules for heavy quarks are usual propagators and vertices
for nonrelativistic particles where $1/m \ra 0$. If the heavy quark
momentum is $p_\mu=(m_Q+\omega, \; \vec{p}\,)$ then
$$
G(p)\;=\;\lim_{m\ra\infty} \, \delta_{\al\be}\,
\frac{1}{\omega-i\epsilon-\frac{\vec{p}^{\,2}}{2m} }\, \delta_{ij}\;=\;
\frac{\delta_{\al\be}}{\omega-i\epsilon }\, \delta_{ij}
$$
\beq
\Gamma_\mu\;=\; g_s \frac{\lambda^a_{\al\be}}{2} \delta_{\mu 0}\;.
\label{57}
\eeq
Here $a$, $\al$, $\be$ are color indices and spinor indices
$i,j$ take values $1$ or $2$. (It is often advantageous
to keep the nonrelativistic
term $\vec{p}^{\,2}/2m_Q$ in the propagator as an infrared regulator.) 
These rules follow
immediately from the static Lagrangian
$$
{\cal L}_Q\;=\; \varphi_Q^+ iD_0 \varphi_Q\;, \qquad
D_0=\partial_0-i A_0^a \frac{\lambda^a}{2}\;.
$$
The same nonrelativistic system can be considered in an arbitrary
moving frame, where one can write
\beq
{\cal L}_v\;=\; \varphi_v^+ i(vD) \varphi_v\;.
\label{39a}
\eeq
Instead of the nonrelativistic spinor $\varphi_Q$ one then considers the
``bispinor'' $\varphi_v(x)= \frac{1+\not\!{\:v} }{2}\tilde Q(x)$; the
propagator is written as
\beq
\frac{1+\not\!\!{\,v} }{2}
\frac{m_Q\!+\!\not\!\!{\,p}-\not\!\!{\,k}}
{m_Q^2\!-\!(p\!-\!k)^2\!-\!i\epsilon}
\frac{1\!+\!\not\!\!{\,v} }{2}
\;\longrightarrow \;
\frac{1\!+\!\not\!\!{\,v}}{2}  \frac{1}{vk\!-\!i\epsilon}
\label{60}
\eeq
$$
\Gamma_\mu\; =\; g_s \frac{\lambda^a}{2} v_{\mu}\;,
$$
where $k=p\!-\!m_Qv$. Of course, such a generalization can be
useful only if the initial and final state hadrons have
different velocities, $\vec v \!\ne \!\vec v\,'$. In that case the external
(`weak') source carries a large momentum $\vec{q} \sim m_Q \,\delta
\vec{v}\,$.

\subsubsection{Subtleties of actual QCD}

It must be noted, however, that the field-theoretic description of
processes where the heavy quarks change velocity becomes subtle if the
quantum radiative corrections are really incorporated. Any change in the
velocity of a static source leads to actual radiation of real hadrons
with momenta $\vec{k}$ all the way up to $m_Q$:
\beq
\frac{{\rm d}w}{{\rm d}\omega}
\;\sim \frac{\alpha_s(\omega)}{\omega} \: (\vec v\,'\!-\! \vec v)^2
\;;
\label{62}
\eeq
here $\omega$ denotes the radiated energy. This means that with
infinitely heavy quarks one would actually observe radiation of
light hadrons off the heavy flavor hadrons with arbitrary large
energies.  Without an ultraviolet cutoff any exclusive transition
probability would vanish being suppressed by the more or less universal
factor (the square of the non-Abelian analogue of the nonrelativistic
Sudakov formfactor)
\beq
S \;\sim\;  {\rm e}\,^{-\frac{4\as}{9\pi}(\Delta \vec v)^2
\ln{\frac{\mu}{\epsilon}} }\;.
\label{64}
\eeq
Here $\mu$ denotes the ultraviolet cutoff and $\epsilon$ determines the
energy resolution for the final hadronic state (for the
validity of the perturbative expansion $\epsilon$ must be much large
than $\Lam$; taking it around the typical hadronic scale gives the
estimate of the overall perturbative suppression of the exclusive
transition probability). This factor has a meaning of the probability
of the heavy color source {\it not to emit} gluons in the interval of
energies between $\epsilon$ and $\mu$ in the act of acceleration.

Beyond the small velocity approximation the factor $(\Delta \vec v)^2$
in Eq.\,(\ref{64}) must be replaced by 
$\frac{3}{2}\left(\frac{1}{|\Delta\vec{v}\,|}
\ln{\frac{1+|\Delta\vec{v}\,|}{1-|\Delta\vec{v}\,|}}-2\right)$ which can 
be read off the
expression for the intensity of electromagnetic radiation in classical
electrodynamics:
\beq
\frac{1}{\omega}\frac{{\rm d}I(\omega)}{{\rm d}\omega} \;=\;
\frac{\alpha}{\pi}
\left(\frac{1}{|\Delta\vec v\,|}
\ln{\frac{1\!+\!|\Delta\vec v\,|}{1\!-\!|\Delta\vec v\,|} } 
-2 \right) \frac{1}{\omega}
\;\,.
\label{new43a}
\eeq
It is remarkable that this law is not modified by quantum corrections
\cite{landaul} as long as $\omega$ remains much smaller than the masses
of all charged particles and the effects of their virtual production
can be neglected.

In the non-Abelian theories like QCD the situation is different, even in
the dipole approximation $|\vec{v}\,| \ll 1$.
The non-Abelian QCD dipole radiation was considered in detail in
Ref.~\cite{dipole}. In contrast with QED where the coupling for the soft
photon
radiation is exactly $\alpha_{\rm em}(0)$ and is not renormalized by
quantum corrections, in QCD the dipole radiation is governed by the 
effective
coupling $\alpha_s^{(d)}(\omega)$ which runs with the energy scale:
\beq
\frac{\alpha_s^{(d)}(\omega)}{\pi}\;=\;
\frac{\bar\alpha_s\left({\rm e}^{-5/3+\ln{2}} \omega\right)}{\pi}
- N_c \left(\frac{\pi^2}{6} -\frac{13}{12} \right)
\left(\frac{\as}{\pi}\right)^2 \,+\,{\cal O}\left(\as^3\right)\;,
\label{new43b}
\eeq
with $\bar\alpha_s$ the $\overline{\rm MS}$ coupling.
It can be shown that the second (conformal) term in the ${\cal
O}(\alpha_s^2)$ part of the coupling must equal twice the
corresponding part of the so-called cusp anomalous dimension of the
Wilson lines (at small angle) 
investigated in detail in the mid 80's \cite{radkorch}. The
dipole radiation coupling governs the normalization point evolution of
the heavy quark masses and a number of heavy quark operators in the
effective low-energy theory.

As any effective coupling, $\alpha_s^{(d)}(\omega)$ is not a purely
perturbative object at arbitrary precision. It has a power-suppressed
nonperturbative component at arbitrarily large energy $\omega$ which,
for example, depends on the particular type of the heavy flavor hadron
which is accelerated. (Pure perturbation theory does not depend on the
light cloud surrounding heavy quark  for actual hadrons.) Using the OPE
approach it was shown \cite{dipole} that such nonperturbative effects in
the dipole radiation fade out at least as $1/\omega^3$, and for the
ground state hadrons ($B$ mesons) it was estimated that
\beq
\left( \delta \alpha_s^{(d)}(\omega) \right)_{\rm nonpert}
\approx -\left(\frac{0.6\GeV}{\omega}\right)^3\;.
\label{new43c}
\eeq

In general, the perturbative factors suppressing the velocity-changing
transition amplitudes of infinitely heavy quarks are universal and
related to the cusp anomalous dimensions of Wilson lines describing
propagation of massive color objects. They appear in the
renormalization of bent Wilson lines; the cusps correspond to
instant changes in velocity and thus lead to new ultraviolet
divergences. The dedicated discussion can be found in the original
publications \cite{radkorch}. The nonperturbative aspects of
factorization in the radiative effects have not been
carefully studied yet.

\section{Basic Parameters of the Heavy Quark Expansion}

\subsection{The heavy quark mass}

\subsubsection{What is $m_Q$?}

In quantum field theory the object we begin our work with is the
Lagrangian formulated at some high scale $M_{ 0}$.
The mass $m_0$  is a parameter in this
Lagrangian; it enters on the same footing as, say,
the bare coupling constant  $\alpha_s^{(0)}$ with the only difference being
that it
carries dimension. As with any other
coupling, $m_0$ enters all observable quantities in a certain
combination with  the
ultraviolet cutoff $M_{0}$, which is universal for a renormalizable
theory.

The mass parameter $m_0$ by itself is not observable, like $\as^{(0)}$.
For calculating observable quantities   at the scale $\mu \ll  M_{0}$
it is usually convenient to relate $m_0$ to some  mass  parameter
relevant to the scale $\mu$. Integrating out momentum scales above
$\mu$ converts $\as^{(0)}$ into $\as(\mu)$ -- and likewise $m_0$ into
$m(\mu)$.  Such $m(\mu)$ is not something absolute since depends on
$\mu$. It is either used on the same footing as $\as(\mu)$ or, in the
final expressions, is eliminated in favor of some suitable observable
mass.  For example, in quantum  electrodynamics (QED) at low energies
({\it i.e.\ }$E\ll m_e$) there is  an obvious ``best" candidate: the
actual mass of an  isolated  electron, $m_e$. In  the perturbative
calculations it is determined  as the position of  the pole in the
electron Green function (more exactly, the beginning of the cut). The
advantages are evident:  $m_e$ is  gauge-invariant and experimentally
measurable.

The analogous  parameter for heavy quarks in QCD is
referred to as the pole quark mass, the position  of the pole of the
quark Green function. Like $m_e$ it is gauge  invariant.  Unlike QED,
however, the quarks do not exist as isolated objects (there are no
states with the quark quantum numbers in the  observable spectrum, and
the quark Green function beyond a given order has neither a pole nor a
cut).  Hence, $m^{\rm pole}$ cannot be directly measured;  $m^{\rm
pole}$ exists only as a theoretical construction.

In principle, there is nothing wrong with using $m^{\rm pole}$
{\em in perturbation theory} where it naturally appears
in  the Feynman graphs
for the quark Green functions, scattering amplitudes and so on. It
may or may not be convenient, depending on concrete goals.

The pole mass in QCD is perturbatively infrared stable, order by
order, like in QED (the formal proof was given recently in 
\cite{kronf}). 
It is well-defined to every given
order in perturbation theory. One cannot define it to all orders,
however; the sum of the series does not converge to a definite number.
In a sense, the pole mass is {\em not}
infrared-stable nonperturbatively.
Intuitively this is clear: since the quarks are
confined in the full theory, the best one can do  is to define the
would-be pole position with  an intrinsic uncertainty of order 
$\Lam$ \cite{gurman}.

Based on experience in QED or ordinary QM, non-existence of
$m_Q^{\rm pole}$ may seem counter-intuitive.
Employing perturbation theory, we start with the free quark
propagator
$$
G(p)= \frac{1}{m_Q(\mu)-\!\not\!\!{\,p} }
$$
($m_Q(\mu)$ is the parameter entering the Lagrangian)
which has a pole at $p^2= m_Q^2(\mu)$ and, therefore, describes a
particle with mass $m_Q(\mu)$. Accounting for the gluon exchanges to
the first order in $\as$ adds the diagram Fig.~2 and the pole moves
to $p^2\simeq \left(m_Q(\mu)\!+\!\frac{4\as}{3\pi}\mu\right)^2 $ 
describing now a particle with the mass $m_Q^{(1)}\simeq
 m_Q(\mu)+\frac{4\as}{3\pi}\mu\,$, {\it etc}. In any order of 
perturbation theory we see a quark pole and the corresponding particle
with certain mass, differing from the mass in the Lagrangian. This mass
is just the ``pole'' mass. Clearly, it depends on the order of
perturbation theory one considers, and on a concrete version of
the employed expansion:
\beq
m_Q^{(k)} \;=\; m_Q(\mu )\;\sum_{n=0}^{k}\,
C_n\left(\frac{\mu}{m}\right) \,\left(\frac{\as (\mu )}{\pi}\right)^n
\; , \;\;\; C_0=1\;.
\label{100}
\eeq
It is tempting to define the `actual' pole mass as a sum of the series
(\ref{100}).
It
appears, however, that the sum does not converge to a reasonable number
and cannot be defined with the necessary accuracy in a motivated way,
but suffers from an irreducible uncertainty of order $\Lam$.

Before explaining the origin of this perturbative uncertainty, let us
remark that such a definition of the quark mass is nothing but equating
it with the lowest eigenvalue of the (perturbative) Hamiltonian in the
sector with the number of heavy quarks $1$. The eigenvalues of the
Hamiltonian, however, are {\it not} short-distance quantities, and
their calculation in the perturbative expansion is not adequate. This
is in the nice correspondence with the observed fact that no isolated
heavy quark exists in the spectrum of hadrons which contains instead
$B,\,B^{**},\, \Lambda_b,\, B_s\,$\ldots with masses differing
by amount ${\cal O}(\Lam^1)$.

The physical reason behind the perturbative instability of the long-distance
regime is the growing of the interaction strength $\as$. One can
illustrate this instability in the following
way.  Consider the energy stored in the
chromoelectric field in a sphere of radius $R \gg 1/m_Q$
around a static color source,
\beq
\delta {\cal E}_{\rm Coul} (R) \propto
\int _{1/m_b \leq |x| < R} d^3x {\vec E}^{\,2}_{\rm Coul}
\propto {\rm const} - \frac{\as (R)}{\pi} \frac{1}{R}\, .
\label{102}
\eeq
This energy is what one adds to the bare mass of a heavy particle to
determine what will be the mass including QCD interactions.
The definition of the pole mass amounts to setting
$R \ra \infty$; {\it i.e.},
in evaluating the pole mass one undertakes to integrate the
energy density associated with the color source over
{\em all} space assuming that it has the Coulomb form.
In real life the color interaction becomes strong at
$R_0 \sim 1/\Lam$; at such distances the
chromoelectric field
has nothing to do with the Coulomb tail. Thus, one cannot include the
region beyond $R_0$ in a meaningful way. Its contribution
which is of order $\Lam$, thus, has to be considered
as an irreducible uncertainty which is power-suppressed relative
to $m_Q$,
\beq
\frac{\delta_{\rm IR}m_Q^{\rm pole}}{m_Q}\; = \;
{\cal O}\left(\frac{\Lam}{m_Q}\right)\; .
\label{3.6}
\eeq

\thispagestyle{plain}
\begin{figure}[hhh]
 \begin{center}
 \mbox{\epsfig{file=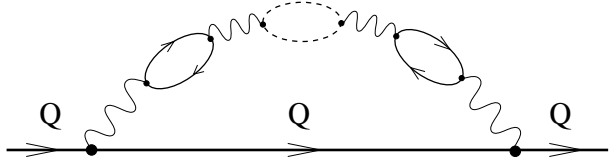,width=8.cm}}
 \end{center}
\caption{ \small
Perturbative diagrams leading to renormalization of the heavy quark
mass. The contribution of the gluon momenta below $m_Q$ expresses the
classical Coulomb self-energy of the colored particle. The number of
bubble insertions into the gluon propagator can be arbitrary
generating corrections in all orders in $\as$. The factorial growth
of the coefficients
produces the IR renormalon uncertainty in $m_Q^{\rm pole}$ of order
$\Lam$.}
\end{figure}

Exactly this behavior is traced formally in perturbation theory. In
the nonrelativistic
regime where the internal momentum $|\vec{k}\,|\ll m_Q$ the
expression for the diagram Fig.~2 is simple,
\beq
\delta m_Q \sim - \frac{4}{3}
\int \frac{d^4k}{(2\pi )^4 i k_0} \frac{4\pi \alpha_s}{k^2} =
\frac{4}{3}
\int \frac{d^3\vec k}{4\pi ^2} \frac{\alpha_s}{\vec k^2}\, .
\label{3.2}
\eeq
The latter expression is $\frac{1}{2}V(0)$ with $V(R)$ the usual
Coulomb potential between two quarks.
The running of the coupling is generated by dressing the
gluon propagator by virtual pairs and leads to
\beq
\delta m_Q \simeq
\frac{4}{3}
\int \frac{d^3\vec k}{4\pi ^2} \frac{\alpha_s (\vec k^2)}{\vec k^2}\; .
\label{3.2a}
\eeq
Since
\beq
\alpha_s(k^2) =
{\alpha_s(\mu ^2)}\left\{1 + \frac{\alpha_s(\mu ^2)}{4\pi} b
\,\ln{\frac{k^2}{\mu ^2}}\right\}^{-1} , \; \; b = \frac{11}{3} N_c -
\frac{2}{3}n_f \, ,
\label{run}
\eeq
we can expand
$\alpha_s(k^2)$ in a power series in $\alpha_s(\mu ^2)$ and easily
find  the $(n\!+\!1)$-th order contribution to $\delta m_Q\,$,
\beq
\frac{\delta m_Q^{(n+1)}}{m_Q} \sim \frac{4}{3}\,
\frac{\alpha_s(\mu)}{\pi} \, n!  \left(
\frac{b \alpha_s(\mu)}{2\pi}\right) ^n \; .
\label{3.4}
\eeq
The coefficients grow factorially and contribute with the same sign.
Therefore, one cannot define the sum of these contributions even using
the trick with the Borel transformation.  The best one can do is to
truncate the series judiciously. An optimal truncation leaves us with
an irreducible uncertainty $\sim {\cal O}(\Lam)$
\cite{pole,bbpole}. The above perturbative corrections are example of
the so-called infrared renormalons \cite{renorm,recent}.

This uncertainty can be quantified. A formal Borel resummation of such
non-summable series leads to the result which literally has an imaginary
part which can be taken as a measure of the uncertainty. The
imaginary part for the series Eq.~(\ref{3.4}) is
\beq
\Im m_Q^{\rm pole}\;=\; \frac{8\pi}{3b}{\rm e}\,^{5/6}\,
\Lam^{\overline{\rm MS}}\;.
\label{103}
\eeq
It became conventional to assign the formal imaginary part divided by
$\pi$ to the irreducible uncertainty. Even with this minimal choice
\beq
\delta m_Q^{\rm pole}\;=\; \frac{1}{\pi}
\left| \Im m_Q^{\rm pole}\right| \;=\;
\frac{8}{27}{\rm e}\,^{5/6}\,
\Lam^{\overline{\rm MS}}\;\simeq \;0.7 \Lam^{\overline{\rm MS}}\;.
\label{103a}
\eeq

Thus, the perturbative
expansion {\em per se} anticipates the onset of the nonperturbative
regime (the impossibility of locating the
would-be quark pole to accuracy better than $\Lam$).  Certainly, the
concrete numerical value of the uncertainty
in $m^{\rm pole}$
obtained through renormalons is not trustworthy.
The renormalons do not represent the dominant
component of the infrared dynamics.
However, they are a clear indicator of the presence of
the
power-suppressed nonperturbative effects, or infrared instability of
$m^{\rm pole}$; the very fact that there is a correction ${\cal
O}(\Lam /m_Q)$ is beyond any doubt.

It is worth noting that the pole mass was the first example
where a quantity which is perturbatively infrared-stable was shown
not to be stable nonperturbatively at the level $\Lam^1$.
The observation of Refs.~\cite{pole,bbpole} gave impetus
to dedicated analyses of other perturbatively infrared-stable
observables
in numerous hard processes  without OPE, in particular, in  jet
physics. Such nonperturbative infrared contributions, linear in
$\Lam /Q$ were indeed  found shortly after
in thrust and many other jet characteristics (for a review and a
representative list of references see e.g. \cite{recent}).

Since it is impossible to relate $m_Q(\mu)$ and $m_Q^{\rm pole}$ to the
necessary accuracy, it is clear that either $m_Q^{\rm pole}$ or
$m_Q(\mu)$ must be irrelevant for the $1/m_Q$ expansion, for example,
for calculating the decay widths of heavy flavors which depend on
high powers of the quark masses.
Which mass must be used then? This question was formulated and
answered in the early 1994 in Ref.~\cite{pole}. In agreement with the
qualitative discussion given above, the answer is: $m_Q^{\rm pole}$ is
irrelevant and must be replaced everywhere by $m_Q(\mu)$. For the
inclusive decay widths, this can be understood as the perturbative
counterpart of the QCD theorem \cite{buv} stating that the mass gap $\sim
{\cal O}(\Lam^1)$ differentiating the hadron mass from the quark mass, 
does not affect the width -- the pole mass dressed by soft interactions 
is the perturbative analogue of the hadron mass.
The $1/m_Q$ infrared renormalon uncertainty in the pole mass 
disappears from the perturbative corrections when the width is
expressed in terms of the short-distance heavy quark mass; it is 
present when one attempts to compute the width through the pole mass.

Another transparent way to illustrate irrelevance of the pole mass is
to vary the number of space dimensions, {\it viz.\ }to descend to the (2+1)
theory. In $D\!=\!3$ the (pole) mass logarithmically diverges in the
infrared, cf.\ Eq.\,(\ref{3.2}). The OPE, at the same time guarantees
that the total widths are infrared safe in any dimension, so this
infrared part cannot affect the width. The infrared divergence of the
mass in $D\!=\!3$ is simply reflection of the Coulomb potential
$\ln{R}$ growing at large $R$. It is evident that total decay
probabilities of a heavy quark must know nothing about the behavior of 
the interaction at infinite distance.

In the OPE,
the infrared part of the pole mass is not related to any local operator
$\bar{Q} O_i Q$, and does not enter any observable calculable in the
short-distance expansion. Since this infrared piece does not enter
observables, it cannot, in turn, be determined experimentally. The
numerical instability of various attempts to pinpoint the value of
$m_Q^{\rm pole}$ is a result of using the perturbative expansion for the
effects originating from the nonperturbative domain.
The above facts were later illustrated in Ref.~\cite{bbz} in
a concrete model for the higher-order corrections to the semileptonic
widths obtained via the so-called bubble resummation of the one-loop
perturbative diagrams. 

It is important to note that the irrelevance of the pole mass goes
beyond the problems with the infrared renormalon contributions illustrated
above. Even if there existed some way to define reasonably the sum of
the pure perturbation series for $m_Q^{\rm pole}$, or asymptotic states
with single-quark quantum numbers with finite energy existed in a
strong-interaction theory like QCD, this mass still would have been
inadequate for constructing the effective field theory, to the extent
the difference with $m_Q(\mu)$ cannot be neglected numerically.

Summarizing, the pole mass {\em per se} does  not appear in  OPE for
infrared-stable quantities.  Such expansions  operate with the
short-distance (running)  mass. Any  attempt to
express the  OPE-based results in terms of the pole mass creates a
problem   making the Wilson coefficients ill-defined theoretically and
poorly convergent numerically.

A properly constructed perturbative treatment suitable for the
$1/m_Q$ expansion
incorporates only gluons with $|\vec{k}| \gsim \mu$ which are not
`resolved' and are included into the heavy quark field wavefunction
corresponding to the heavy quark field $Q_{(\mu)}(x)$ normalized at the
scale $\mu$. The normalization point $\mu$ can be changed: descending
from $\mu$ down to $\mu_1<\mu$ one has to integrate out newly-unresolved
gluons with $\mu_1\!<\!|\vec{k}| \!<\! \mu$. For example, the Coulomb field
associated with such gluons increases the mass of the quark by the
amount
\beq
\delta m_Q=
\int_{\mu_1<|\vec{k}\,|<\mu} \frac{d^3\vec k}{4\pi ^2} \,\frac{4}{3}\,
\frac{\alpha_s (\vec k^2)}{\vec k^2}\; .
\label{104}
\eeq
The pole mass clearly appears when $\mu_1\!\to \! 0$. From the OPE point of
view, it is an attempt to construct an effective theory with the
normalization scale $\mu\!=\!0$ formulated, nevertheless, still in terms
of quarks and gluons.  Speaking theoretically, one can imagine a limit
of small $\mu$ which would correspond to integrating out all modes down
to $\mu\!=\!0$ in evaluation of the effective Lagrangian. It would be
nothing but constructing the $S$-matrix of the theory from which one
can directly read off all conceivable amplitudes. Clearly, it could
have been formulated in terms of physical mesons and baryons but not of
quarks and gluons.

In the Wilson OPE one uses $m_Q(\mu)$ with $\Lam \ll \mu\ll m_b$. 
We illustrate later that just such a mass can be accurately measured in
experiment. It {\it is} $\mu$-dependent:
\beq
\frac{{\rm d}m_Q(\mu)}{{\rm d}\mu}\; =\; -\;\frac{16}{9}
\frac{\as(\mu)}{\pi}\;- \;
\frac{4}{3} \frac{\as(\mu)}{\pi} \frac{\mu}{m_Q}
\; + \; {\cal O} \left( \as^2, \as\frac{\mu^2}{m_Q^2}\right)\;;
\label{112}
\eeq
the higher order perturbative corrections were computed recently
\cite{dipole}.
There are different schemes for defining $m_Q(\mu)$ (similar to
renormalization schemes for $\as$), and the coefficients above are
generally different there. As long as a concrete scheme is adopted,
there is no ambiguity in the numerical value of $m_Q(\mu)$.
Instead of the HQET parameter $\La$ in QCD one has
$\La(\mu)=\lim_{m_Q\ra \infty} M_{H_Q}\!-\!m_Q(\mu)$. The value of
$\La(\mu)$ is of the hadronic mass scale if $\mu$ does not scale with
$m_Q$.

There exists a popular choice of a short-distance mass, the so-called
$\overline{\rm MS}$ mass $\bar{m}(\mu)$. The $\overline{\rm MS}$ mass
is not a parameter in the effective
Lagrangian;
rather it is a certain {\em ad hoc} combination of the  parameters
which is
particularly convenient in the perturbative calculations using
dimensional
regularization. Its relation to the perturbative pole mass 
since recently is known already to
three loops \cite{melnikov}:
\beq
m_Q^{\rm pole}\;=\;\bar{m}_Q(\bar m_Q)\left\{1+\frac{4}{3}
\frac{\as(\bar m_Q)}{\pi} + (1.56\,b
-3.73)\left(\frac{\as}{\pi}\right)^2
+...\, .
\right\}
\label{m15}
\eeq
At $\mu \gsim m_Q$ the $\overline{\rm MS}$ mass coincides,
roughly speaking,
with the running Lagrangian  mass seen at the scale $\sim
\mu$. However, it becomes rather meaningless at  $\mu  \ll  m_Q$:
\beq
\bar m_Q(\mu) \;\simeq\;\bar{m}_Q(\bar m_Q)\left\{1+\frac{2\as}{\pi}
\ln{\frac{m_Q}{\mu}} \right\}
\; .
\label{114}
\eeq
It logarithmically diverges when $\mu/m_Q \ra 0$.
For this reason $\bar m(\mu)$ is not
appropriate
in  the  heavy quark theory where  the possibility of
evolving down to a low normalization  point,  $\mu \ll  m_Q$,
is crucial. Otherwise, for example, $M_{H_Q}\!-\!\bar m_Q \propto m_Q$ and
does not stay constant in the heavy quark limit.

The reason for this IR divergence is that the $\overline{\rm MS}$
scheme technically attempts to determine the (perturbative) running of all
quantities by their divergences calculated when the space-time
dimension approaches $D\!=\!4$.  The divergence can emerge only at $k \ra
0$ or $ k\ra \infty$. For the mass, the IR divergences are absent, and
the dimensional regularization is sensitive to the UV divergence
at $k\gg m_Q$. Studying only $1/(D\!-\!4)$ singularities, it is
unable to capture the change of the regime at $\mu \lsim m_Q$ and
assumes the same running in this domain. The
actual running below $m_Q$ is slower.

The properly defined short-distance masses always exhibit an
explicit linear $\mu$-dependence similar to Eq.~(\ref{112}) at $\mu
\ll m_Q$.  The perturbative pole mass, order by order, would correspond
to the limit $\mu \ra 0$. However such a limit does not exist.

Since $m_Q^{\rm pole}$ does not exist as a well-defined mass parameter,
a different, short-distance mass must be used. The normalization point
$\mu$ can be arbitrary as long as $\mu \gg \Lam$. It does not
mean, however, that all masses  are equally
practical,  since the  perturbative series are necessarily truncated
 after a few  first  terms. Using an inappropriate scale makes
numerical approximations bad. In particular, relying on
$\bar{m}_Q(m_Q)$
in treating the low-scale observables can be awkward. The
pedagogical
example illustrating this point can be found in
Refs.~\cite{rev,varenna}.

Needless to say, it is the
high-scale masses that appear directly in the processes at high
energies. In particular, the inclusive width $Z\to b \bar b$ is
sensitive to $m_b(M_Z)$;  using $\overline{\rm MS}$  mass normalized at
$\mu\!\sim \!M_Z$ is appropriate here.  On the contrary, the inclusive
semileptonic decays $b\ra c\,\ell \nu$ are rather low-energy in this
respect \cite{five}; this is true even for $b\to u$.

The construction of the running low-scale heavy quark mass suitable for
the OPE to any order in perturbation theory can be done in a
straightforward way. However, the Wilsonian approach implies
introducing a cutoff on the momenta of the gluon fields. Since gluon
carries color, its momentum is not a gauge-invariant quantity, and such
a mass typically looks not gauge-invariant. More accurately,
obtaining the same $m_Q(\mu)$ requires somewhat different
cutoff rules in different gauges.

Even though this is not a real
problem for the theory, it is often viewed as a disadvantage. To get
rid of this spurious problem, a manifestly gauge-invariant definition
of the running mass was suggested in \cite{five,blmope} which is
formulated only in terms of observables. The idea is to
explicitly subtract from the pole mass the infrared pieces it
contains. These are uniqually determined by the so-called
small velocity (SV) sum rules in the heavy quark limit discussed in
Sect.~4. They are
moments of the structure functions of the infinitely heavy quark
in the process where its velocity changes by a small amount. In the
presence of hard gluons these moments diverge in the ultraviolet, and
must be cut off at some energy $\mu$. This cutoff enters then as the
normalization point for the heavy quark mass.

Such defined mass is
convenient for the OPE in the $1/m_Q$ expansion since more or less
directly enters many relevant processes, {\it e.g.\ }heavy flavor
transitions or the threshold heavy flavor production. Its relation to
the $\overline{\rm MS}$ mass is known with enough accuracy
\cite{dipole}. To avoid ambiguities we always use this definition unless
other convention is indicated explicitly.

It is important to note the following fact. In the relativistic theory
for a particle with mass $m$ one always has $p^2\!=\!m^2$, that is,
$E\!=\!\sqrt{m^2\!+\!\vec{p}^{\,2}}$. In the nonrelativistic expansion,
therefore,
\beq
E\;=\; m\,+\,
\frac{\vec{p}^{\,2}}{2m}\,-\,\frac{\vec{p}^{\,4}}{8m^3}\:+\;...\;;
\label{m20}
\eeq
all coefficient are fixed in terms of powers of $m$ by Lorentz
invariance. In applications to heavy quarks it
is advantageous to use such a Wilsonian cutoff which preserves
usual QM properties for the price of apparently violating Lorentz 
invariance.
This is justified since the problem from the very beginning has
physically preferred frame -- the one in which the heavy quark is at
rest. (More detailed discussion can be found
in \cite{blmope}.) Then the mass parameters in
Eq.~(\ref{m20}) generally become different:
\beq
{\cal H}_Q\;=\; m_0 \varphi_Q^+ \varphi_Q \: - \:
\varphi_Q^+ A_0 \varphi_Q \: + \: \frac{1}{2m_2}\,
\left((i\vec D)^2 +
c_G \vec\sigma \vec B\right)\; - \; ...
\label{m22}
\eeq
even for a ``quasifree'' quark in the effective theory. This phenomenon
is well known in the solid state physics. Of course, the
difference between $m_i$ appears only due to perturbative corrections:
$$
m_i(\mu)\,-\, m_k(\mu) \;=\; {\cal O} (\alpha_s \mu)\;.
$$
Moreover, the differences can be calculated perturbatively and are
completely free from any infrared effects existing below the scale $\mu$. For
example, in our scheme $m_0(\mu) \simeq m_2(\mu)+\frac{4}{9}
\frac{\as(\mu)}{\pi}$. We use the mass $m_2(\mu)$, the mass that enters
the kinetic energy operator $\vec{p}^{\,2}/2m$; it is the most
relevant mass for heavy quark transitions.

This calculable difference between different ``masses'' for the same
quark must be properly accounted for in the OPE analysis. Probably, the
most obvious place where it plays a role is the $\bar{Q} Q$ threshold
physics.
While in the short-distance expansion
the free quark threshold starts at $2 m_0(\mu)$, the
propagation of heavy quarks or the bound state dynamics are actually
determined by $m_2(\mu)$. The shift in the position of the threshold
which serves as a reference point for energy in the nonrelativistic
system must be properly taken into account.

Concluding this section, let us make a side remark
concerning
the $t$ quark mass \cite{rev}. The peculiarity of the $t$ quark is that
it has a significant width $\Gamma_t \!\sim 1\! \GeV$ due to its weak
decay.  The perturbative position of the pole in the propagator is,
thus, shifted into the complex plane by $-\frac{i}{2} \Gamma_t$.  The
finite decay width of the $t$ quark introduces a {\em  physical}
infrared cutoff for the infrared QCD effects \cite{khoze}. In
particular, the observable decay characteristics do not have ambiguity
associated with the uncertainty in the pole mass discussed above. The
uncertainty cancels in any physical quantity  that can be measured.
That is not the case, however, in the position of the pole of the
$t$-quark propagator in the complex plane (more exactly, its real
part). The quark Green function is not observable there, and one would
encounter the very same infrared problem and the same infrared
renormalon. The latter does not depend on the absolute value of the
quark mass (and whether it is real or have an imaginary part). Thus, in
the case of top, one would observe an intrinsic -- but artificial  --
infrared renormalon uncertainty of several hundred $\MeV$ in attempts
to relate the peak in the physical decay distributions to the position
of the propagator singularity in the complex plane.

\subsubsection{The numerical values of $m_c$ and $m_b$}

The mass of the $c$ quark at the scale $\sim m_c \sim 1\GeV$ can be
obtained from the charmonium sum rules \cite{SVVZ}, $\bar{m}_c(m_c)
\simeq 1.25 \GeV$. The result to some extent is affected by the value
of the gluon condensate.  To be safe, we conservatively ascribe a rather
large uncertainty,
$$
m_c(m_c) = 1.25 \pm 0.1 \, \mbox{GeV}\, .
$$
There are reasons to believe that the precision
charmonium sum rules actually determine the charmed quark
mass to a better accuracy.

An accurate measurement of $m_b$ is possible in the $\bar bb$
production in $e^+e^-$ annihilation. Since we want to know $m_b$
with comparable or better absolute precision, both the data and
calculations, at first glance, must have increased accuracy. The data
are available, however, only below and near the threshold. Certain
integrals (moments) of the cross section over this domain are
particularly sensitive to the low-scale mass $m_b(\mu)$ with $\mu$ in
the interval $1$ to $2 \GeV$.

Using
dispersion relations
\beq
\Pi_b(q^2) \; =\; \Pi_b(0)\;+\; \frac{q^2}{2\pi^2}
\,\int\,\frac{ds\, R_b(s)}{s(s\!-\!q^2)}
\label{m25}
\eeq
one evaluates the polarization operator $\Pi_b(q^2)$ (and its
derivatives) induced by the vector currents $\bar b\gamma_\mu b$,
in
the  complex $q^2$  plane at an adjustable distance $\Delta$ from the
threshold. Such  quantities are proportional to weighted integrals
over
the experimental cross  section; the integrals are saturated  in the
interval $\sim \Delta$ near threshold, and are very sensitive to the
mass $m_b(\Delta)$.

A dedicated analysis of this type was first carried out by 
Voloshin \cite{volmb} who considered a
set of relatively high derivatives of $\Pi_b$ at $q^2\!=\!0$. On the
phenomenological side they  are  expressed through moments of
$R_b(s)$,
$$
\frac{2\pi^2}{n!}\, \Pi_b^{(n)}(0) \; =\; I_n\; = \;
\int\,\frac{ds\,R_b(s)}{s^{n+1}}
\simeq
$$
\beq
M_{\Upsilon(1S)}^{-2(n+1)}\, \int \: ds\:R_b(s)\,
\exp\left\{ -(n\!+\!1)\left(s\!-\!4M_{\Upsilon(1S)}^2\right)\right\}\, ,
\label{m26}
\eeq
while the theoretical expressions for the very same moments
are given in terms of the $b$ quark mass and $\alpha_s$.
The relevant momentum scale here is $\mu \sim m_b/\sqrt{n}$.
Considering small-$n$ moments $I_n$ one
would determine $m_b$ at the scale of the order $m_b$.
The small-$n$ moments  are contaminated by the contribution of
$R_b$ above the open beauty threshold where
experimental data  are poor. Increasing $n$
shrinks the interval of saturation and, thus, lowers the effective
scale. On the other hand, we
cannot go to too high values of $n$ where  infrared
effects
(given, first, by the gluon condensate) explode. There is still enough
room to keep the gluon condensate small and, simultaneously,
suppress
the domain above the open beauty  threshold.  In the fiducial
window,
$\mu$ must be large enough to ensure control over the QCD
corrections. The latter requires a nontrivial summation of
enhanced Coulomb terms unavoidable in the nonrelativistic
situation. As known from textbooks, the part of the perturbative
corrections to the polarization operator, associated with the potential
interaction, is governed by the parameter $\alpha/|\vec{v}\,|$
rather than by  $\alpha$ {\em per se}.

Let us briefly illustrate why the moments $I_n$ determined in this way 
pinpoint the running mass $m_b$. Naively, the cut in $\Pi_b(q^2)$ 
starts in the perturbative calculations at $2m_b^{\rm pole}$, which 
seems to determine the strongest dependence on the $b$ quark mass. 
However, the $\bar{b}b$ production is also affected by the potential 
interaction of $b$ and $\bar{b}$; the latter even generates a number of 
bound states below the $b\bar{b}$ threshold. The gluon exchanges 
increas $m_b^{\rm pole}$ and thus tend to suppress $I_n$ (we 
keep the short distance mass fixed). However, the Coulomb effects 
enhance the moments both owing to the emerging bound states and due to 
attraction above the threshold. 

The largest infrared contribution to
the pole mass is linear in the gluon momentum, Eq.~(\ref{3.2}). Its 
cancellation can be understood on the example of QED with the Abelian 
gauge interaction. 
The
expression for the mass shift is simply 
self-interaction $\frac{1}{2} V_{\rm IR}(0)$ where $V_{\rm IR}$ is the
heavy quark potential mediated by the gauge interactions with momenta
below certain $\mu \ll m_b$ \cite{gurman,pole}. Yet the mass of the
$\bar{b}b$ system includes also the same Coulomb interaction between
quark and antiquark.  Since for the colorless $\bar{b}b$ state the sum
of color ``charges'' is zero, these effects cancel each other for the
quanta with wavelength less than the interquark spacing $r$. Therefore,
for the Fourier transform of the potential $V(\vec{q}\,)$ in terms of
which
\beq V(0) \;=\; \int\; \frac{{\rm d}^3 \vec{q}}{(2\pi)^3}\:V(\vec{q}\,)
\;,
\label{3.20}
\eeq only the components with $|\vec{q}\,| \gsim 1/r$
contribute. The softer exchanges are suppressed by powers of the
multipole factor $\vec{q}^{\,2} r^2$.

This, of course, automatically emerges in all calculations. Let us
single out the effect of gluon exchanges with $|\vec{q}\,| < \mu$:
\beq
V_{\rm IR}(r) \;=\; - \int_{|\vec{q}\,| < \mu}
\; \frac{{\rm d}^3 \vec{q}}{(2\pi)^3}\:V(\vec{q}\,)\:
{\rm e\,}^{-i\vec{q}\vec{r}}\;=\;
-V_0\,+\, \frac{1}{6} r^2 \mu^2 \,V_2 \;-\; ... \;,
\label{3.22}
\eeq
$$
V_0\;=\; \int_{|\vec{q}\,| < \mu}
\; \frac{{\rm d}^3 \vec{q}}{(2\pi)^3}\:V(\vec{q}\,)\;,\qquad\;
V_2\;=\; \int_{|\vec{q}\,| < \mu}
\; \frac{{\rm d}^3
\vec{q}}{(2\pi)^3}\:\frac{\vec{q}^{\,2}}{\mu^2}\,
V(\vec{q}\,)\;, \qquad ...
$$
(the minus sign reflects the fact that the second particle is an
antiquark).  
If quarks reside at distances much smaller than $1/\mu$, the soft
potential is just a constant. Its sole role is only to shift the energy
of all $\bar{b}b$ states by a constant amount $-V_0$. It does not
affect wavefunctions and, therefore, does not modify the coupling of
the virtual photon in $\rm e^+e^-$ annihilation to these states.

However, a constant potential $A_0$ cannot change the energy of
the neutral system, whether the field is classical or quantum. It means
that just the opposite shift is present in the sum of masses of $b$ and
$\bar{b}$ renormalized by the same gauge interaction, which is 
self-manifest in Eq.~(\ref{3.2}). The above arguments show that the 
cancellation of the infrared effects in the quark mass is a nontrivial 
consequence of the gauge nature of QCD interactions, the fact 
explicit in the OPE.

The above reasoning, while illustrating the cancellation of the infrared
effects present in the pole mass, is too simplified in many aspects. A 
more dedicated discussion can be found in Ref.~\cite{varenna}.

Over the last few years the perturbative decsription of the moments 
$I_n$ has been significantly improved and now includes the 
next-to-next-to-leading (NNLO) terms. The corresponding analysis of the 
data was performed by a few groups \cite{mbnnlo}, differing in 
technical details; the values of $m_b(1\GeV)$ were found around 
$4.57\GeV$ with the estimated uncertainty $40$ to $60\MeV$. In 
what follows we accept the value of the running $b$ quark mass
\beq
m_b(1\GeV) = 4.57\pm 0.05\GeV
\label{mbnum}
\eeq

\subsection{ $\mu_\pi^2$ and $\mu_G^2$}

The heavy quark masses $m_c$, $m_b$, being the key parameters in the
HQE, are to a large extent `external' to the properties of the
effective low energy theory itself. There are two nonrelativistic heavy
quark
operators in the Hamiltonian; their expectation values Eqs.\,(\ref{34}) in
the heavy meson $B$ play a key role in many applications. In contrast
to $m_Q$, they are determined by the QCD dynamics itself. We cannot yet
calculate theoretically their values from first principles of QCD since
it would require more or less exact solution of QCD in the strong
coupling regime. Instead, we can try to measure them extracting from
known properties of hadrons.

The value of $\mu_G^2$ is known: since
$$
\frac{1}{2m_Q} \bar{Q}
\frac{i}{2}
\sigma_{\mu\nu} G^{\mu\nu} Q
$$
describes the interaction of the heavy quark spin with the light cloud
and causes the hyperfine splitting between $B$ and $B^*$,
\beq
\mu_G^2\; \simeq\;\frac{3}{4}\:2m_b(M_{B^*}-M_B) \;\simeq\;
\frac{3}{4}\,(M^2_{B^*}-M^2_B)\;\approx\; 0.36\GeV^2\;.
\label{p7}
\eeq
In actual QCD $\mu_G^2$ logarithmically depends on the normalization
point; usual one-loop diagrams yield
\beq
\mu_G^2(\mu')\; \simeq \;
\left(
\frac{\as(\mu')}{\as(\mu)}
\right)^{\frac{3}{b}}\,  \mu_G^2(\mu)\;\;.
\label{140}
\eeq
In the mass relation given above the operator is normalized at the
scale $\mu \simeq m_b$:
\beq
\frac{1}{m_Q} {\cal H}_1^{\rm spin}\;=\; -C_G(\mu)\frac{\left(
\bar{Q} \frac{i}{2}\sigma G Q\right)_\mu }{2m_Q}\;, \qquad
C_G\simeq
\left(
\frac{\as(m_Q)}{\as(\mu)}
\right)^{\frac{3}{b}}\: .
\label{141}
\eeq
Evolving perturbatively to the
normalization scale $\mu \sim 1\GeV$ slightly enhances the
value of $\mu_G^2(\mu)$, but this
effect is numerically insignificant, and we usually neglect it.

The kinetic expectation value $\mu_\pi^2$ has not been directly measured 
yet. It
enters various distributions in semileptonic decays or processes of the
type $ b \ra s+\gamma$. 
First  attempts to extract it from the semileptonic
distributions were reported 
\cite{chern,gremmsav}, 
however the analysis is vulnerable at some points, and so far the 
outcome is inconclusive.

A model-independent lower bound was established
\cite{motion,volpp,vcb,optical}
\beq
\mu_\pi^2\;>\;\mu_G^2\approx 0.4 \mbox{GeV}^2
\label{p11}
\eeq
which  constrained  possible values of $\mu_\pi^2$. One of its possible 
derivations will be given below in Sect.~4 (see also Sect.~5.1). 
Here we instead mention its physical interpretation.

Unlike quantum-mechanical pure potential problems, in QCD the heavy
quark kinetic operator is expressed in terms of the {\em covariant}
derivatives $\pi_j\!=\!i D_j \!=\! i\partial_j \!+\!A_j$ 
which include the vector
potential $A_j$. The latter lead to non-commutativity of
different spatial components of the  momentum operator in the presence
of the chromomagnetic field $\vec{B}$, 
\beq 
[\pi_j,\pi_k]\;=\; iG_{jk}\;=\; - i\epsilon_{jkl} B_l 
\;\simeq\; 
-\frac{2}{3} i \epsilon_{jkl} \; j_l \cdot 0.4\GeV^2 \;. 
\label{p20} 
\eeq 
This
non-commutativity immediately leads to the lower bound on the
expectation value of $\vec{\pi}^{\,2}$ \cite{motion}, in full analogy
with the uncertainty principle in quantum mechanics. 
Formally, it follows from the positivity of the Pauli
Hamiltonian \cite{volpp}:
$$
\frac{1}{2m}(\vec\sigma\,i\vec D\,)^2 =
\frac{1}{2m}\left((i\vec D)^{\,2}-\frac{i}{2}\sigma G\right) > 0 \;.
$$
More physically, it means the Landau
precession of a colored, i.e.\ ``charged" particle in the
(chromo)magnetic field, Fig.~3. Hence, one has $\aver{p^2}\ge |\vec B|$.
Literally
speaking, in the $B^*$ meson the
quantum-mechanical
expectation value of the chromomagnetic  field is suppressed,
$\aver{B_z}=-\mu_G^2/3$. It
completely vanishes in the $B$ meson. However, the essentially
non-classical nature of the `commutator' $\vec{B}$ proportional to the 
spin operator of the light clound
(e.g.  $\aver{\vec{B}^{\,2}} \ge 3 \aver{\vec{B}}^{2}$), in turn, 
enhances the bound  which then takes
the same form as in the external classical field.

\thispagestyle{plain}
\begin{figure}[hhh]
 \begin{center}
 \mbox{\epsfig{file=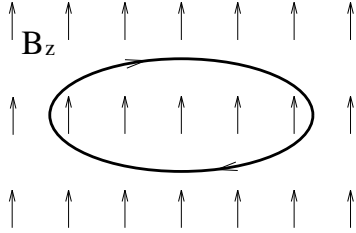,height=3.cm}}
 \end{center}
 \caption{ \small
Landau precession of a free charged particle in the magnetic field. The
average of the momentum square is bounded from below even in the absence
of binding potential. This illustrates the physical meaning of the
inequality between $\mu_\pi^2$ and $\mu_G^2$.
}
\end{figure}

It is worth emphasizing that the inequality between the kinetic and
chromomagnetic expectation values takes place for
$\mu_\pi^2$ normalized at any point $\mu$, provided
$\mu_G^2$ is normalized at the same point.
For large $\mu$ it becomes uninformative; so, it is in our best
interests to use it at $\mu$ = several units $\times \Lambda_{\rm
QCD} \lsim 1 \GeV$.

The $B$ meson average of the kinetic operator was estimated using the
technique of the QCD sum rule in \cite{pp,ppnew} and in \cite{neubp},
with quite different results $(0.5\pm 0.15) \GeV^2$ and $0.1\pm
0.05\GeV^2$, respectively. A dedicated discussion of the subtleties
inherent to those analyses can be found in Ref.\,\cite{rev}, Sect.~6.
The discreapancy roots to a different treatment of the contributions
of the excited states which are a background in applications of the
QCD sum rules, and the actual result seemingly lies somewhere in
between the two extremes, probably closer to the upper value. 
It was later argued in Ref.~\cite{lubl} based
on the analysis of the QCD sume rule approach to the three-point correlators 
in the harmonic oscillator, that using the same continuum threshold for
the two- and three-point correlators (which is routinely assumed in QCD 
sum rules) leads to
a systematic bias, and that it is advantageous to use a lower continuum
threshold for the three-point functions. Applying this prescription to
modify the analysis of Ball {\it et al.}, the authors got a value
close to $0.4\GeV^2$. 

It is important to note, however that the normalization point
dependence of the kinetic expectation value was not treated
consistently. In fact, the above papers attempted to determine the HQET
parameter $-\lambda_1$ rather than the properly defined QCD expectation
value $\mu_\pi^2$ dependent on the normalization point. In simple QM
models with fixed numbers of constituents in a hadron, one would have
them equal, $-\lambda_1 \!=\! \mu_\pi^2$. However, in actual QCD
$-\lambda_1$ can be thought of as the kinetic expectation value
$\mu_\pi^2(\mu)$ ($\mu$ is the normalization point of the operator) from
which ``all perturbative'' ($\mu$-dependent) pieces are subtracted. In 
this respect $-\lambda_1$ is a close relative of the pole mass, and
likewise cannot be defined consistently in reality. Technically, it
corresponds to extrapolating the normalization point in $\mu_\pi^2(\mu)$
down to zero momentum, which leads to the usual conceptual problems.
Related questions are discussed in more detail later in Sect.~4.2 and
in Ref.~\cite{varenna}, Sect.~3.
In practical terms, since the above mentioned QCD sum rules analyses were
performed at the level of one-loop ${\cal O}(\alpha_s)$ 
perturbative corrections, it is more
or less sufficient to use identification $-\lambda_1 \to -\lambda_1 +
\frac{4\alpha_s}{3\pi} \mu^2 \simeq \mu_\pi^2(\mu)$. This
suggests adding about $0.15\GeV^2$ to the quoted values of $-\lambda_1$
to arrive at the QCD kinetic expectation value normalized at the
canonical scale around $0.7\:\mbox{to}\:1\GeV$.

To summarize, the value of the kinetic expectation value $\mu_\pi^2$ in
$B$ mesons determined from the QCD sum rules is about
\beq
\mu_\pi^2 \simeq 0.5\pm 0.15\GeV^2\;,
\label{ppvalue}
\eeq
however the literally quoted intrinsic uncertainty of the method may be
underestimated here. It is noteworthy that this estimate seems to be in a
good agreement with the bound Eq.\,(\ref{p11}). More details regarding 
various subtleties in the
kinetic operator can be found in Ref.\,\cite{varenna}.

\subsection{Heavy quark potential}

We discuss here briefly some
aspects related to the heavy quark
potential in QCD. Strictly speaking, it refers to a somewhat different
situation where two heavy quarks (actually, a quark and an antiquark)
are present simultaneously. Heavy quark potential attracted recently
attention in connection to pair production of $t\bar{t}$ and $b\bar{b}$
near the threshold, and in particular in view of its apparent connection to the
problem of the heavy quark mass illustrated in Sect.~3.1.

A closer look at the notion of the heavy quark potential reveals certain
subleties. As a matter of fact, it is not completely clear what exactly
must be called the potential between heavy quarks in actual QCD.

The original notion of the potential refers to the interaction of
infinitely heavy (static), or completely nonrelativistic heavy
particles, which is {\em instantaneous}. The most familiar example is the
electromagnetic interaction of heavy charges. The Hamiltonian
of such a potential system is given by
\beq
{\cal H} = \sum_i \frac{(i\vec\partial\,)^2}{2m_i} +
V\left(\vec{r}_i\!-\!\vec{r}_j\right)\;,
\label{pot5}
\eeq
where $m_i$ are masses of particles and the potential $V$ is a function
of their instant coordinates. Taking the limit $m_i\!\to\!
\infty$ (at fixed $\vec r_i$, which corresponds to semiclassically high
excitation numbers of the quantum system in Eq.\,(\ref{pot5}))
eliminates quantum uncertainties in the coordinates and allows to
measure the potential directly as the position-dependent energy of the 
infinitely slowly
moving collection of particles. This is well known for QED where the
potential of the charges $q_1$ and $q_2$ (in units of electron charge)
is given by
\beq
V^{\rm QED}(R) = \alpha_{{\rm em}}(0) \,\frac{q_1 q_2}{R}
\;.
\label{pot7}
\eeq
This expression is exact in the absense of light charged particles; the
known quantum corrections appear only if other matter
fields are not much heavier than the scale $1/R$.

The definition of the similar quantity in a non-Abelian theory like QCD
is more tricky. First, one has to limit consideration to the
systems in the colorless state. The heavy quark potential is defined
only between $Q$ and $\bar{Q}$ in the color singlet case; even there the
requirement of gauge invariance is not trivial. The color of the
individual heavy quark remains fully quantum in nature and changes
through the interaction with gluons. In contrast to usual coordinates,
the limit $m_Q\!\to \!\infty$ does not make color variables describing the
state of the heavy quark semiclassical.

To avoid this problem the heavy quark potential is defined {\it via} the
vacuum expectation value of the long Wilson loops:
\beq
V(R) = -\lim_{T\to\infty}\, \frac{1}{T} \ln{\langle{\rm Tr}
\:{\cal P}\, {\rm exp}
\left(
i\oint_{C(R,T)} A_\mu {\rm d}x_\mu
\right)}
\rangle \;,
\label{pot11}
\eeq
where the rectangular contour $C(R,T)$ spans distance $R$ and $T$ in the
space and time directions, respectively. (It is usually assumed to be in
the Euclidean space as in Eq.~(\ref{pot11}).)  This definition is
intuitively clear, since Wilson lines describe propagation of the
infinitely heavy quarks. Moreover, such Wilson loops are readily
computed in the $U(1)$ gauge theory (free QED) and, of course,
reproduce Eq.~(\ref{pot7}) (with $q_1\!=\!-q_2$).
In QCD the perturbative expansion of $V(R)$ has been computed through
order $\alpha_s^3$ \cite{schroder}. Usually the potential in the
momentum representation is considered:
\beq
V(\vec{q}\,) = \int {\rm d}^3 \vec R \; V(\vec R) \; {\rm
e}\,^{-i\vec{q}\vec{R}}
= -\frac{4}{3} \frac{4\pi\alpha_s}{\vec{q}^{\:2}}
\left\{1+ \left(\frac{31}{3}-\frac{10}{9}n_f\right) \frac{\alpha_s}{4\pi} 
+ c_3
\left(\frac{\alpha_s}{4\pi}\right)^2 +\ldots \right\}\;.
\label{pot13}
\eeq

There are important peculiarities in thus defined interaction in the
$Q\bar{Q}$ system; they were first analyzed by Appelquist {\it et al.}\ 
already in the late 70's \cite{appel}. Due to gluon self-interaction,
the potential in higher orders contains diagrams of the type shown in
Figs.~4. Here the dashed line denotes the Coulomb quanta (they mediate
instantaneous interaction in the physical Coulomb gauge), while the wavy
lines are used for actual transverse gluons. These diagrams signal the
real propagation in time of transverse gluons. This means that at this level 
the problem ceases to be a two-body one, and includes more full-fledged
dynamical degrees of freedom. 

Moreover, while the first diagram Fig.~4a with the rung gluon appearing
in order $\alpha_s^3$ safely converges at the gluon momenta $\sim \!1/R$,
including additional Coulomb exchanges, as in Fig.~4b leads to
infrared divergence. With one exchange it is logarithmic; the 
formal degree of the infrared divergence increases
with adding extra Coulomb quanta between the emission and absorption of
the transverse gluon. The physics behind this infrared behavior was
discussed in Ref.~\cite{appel}: emission of the soft transverse gluon
changes the overall color of the $Q\bar{Q}$ pair and, therefore modifies
the interaction energy between them. The energy shift
associated with the exchange of the transverse gluon depends
nonanalytically on the energy denominator (in the language of
non-covariant time-ordered perturbation theory), since the gluon can be
arbitrarily soft. Formally expanding the exact result in $\alpha_s$
includes expanding in this change of the Coulomb energy proportional to
$\alpha_s/R$, therefore one
obtains increasing infrared singularities. These arguments suggested
that resummation of the Coulomb exchanges in Fig.~4b would render these
diagrams finite, however the effective infrared cutoff is of the order
of $\alpha_s/R$ and, thus the potential is not infrared finite {\em
perturbatively} containing terms $\sim 1/R\cdot \ln{\alpha_s}$ starting
order ${\cal O}\left(\alpha_s^4\right)$.

\thispagestyle{plain}
\begin{figure}[hhh]
 \begin{center}
 \mbox{\epsfig{file=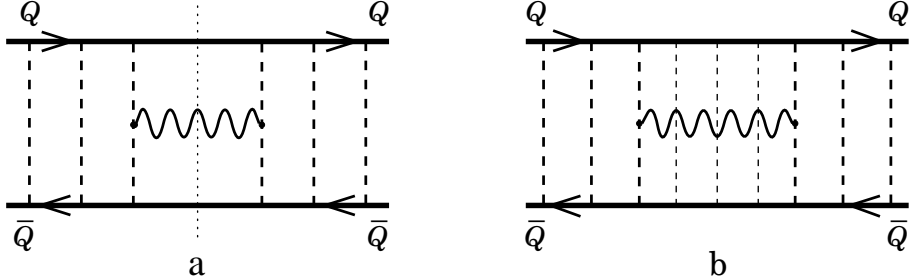,width=12cm}}
  \end{center}
 \caption{ \small
Diagrams for heavy quark potential in QCD. Dashed lines are
instantaneous Coulomb exchanges, transverse gluons propagating in time 
are shown by wavy lines. \newline
a)  The convergent diagrams with the transverse gluon as a rung. 
\newline
b) Adding more Coulomb exchanges inside the ladder with the transverse gluon 
leads to infrared divergence in perturbation theory.
}
\end{figure}

This infrared singularity emerges since the external momentum scale
$1/R$ does not fix all (spacelike) momenta in the problem. In the static
situation the low-momentum tail in the propagation of  very low-momentum
(ultrasoft) gluons is cut off only by the interaction energy of the
quarks, which appears itself only in the perturbative expansion;
therefore it turns out to be suppressed only by powers of $\alpha_s$.
As a result, beyond order $\alpha_s^3$ the heavy quark potential,
whatever it means, is not a short-distance quantity even at arbitrary
small space separation between the heavy quarks.\footnote{This situation
is not peculiar to non-Abelian gauge theories. Similar long-range
propagation of light charged particles occurs also in QED if their
masses are well below $1/R$.}
Of course, this becomes
possible only since time interval used to define $V(R)$ is taken
infinitely large. Should one use instead finite-$T$ Wilson loops, the
time interval would place the final infrared cutoff in all gluon
exchanges.

This purely perturbative analysis shows that there is no direct analogue
of the potential between heavy quarks in QCD. The $Q\bar{Q}$ system
incorporating all gluon interaction is not a two-body system but
includes actual propagation of gluons with energy small compared to
$1/R$. The interaction then cannot be universally described by an
instantaneous potential and is intrinsically non-local in time. An
attempt to integrate out in one way or another the ``extra'' degress of
freedom does not yield in general a meaningful analogue of the
potential: the result would not be universal but rather
process-dependent. Thus, the prescription (\ref{pot11}) based on Wilson
loops in QCD does not yield the heavy quark potential in its
conventional understanding.

In practice, this may not pose serious problems in the perturbative
computations. Through order $\alpha_s^3$ all diagrams including Fig.~4a
are convergent and saturated by the gluon momenta of order $1/R$. The
above nonlocalities in time are then revealed only at time intervals
smaller than $R$. In many applications with nonrelativistic heavy quarks
the scale $1/R$ is still much smaller than the characteristic dynamic
energies. In the spirit of the Wilsonian OPE such exchanges then can be
integrated out to yield a local in time effective potential. Yet it must
be remembered that this can be done only up to a certain order in
$\alpha_s$, depending on the concrete problem.

As discussed above, the heavy quark potential $V$ (from now one we
assume that it is defined via Wilson loops, Eq.~(\ref{pot11})) is not an
infrared finite object already in the perturbative expansion. The
situation can be better for the {\it integral} of $V(\vec{q}\,)$ over
spacelike momentum:
\beq
V(0) = \int \frac{{\rm d}^3 \vec q}{(2\pi)^3} \: V(\vec{q}\,)
\;.
\label{pot21}
\eeq
In order to get a finite result we need to assume an ultraviolet cutoff
at a certain scale $\mu$, for example, adding the step-function
$\theta (\mu^2\!-\!\vec{q}^{\,2})$ to the integrand. It was mentioned in
Sect.~3.1 that in electrodynamics such an integral of the 
potential
equals minus double of the contribution to the mass of the static source
coming from the same range of gluon momenta. In QCD the perturbative
diagrams for both $m_Q$ and $V(\vec{q}\,)$ are more complicated. Yet the
same relation holds in order ${\cal O}(\alpha_s^2)$ as well. Most simply
this is seen in the Coulomb gauge where all the effect to this order
reduces to dressing the propagator of the Coulomb quanta. (An
alternative discussion can be found in Ref.~\cite{beneke}.) There are
reasons to believe that such a relation may hold to all orders in
perturbation theory: the infrared contribution to $V(0)$ in
Eq.~(\ref{pot21}) equals minus twice the same contribution to the pole
mass of a static source.

Indeed, let us imagine we were able to introduce in some way an ansemble
of gauge field configurations where the modes with momenta much larger
than a certain scale $\mu$ are practically absent. This field-theoretic
system would not need regularization, and everything can be expressed in
terms of the bare parameters, including the bare quark mass $m_Q^{(0)}$.
At $R\!\to\! 0$ the Wilson loop will approach the free value $N_c$ thus
yielding $V(0)\!=\!0$. This is clear on the physical grounds: $Q\bar{Q}$
form a dipole with the infinitesimal dipole moment, and its interaction
with any soft gluon field vanishes as $R$ goes to $0$. It is important
at this point that our gauge ansemble explicitly includes only soft
modes. Otherwise, as in full QCD, the modes with $|\vec{k}\,| \sim 1/R$
generate growing attractive potential at arbitrary small $R$.

Next we note that $V(R)$ is traditionally determined up to a constant;
in perturbation theory  (in four dimensions) the potential is actually
defined as
\beq
U(R) = V(R)-V(\infty)
\;;
\label{pot23}
\eeq
we assign $U$ to this ``standard'' potential to distingush it from
$V(R)$ which has a precise meaning in a finite theory. While $U(R)$ by
definition vanishes at $R\to \infty$, $V(\infty)$ does not and reflects
nontrivial interaction with the gluon field. In the momentum
representation $V(\vec{q}\,)$ contains explicitly the term
$V(\infty)\,\delta^3(\vec{q}\,)$, which is discarded in the usual
perturbative computations. Since at $R\!\to \!\infty$ the $\bar{Q}$ and $Q$
lines are well separated, their interaction (at least in perturbation
theory) must vanish as $1/R$, and the value of the Wilson loop is simply
given by the mass renormalization of each static source:
\beq
V(\infty) = 2\left(m_Q^{{\rm Ren}} - m_Q^{(0)}\right)
= 2\delta m_Q
\;.
\label{pot25}
\eeq
Therefore, for the ``ordinary'' potential we have
\beq
U(0) = \int\frac{{\rm d}^3 \vec q}{(2\pi)^3} \: V_{\rm reg}(\vec{q}\,)
= -2\delta m_Q
\;.
\label{pot27}
\eeq
Here $V_{\rm reg}$ is the usually computed regular part of 
$V(\vec{q}\,)$ not containing
self-energy diagrams yielding $\delta^3(\vec{q}\,)$.

The above relation for the infrared contributions to the mass and the
potential
\beq
\int\frac{{\rm d}^3 \vec q}{(2\pi)^3} \: V_{\rm reg}^{{\rm
IR}}(\vec{q}\,)
= -2\delta_{{\rm IR}} m_Q
\label{pot29}
\eeq
at first may seem strange: What happens to the perturbatively
infrared-singular contributions of Fig.~4b which, at order
$\alpha_s^4$,
behave like $\frac{\alpha_s^4}{\vec{q}^{\,2}}
\ln{\frac{\vec{q}^{\,2}}{\epsilon}}$, with $\epsilon$ an infrared cutoff
in the ``rung'' gluon momentum? The corresponding contributions are
absent from the mass since the Coulomb exchanges are instantaneous.

The subtlety resides in the necessity to introduce an upper citoff $\mu$
on gluon momenta in $V^{{\rm IR}}$. This not only cuts off
$V(\vec{q}\,)$ at $\vec{q}^{\:2} \gg \mu^2$, but also modifies
$V(\vec{q}\,)$ at $|\vec{q}^{\,}| \sim  \mu$ and must lead to an
additional term residing around $|\vec{q}^{\,}| \sim  \mu$ which
contains a similar term $\frac{\alpha_s^4}{\mu^{2}}
\ln{\epsilon}$ with the opposite sign. They do not, of course, eliminate
dependence on $\epsilon$ in the resulting $V^{{\rm IR}}(\vec{q}\,)$, but
have to cancel it when integrated over all $\vec{q}$.

This requirement has a transparent meaning: the amplitudes of emitting a
very soft gluon out of the $Q\bar{Q}$ system where initially $Q$ and
$\bar Q$ are set to be at arbitrarily small separation $R$, must vanish
unless virtual gluons with $\vec{k} \sim 1/R$ are allowed to resolve $Q$
and $\bar Q$. Therefore, with all gluon fields softer than a certain
$\mu$, the sum of all diagrams for the emission {\it amplitude}
(given by the part of the graphs in Figs.~4 to the left from the 
transverse gluon) must vanish upon integrating over $\vec q\,$.

Relying on relation (\ref{pot29}), one can try to define
perturbatively a certain running heavy quark mass which is free from the
leading renormalon uncertainty $\sim \Lambda_{\rm QCD}$.\footnote{This
idea was first discussed myself in 1996 and later independently
advocated by M.~Beneke \cite{beneke}.}  
One defines 
\beq
m_Q^{\rm PS}(\mu) = m_Q^{\rm pole} +
\frac{1}{2} \int_{|\vec{q}\,|< \mu} \frac{{\rm d}^3 \vec q}{(2\pi)^3}
\: V_{\rm reg}(\vec{q}\,)
= m_Q^{\rm pole} + \frac{1}{\pi}\int_0^{\infty} \!\!{\rm d}R\; V(R)
\!\left[ \frac{\sin{\mu R}}{R} \!-\! \mu\cos{\mu R}
\right]
,
\label{pot31}
\eeq
where simply the literal $V(\vec{q}\,)$ computed to a certain
order in perturbation theory without an explicit cutoff is used, and 
the pole
mass is taken to the same order in $\alpha_s$. The mass
$m_Q^{\rm PS}(\mu)$ is known as the ``potential-subtracted'' mass
\cite{beneke}. Although purely soft gluon configurations would generate
`an infrared' part of the potential $V(\vec{q}\,)$ literally different
from $V(\vec{q}\,) \theta (|\vec{q}\,|\!-\!\mu)$ used in
Eq.\,(\ref{pot31}), this ansatz 
is hoped to include all (perturbative) infrared contributions
originating from the domain sufficiently below $\mu$, and simply
subtracts some additional {\it ad hoc} perturbative pieces.

A technical problem would arise here starting order $\alpha_s^4$ --
since $V(\vec{q}\,)$ becomes IR singular here, this routine literally
fails. According to the previous discussion, the weight itself with
which the infrared contributions are to be subtracted apparently gets
modified by the explicit cutoff, should we implement it in practice.
Therefore, they are not properly subtracted from the pole mass by
prescription (\ref{pot31}). It is conceivable that this mismatch exists
already at order $\alpha_s^3$, but is not revealed explicitly due to
infrared convergence of $V$ to this order.

It should be noted that $V(0)$, although infrared finite in perturbation
theory, is not a genuinely short-distance quantity (nor its
$\mu$-dependence) for which, for instance, usual OPE can be applied.
Eq.~(\ref{pot31}) also shows that at arbitrary large $\mu$ it includes
contribution from the potential $V(R)$ at large distances $R$. The
nonperturbative definition of such an object therefore remains unclear.
These interesting questions fall outside the scope of the present
review.

\section{Heavy Quark Sum Rules and Exact Inequalities 
in the 
Static Limit $m_Q \!\to \!\infty$}

Discussing the basic parameters of the heavy quark theory and the
properties of the actual $b\to c$ transition it is often advantageous to
resort to a theoretical limit when both quark masses are asymptotically
large. This usually greatly simplifies the problem since eliminates
proliferating effects of various power corrections in $1/m_Q$. In this
limit physics of transitions between heavy quarks without change of
the heavy hadron velocity is trivial: spin degrees of freedom are
decoupled, only quasi-elastic transitions (those which do not change the
light cloud, and the heavy quark spin can be flipped only by the weak current) 
occur between the respective
members of the heavy quark symmetry multiplets. The transition
amplitudes of heavy quark currents between the ground and excited states
vanish in the heavy quark limit.

Nontrivial physics emerges when the external weak current changes the 
velocity of the heavy flavor hadron. The most simple and instructive are
transitions which occur in the lowest, linear in $\Delta \vec{v}$ order.
This so-called small velocity (SV) limit introduced by Shifman and 
Voloshin in the mid 80's as a theoretical tool for studying heavy flavor 
amplitudes, is also relevant for actual
semileptonic decays of $b$ to charm \cite{SV}: the typical velocity of
charmed final state hadrons is rather small.

Physics of semileptonic transitions in the SV regime is described by the
corresponding SV structure functions of the heavy mesons (or baryons;
here we explicitly formulate our consideration assuming the initial
hadron is the ground state meson): 
\beq 
W_{\mu\nu} (q_0, \vec{q}\,)=
\frac{1}{2\pi} \Im \frac{1}{2M_{H_Q}}  
\matel{H_Q}{\int \; {\rm d}^3x\,{\rm d}x_0\; 
{\rm e}\,^{i\vec{q}\vec x -iq_0x_0}\:  
iT\{J_\mu(x),\, J_\nu^\dagger(0)\}  }{H_Q} 
\label{sr11} 
\eeq 
and we retain only terms through
order $\vec{v}^{\,2}$. $W_{\mu\nu}$ is normally expanded over invariant
tensor structures. Heavy quark and SV limit lead to decrease in the
number of  structures for arbitrary weak currents -- there is only two
independent structure functions.\footnote{Strictly speaking, the second
one describing the antisymmetric Lorentz structure in the infinite mass
limit would require considering the nonforward scattering amplitude with
the overall change of velocity, and $H_Q$ carrying spin like the 
ground-state vector mesons. Alternatively, one can obtain its analogue
considering the usual structure functions, for example, with vector
currents to order $1/m_Q^2$ even at zero recoil, or taking more 
complicated weak vertices with covariant derivatives.} 
We do not present here the corresponding formalism, but instead resort
to an alternative consideration phrased in terms of individual
transition amplitudes  from the ground to the excited states. The
analogies of the moments of the SV structure functions in this language
are the so-called heavy quark sum rules. 

\subsection{Sum rules}

The heavy quark sum rules we discuss are the following: 
\bea
\varrho^2-\frac{1}{4} &=& 
 2\sum_m\;|\tau_{3/2}^{(m)}|^2 \,\;\;\;+\;\;\;
\sum_n\;|\tau_{1/2}^{(n)}|^2 \, ,
\label{bj}
\\
\frac{1}{2}\,\: &=& 
 2\sum_m\;|\tau_{3/2}^{(m)}|^2 \,\;\;\;-\:
 2\sum_n\;|\tau_{1/2}^{(n)}|^2 \, ,
\label{ur}
\\
\frac{\La}{2}\; &=& 
2\sum_m \epsilon_m|\tau_{3/2}^{(m)}|^2 \;+\;\;\;
\sum_n \epsilon_n|\tau_{1/2}^{(n)}|^2\, ,
\label{vol}
\\
\overline \Sigma\; &=& 
2\sum_m \epsilon_m|\tau_{3/2}^{(m)}|^2 \;-\;
2\sum_n \epsilon_n|\tau_{1/2}^{(n)}|^2\, ,
\label{ur2}
\\
\frac{\mu_\pi^2}{3} &=&
2\sum_m \epsilon_m^2|\tau_{3/2}^{(m)}|^2 \;+\;\;\;
\sum_n \epsilon_n^2|\tau_{1/2}^{(n)}|^2\, ,
\label{srmupi}
\\
\frac{\mu_G^2}{3} &=&
2 \sum_m \epsilon_m^2|\tau_{3/2}^{(m)}|^2 \;-\;
2 \sum_n \epsilon_n^2|\tau_{1/2}^{(n)}|^2 \, ,
\label{pig}
\\
\frac{\rho_D^3}{3} &=& 
2\sum_m \epsilon_m^3|\tau_{3/2}^{(m)}|^2 \;+\;\;\;
\sum_n \epsilon_n^3|\tau_{1/2}^{(n)}|^2\, ,
\label{fourth}
\\
-\frac{\rho_{LS}^3}{3} &=&
2 \sum_m \epsilon_m^3|\tau_{3/2}^{(m)}|^2 \;-\;
2 \sum_n\epsilon_n^3|\tau_{1/2}^{(n)}|^2 \, ,
\label{fourthls}
\eea
a sequence which, in principle, can be continued further. Here
$\epsilon_k$ is the excitation energy of the $k$-th intermediate state
(``$P$-wave states" in the quark-model language),
$$
\epsilon_k=M_{H_Q^{(k)}}-M_{P_Q}\;  ,
$$
while the functions $\tau_{3/2}^{(m)}$ and $\tau_{1/2}^{(n)}$ describe
the transition amplitudes of the ground state $B$ meson to these
intermediate states. We follow the notations of  Ref.~\cite{isgw},
\beq
\frac{1}{2M_{H_Q}}\langle H_Q^{(1/2)} |A_\mu |P_Q\rangle
= - \tau_{1/2} \,(v_1\!-\! v_2)_\mu\, ,
\eeq
and
\beq
\frac{1}{2M_{H_Q}}\langle H_Q^{(3/2)} |A_\mu |P_Q\rangle
= -\frac{1}{\sqrt{2}}\,i\,\tau_{3/2} \,\epsilon_{\mu\alpha\beta\gamma}
\,\varepsilon^{*\alpha} \,v_2^\beta \,v_1^\gamma\, ,
\eeq
where 1/2 and 3/2 mark the quantum numbers of the light cloud in
the intermediate states, $j^\pi = 1/2^+$ and $3/2^+$, respectively,
and $A_\mu$ is
the axial current. Furthermore, the slope parameter $\rho^2$ of
the Isgur-Wise function is defined as
\beq
\frac{1}{2M_{P_Q}}
\matel{P_Q(\vec v)}{\bar Q \gamma_0 Q}{P_Q}=
1-\varrho^2\,\frac{\vec v^{\,2\!}}{2} +{\cal O}(\vec v^{\,4})\, .
\label{slope}
\eeq

Equation (\ref{bj}) is known as the Bjorken sum rule \cite{BJSR}.
Superconvergent sum rules (\ref{ur}) and (\ref{ur2}) are new. 
Equation (\ref{vol}) was obtained by Voloshin \cite{volopt}. The
expression for $\mu_\pi^2$ is the BGSUV sum rule \cite{third}. The next
one was derived in Ref.~\cite{rev}, as well as Eq.~(\ref{fourthls}).
The last two sum rules are obtained along the same lines. 
The sum rule for the Darwin term  $\rho_D^3$ 
was first presented in \cite{pirjol}.

Let us illustrate the simple derivation of the sum rules for 
$\mu_\pi^2$ and $\mu_G^2$. Since in the heavy quark limit 
$\bar QQ$ pairs are not produced, we will use the
quantum-mechanical language with respect to $Q$  (but not the  light
cloud, of course).  Moreover, as discussed in Sect.~1.1 it is convenient, 
at the first stage to
assume $Q$ to be spinless. The $Q$ spin effects are trivially included
later. Then the  lowest-lying states, the  $S$-wave configurations
corresponding  to $B$ and $B^*$, are  spin-1/2 fermions, with  two spin
orientations of the light cloud. We shall denote them  $|\Omega
_0\rangle $; the  spinor wavefunction of this state is $\Psi_0$. It is
obvious that
\beq
\matel{\Omega _0}{\bar Q (i D_j) (i D_l) Q}{\Omega _0} \equiv
\frac{\mu _{\pi}^2 }{3} \delta_{jl}
\Psi_0^\dagger \Psi_0 - \frac{\mu_G^2}{6}
\Psi_0^\dagger \sigma_{jl} \Psi_0 =
\sum_k \matel{\Omega _0}{ \pi_j}{n}
\matel{k}{\pi_l}{\Omega_0},
\label{p14}
\end{equation}
where  a complete set of intermediate states is  inserted.
They are spin-1/2 states (of the opposite parity with respect  to
$|\Omega _0\rangle $), generically denoted by
$\phi ^{(n)}$, and spin-3/2 states $\chi ^{(m)}$.
We will use the Rarita-Schwinger wavefunctions for the latter, i.e. a
set of three spinors $\chi_l$ obeying the constraint 
$\sigma_i\chi_i\!=\!0$. The normalization
of these spinors is fixed by the sum over polarizations $\lambda$
\begin{equation}
\sum_{\rm \lambda} \chi _i (\lambda)\chi _j^{\dagger}(\lambda) =
\delta _{ij} -
\frac{1}{3} \sigma _i \sigma _j \;\;.
\end{equation}
Defining the reduced matrix elements $a_n$ and $b_m$ as
\begin{equation}
\matel{\phi ^{(n)}}{\pi_j}{\Omega _0} \equiv
a_n \phi ^{(n)\dagger}\sigma _j \Psi _0 \; ,\;  \; \;
\matel{\chi^{(m)}}{\pi_j}{\Omega _0}\equiv
b_m \chi_j^{(m)\dagger} \Psi _0\, ,
\end{equation}
where $\phi ^{(n)}$ and $\chi ^{(m)}$ stand for the states as well as
for their wavefunctions, we get
\begin{equation}
\mu _G^2 \;=\;
-6 \sum_n |a_n|^2 + 2 \sum_m |b_m|^2 \; ,
\label{3.7a}
\eeq
and
\beq
\mu _{\pi}^2 \;=\;
3 \sum_n |a_n|^2 + 2 \sum_m |b_m|^2 \; .
\label{3.7b}
\end{equation}

These expressions can be immediately  generalized to the actual
case of the spin-$\frac{1}{2}$ quarks $Q$. The quantities $a_n$ and $b_m$
are  to be understood as the matrix elements of $\bar b i\vec D
b$
between the $B$ meson  and higher even-parity states.
They are related to $\tau_{1/2}^{(n)}$ and
$\tau_{3/2}^{(m)}$ as follows:
\beq
\tau_{1/2}^{(n)}\;=\; \frac{a_n}{\epsilon_n}\;\;,\;\;\;\;
\tau_{3/2}^{(m)}\;=\;
\frac{1}{\sqrt{3}} \frac{b_m}{\epsilon_m}\;,
\label{p30}
\eeq
and, therefore,
$$
\mu _\pi^2 =
3\left( 2\sum _m \epsilon_m^2|\tau_{3/2}^{(m)}|^2 +
\sum_n \epsilon_n^2|\tau_{1/2}^{(n)}|^2 \right) ,\;\;
\mu_G^2 =
3 \left( 2 \sum _m \epsilon_m^2|\tau_{3/2}^{(m)}|^2 -
2 \sum_n \epsilon_n^2|\tau_{1/2}^{(n)}|^2 \right)
\, .
$$
Note that in atomic physics the combinations analogous to
$\epsilon \,|\tau|^2$ are called ``oscillator strengths''.
 
Relations (\ref{p30}) are most easily obtained from the fact that the SV
amplitudes of transitions to the corresponding states are given by the
overlap $\langle n(\vec{v}\,)|B(\vec{v}\!=\!0)\rangle $ (in the case of
spinless $Q$; for spin-${\small \frac{1}{2}}$ they are given 
by $\matel{n(\vec{v}\,)}{\vec\sigma_Q}{B(\vec{v}\!=\!0)}$). 
Then we use the general relation 
\beq
\state{H_Q(\vec v\,)}\;=\; \state{H_Q(0)} +
\pi_0^{-1}\vec{v}\vec{\pi}\,\state{H_Q(0)}\;+\;{\cal O}(\vec v^{\,2})\;.
\label{v7}
\eeq
which nicely elucidates the meaning of the small velocity sum rules:
the operator $\pi_0^{-1}(\vec v\vec\pi)$ acting on $\state{H_Q}$ is
the generator of the boost along direction $\vec v\,$.
Indeed, to get $\state{H_Q(\vec v\,)}$ one must find the eigenstate of the
Hamiltonian with heavy quark moving with the momentum
$\vec{q}\!=\!m_Q\vec{v}$. The only part which explicitly depends on
momentum comes from  the heavy quark 
Hamiltonian $\frac{\vec{\pi}^{2}}{2m_Q}$ (plus, 
in general, the higher terms in $1/m_Q$). We use the relation 
$\exp{(-i\vec{q}\vec{x}\,)}\,{\cal H}_Q\, \exp{(i\vec{q}\vec{x}\,)}= 
{\cal H}_Q \!+\!
\vec{v}\vec{\pi} \!+\! m_Q\vec{v}^{\,2}/2$ and drop the last term which is a
constant; $A_0$ obviously commutes with $x$. Then Eq.\,(\ref{v7})
represents the first-order perturbation theory
in $\delta {\cal H}\!=\! \vec
v\vec \pi$; the unperturbed Hamiltonian is ${\cal H}_0\!=\!\pi_0$
(further  details can be found in \cite{optical}, Eq.\,(178) and
Sect.~VI).
In the second-quantized notations the very same relation  takes the
form
\beq
\state{H_Q(\vec v)}\;=\; \state{H_Q(0)}\; +\;
\int\; {\rm d}^3\vec{x}\;
\bar Q\,\pi_0^{-1}\vec{v}\vec{\pi}\,Q(x)\state{H_Q(0)}\;+\;
{\cal O}(\vec v^{\,2})\;.
\label{v8}
\eeq

The first four sum rules are obtained differently, using the OPE for the
heavy quark transition operator. The Bjorken sum rule basically states
that the probability of the quark to hadronize into some heavy flavor 
state after weak interaction, is exactly unity. In the different
notations it was stated already in Ref.~\cite{vshqs}. 
The `optical' sum rule for $\La$
follows from the fact that no independent dimension-$4$ heavy quark
operator exists for the forward matrix elements. This means that the
apparent change in the kinematics when passing from the quark level to
actual $B$ mesons must be compensated in the average energy of the final
states by the sums over excited states:
the threshold for quark transition is at 
$q_0 \!=\! \frac{\vec q^{\,2}}{2m_Q}$ while for the hadron the elastic
peak is at $q_0 = \frac{\vec q^{\,2}}{2M_{H_Q}}$, lower by
$\La\,\frac{\vec v^{\,2}\!}{2}$. (The original derivation by Voloshin was
different.) 

The new sum rules (\ref{ur}) and (\ref{ur2}) can be obtained applying the
OPE to the nonforward scattering amplitude where the momenta
flowing in and out the two weak vertices are not equal, so that the
overall change in velocity $\vec{v} \!=\! \vec{v}_1 \!-\!\vec{v}_2$ is
nonzero. They actually require the part antisymmetric in $\vec{v}_1$ and
$\vec{v}_2$.\footnote{This complication is not necessary for the sum
rules for operators 
dimension $5$ and higher ($\mu_G^2$, $\rho_{LS}^3$, {\it etc.}) where
the antisymmetric structure functions appear, for example, in $B^*$ for 
the spacelike vector current vertices, although they are $1/m_Q^2$
suppressed. However, the excited states contribute to these 
structure functions with extra factor of $\epsilon^2$, and thus yield only 
the higher sum rules for local operators.} 
The new parameter of the heavy quark theory $\overline{\Sigma}$ is
unfortunately unknown. 
It is defined as a small velocity elastic transition 
matrix element between the states with explicit spin of light degrees of
freedom:
for the ground-state vector mesons like $B^*$ 
\beq
\frac{1}{2M_{B^*}\!}\matel{H_Q(\vec{v},_{\!}\varepsilon')}{\bar{Q} iD_j Q(0)}
{H_Q(0,\varepsilon)} = 
- \frac{\La}{2}\, v_j \:(\vec\varepsilon\,'^* \vec \varepsilon\,)
+
\frac{\overline\Sigma}{2} \!
\left\{\!(\vec\varepsilon\,'^* \vec v\,)\varepsilon_j
\!-\!
\varepsilon'^*_j(\vec\varepsilon \,\vec v\,)\!\right\}\: 
 + 
{\cal O}\!\left(\!\vec v^{\,2}\!\right)^{\!}
.
\label{newsr7}
\eeq
Alternatively, for spinless heavy quarks $Q$ this would read as 
\beq
\matel{\Omega_0(\vec{v})}{\bar{Q} iD_j Q(0)}{\Omega(0)} = 
-\frac{\La}{2}\, v_j\, \Psi_0^\dagger \Psi_0 - i 
\frac{\overline{\Sigma}}{2} \,\epsilon_{jkl} \,v_k \,
\Psi_0^\dagger \sigma_{l} \Psi_0+ 
{\cal O}\left(\vec v^{\,2}\right)
\;,
\label{newsr9}
\eeq
{\it etc.}  
Let us note that the full (quasielastic) matrix element 
in the l.h.s.\ of Eq.\,(\ref{newsr7})
is not well defined due
to ultraviolet divergences in the static theory, and therefore the
symmetric part proportional to $\La$ depends on regularization. Yet the
antisymmetric part proportional to $\overline{\Sigma}$ is well defined
and finite. It can be directly measured, for example, on the lattices.
We expect $\overline{\Sigma}$ to be about $0.25\GeV$.

Let us briefly outline the unified derivation of the sum rules
Eqs.\,(\ref{bj})-(\ref{ur2}). The OPE approach 
leading to the sum rules is described
in Sect.~5. Here, however,  we consider the nonforward scattering
amplitude similar to Eq.\,(\ref{sr11}) with 
$J_\mu \!=\!J_\nu^\dagger\!=\!\bar{Q}\gamma_0 Q$ assuming that the final 
hadron has nonzero momentum $M_{H_Q}
\vec{\tilde v}\,$:
\beq
T (q_0; \vec{v},\vec{\tilde v})=
\frac{1}{2M_{H_Q}} 
\matel{H_Q(\vec{\tilde v})}{\int \; {\rm d}^3x\,{\rm d}x_0\;
{\rm e}\,^{i\vec{q}\vec x -iq_0x_0}\:
iT\{J_0(x),\, J_0^\dagger(0)\}  }{H_Q(0)}
\label{newsr31}
\eeq
with $\vec{q} \!\equiv\! m_Q \vec{v}$, and we retain only terms through
second order in $\vec{v}$ and $\vec{\tilde v}$. All terms suppressed by
powers of $1/m_Q$ are discarded. As the energy variable we take
$\epsilon=q_0-\left(\sqrt{\vec{q}^{\:2}\!+\!m_Q^2}\!-\!m_Q \right)\simeq
q_0\!-\!\frac{m_Q\vec{v}^{\:2}}{2}$. With this choice $\epsilon\!=\!0$
would correspond to the elastic transition for free heavy quark; the
larger $\epsilon$, the larger is the mass of the intermediate state. 

As explained in Sect.~5, we expand the amplitude  Eq.\,(\ref{newsr31})
in $1/\epsilon$ assuming $m_Q \!\gg \!\epsilon\!\gg\!\Lam$; the
dispersion relation equates the terms in the $1/\epsilon$ expansion with
the moments of the discontinuity of $T$ due to on-shell transitions into 
intermediate states. The latter to the second order in velocities
are the ground state ($B^{(*)}$) and the $P$-wave excitations. The OPE
yields
\bea
\nonumber
-T (\epsilon; \vec{v},\vec{\tilde v}\,)
\!\!\!\!&=&\!\!\!\!
\frac{1}{\epsilon} \:\frac{1}{2M_{H_Q}}
\matel{H_Q(\vec{\tilde v}\,)}{\,\bar{Q} (
1\!-\!\mbox{{\small $\frac{\vec
v^{\,2}}{4}$}}\!+\!\mbox{{\small $\frac{\vec{v}\vec{\gamma}}{2}$}}
\,) Q(0)\,}{H_Q(0)}\\
\nonumber
&+&\!\!\!\!
\frac{1}{\epsilon^2} \: \frac{1}{2M_{H_Q}}
\matel{H_Q(\vec{\tilde v}\,)}{\,\bar{Q} (i\vec{D}\vec{v}\,)
Q(0)}{H_Q(0)}\\
&+&\!\!\!\!
 \frac{1}{\epsilon^3} \: \frac{1}{2M_{H_Q}}
\matel{H_Q(\vec{\tilde v}\,)}{\,\bar{Q} [(i\vec{D}\vec{v}\,)^2
\!-\!
iD_0(i\vec{D}\vec{v}\,)]
Q(0)}{H_Q(0)}
\,+\ldots \;\;\;\;
\label{newsr33}
\eea
Comparing this to the explicit hadronic contributions to the small
velocity moments 
$1/2\pi\, \int {\rm d} \epsilon \, \epsilon^k \,\Im T(\epsilon;
\vec{v},\vec{\tilde v})$ we get the stated sum rules. The structures
antisymmetric in $\vec{v}$ and $\vec{\tilde v}\!-\!\vec{v}$ emerge if $H_Q$
carries spin correlated with the spin of light cloud, and for $B^*$
yield the new sum rules involving the difference between the
$\frac{3}{2}$- and $\frac{1}{2}$-contributions. The quasielastic
transitions do not contribute to this structure for actual
spin-$\frac{1}{2}$ quarks.

The computation proceeds similarly and is even simpler for spinless
heavy
quarks. The difference is only for the zeroth moment, sum rule
(\ref{ur}): for scalar $Q$ the antisymmetric part is absent from the OPE
expression for $I_0$, but the elastic contribution yields it with the
opposite sign.

The fact that the symmetric part of the $D=4$ nonforward elastic matrix element
(the first term in the r.h.s.\ of Eqs.\,(\ref{newsr7})-(\ref{newsr9}))
is given by $\frac{\La}{2}$ follows from equations of motion:
$$
M_{H_Q} ( v\!-\!u)_\mu v_\mu
\matel{H_Q(\vec{v}\,)}{\bar{Q} Q(0)}{H_Q(0)}=
\left.
-(iD_\mu  v_\mu) \,\matel{H_Q(\vec{v}\,)}{\bar{Q}
Q(x)}{H_Q(0)}\right\vert_{x\!=\!0}\;=
$$
$$
\matel{H_Q(\vec{v}\,)}{m_Q \bar{Q}Q(0)- \bar{Q}(iD_\mu 
v_\mu) Q(0)}{H_Q(0)}
$$
($u_\mu=(1,0,0,0)$ is the restframe four-velocity and $Q$ are the
original full QCD fields), or, to order $\vec{v}^{\,2}$
\beq
\left(M_{H_Q}\!-\!m_Q\right) \frac{\vec{v}^{\,2}}{2} \:
\matel{H_Q}{\bar{Q} Q(0)}{H_Q} = - v_k
\matel{H_Q(\vec{v}\,)}{\bar{Q} iD_k Q(0)}{H_Q(0)}
\label{newsr37}
\eeq
which fixes the symmetric term in the above matrix elements. 

The sum rule Eq.\,(\ref{ur}) deserves a special note. The sum of the
differences between $\tau^2_{3/2}$ and $\tau^2_{1/2}$ amounts to a
nonvanishing number which does not depend on strong dynamics. This can 
naturally raise a question of how this can be true, for the $\frac{3}{2}$ and
$\frac{1}{2}$ states are differentiated only by spin-orbital interaction. For
example, the light quark in the meson can be, in principle, arbitrary
nonrelativistic as well, in which limit this interaction can be
switched off.

The point is that in the nonrelativistic case $\tau^2$ are large
scaling like inverse square of the typical velocity of the light
quark, $\tau^2\sim 1/\vec v_{\rm sp}^{\,2}$, and the relativistic spin-orbital
effects must appear at the relative level $\sim \vec v_{\rm sp}^{\,2}$ 
simply 
because spin ceases to commute with momentum to this accuracy. This
then leads to corrections of order $1$ in the first two sum rules 
Eqs.\,(\ref{bj}), (\ref{ur}). In the Bjorken sum rule the 
relativistic constant $1/4$ is obscured by perturbative effects, while 
the sum rule Eq.\,(\ref{ur}) shows this universal relativistic effect
in the most clean way.

As is seen from the preceding analysis, physics of the SV transitions
is described by two ``structure functions'' $W_+$ and $W_-$:
\bea
\nonumber
W_+(\epsilon, \vec v\,) \!\! &=& \!\!
2\vec{v}^{\,2} \sum_m |\tau_{3/2}^{(m)}|^2
\delta(\epsilon\!-\!\epsilon_m)
+\vec{v}^{\,2} \sum_n |\tau_{1/2}^{(n)}|^2
\delta(\epsilon\!-\!\epsilon_n)
\:+\:\xi(\vec{v}^{\,2})\, 
\delta(\epsilon\!+\!\mbox{\small$\frac{\La\vec{v}^{\,2}}{2}
$})\,, \\
W_-(\epsilon)  &=& \;
\,2\;\sum_m |\tau_{3/2}^{(m)}|^2
\delta(\epsilon\!-\!\epsilon_m)\:
-\:2\,\sum_n |\tau_{1/2}^{(n)}|^2
\delta(\epsilon\!-\!\epsilon_n)
\label{newsr39}
\eea
and the corresponding invariant hadronic tensors
\bea
\nonumber
h_+(\epsilon, \vec v) \!\! &=& \!\!
2\vec{v}^{\,2} \sum_m \frac{|\tau_{3/2}^{(m)}|^2}{\epsilon_m\!-\!\epsilon}
+\vec{v}^{\,2} \sum_n \frac{|\tau_{1/2}^{(n)}|^2}{\epsilon_n\!-\!\epsilon}
\:-\:\frac{\xi(\vec{v}^{\,2})}{\epsilon\!+\! 
\mbox{\small$\frac{\La\vec{v}^{\,2}}{2}$}}\,, \\
h_-(\epsilon)  &=&\; 
\,2\;\sum_m \frac{|\tau_{3/2}^{(m)}|^2}{\epsilon_m\!-\!\epsilon}
\,-\,2\, \sum_n \frac{|\tau_{1/2}^{(n)}|^2}{\epsilon_n\!-\!\epsilon}
 \;.
\label{newsr41}
\eea
They directly enter, for example, $1/m$ corrections to various weak
currents at zero recoil. Sum rules  Eqs.\,(\ref{ur}) and (\ref{ur2})
represent the exact low-energy theorems for $h_-(\epsilon)$. 

Let us note that the sum rule (\ref{ur}) provides the rationale for the
fact that pseudoscalar mesons $B$, $D$ are lighter than their hyperfine
partners $B^*\!$, $D^*\!$. If the sum rule for $\mu_G^2$ is dominated by
the low-lying states then $\mu_G^2$ must be of the same sign as the
constant in Eq.~(\ref{ur}), which dictates the sign of the hyperfine
mass splitting.

The sum rules (\ref{bj})--(\ref{fourth}) obviously entail a set of exact
QCD inequalities. They are similar to those which have been with us
since the early eighties \cite{EIQCD}, and  reflect the most general
features of QCD (like the vector-like nature of the quark-gluon
interaction). The advent of the heavy quark theory paved the way to a
totally new class of inequalities among the fundamental parameters. As
with the old ones, they are based on the equations of motion of QCD and
certain positivity properties. All technical details of the derivation
are different, however, as well as the sphere of applications.
The first in the series
is the Bjorken inequality $\varrho^2 \ge 1/4$  \cite{BJSR}. We, in fact, 
have a stronger bound $\rho^2\ge 3/4$. 
Other bounds include
$\mu_\pi^2 \ge \mu_G^2$ and $\rho_D^3 \ge |\rho_{LS}^3|/2$, $\;\rho_D^3
\ge -\rho_{LS}^3$ and $\La\ge 2\overline\Sigma$ which are valid at
arbitrary normalization point. We also have the 
direct inequalities between the parameters of different dimension:
\beq
\mu_\pi^2 \ge \frac{3 \La^2}{4\varrho^2 \!-\!1} \;,\qquad 
\rho_D^3 \ge \frac{3}{8} \frac{\La^3}{\left(\varrho^2 \!-\!1/4\right)^2} 
\;,\qquad
\rho_D^3 \ge 
\frac{(\mu_\pi^2)^{3/2}}{\sqrt{3 (\varrho^2 \!-\!1/4)}} \;.
\label{pplrho}
\eeq
All these inequalities are saturated if only one excited state
contributes. 
A number of additional inequalities involving the energy of the first
$P$-wave excitation $\Delta_1$ is discussed in Ref.~\cite{rev}. 
For further reference, let us also write an obvious consequence of the two
equations for $\mu_\pi^2$  and $\mu_G^2$:
\beq
\mu_\pi^2 -  \mu_G^2 \;=\; 9 \sum_n \:\epsilon_n^2\,
|\tau_{1/2}^{(n)}|^2\; .
\label{p31a}
\eeq

\subsection{Hard QCD and normalization point dependence}

The sum rules (\ref{bj})--(\ref{fourth}) express heavy quark parameters,
including $\La\!=\! M_B\!-\!m_b$, $\,\mu_\pi^2$
and $\mu_G^2$ as the sum of observable quantities, products of the
hadron mass differences and transition probabilities. The observable
quantities are scale-independent. How then, say, 
$\La\!=\! M_B\!-\!m_b$, $\,\mu_\pi^2$ and $\mu_G^2$ 
happen to be $\mu$-dependent?

The answer is that in actual quantum field theory like QCD the sums
over excited states are generally ultraviolet 
divergent when $\epsilon_k \gg
\Lam$; in contrast to ordinary quantum mechanics they are not saturated 
by a few lowest states with contributions
fading out fast in magnitude with the excitation number. The 
contributions of hadronic states
with $\epsilon_k \gg \Lam$ are dual to what we calculate in
perturbation theory using its basic objects, quarks and gluons. The
latter yield the continuous spectrum 
and can be evaluated perturbatively using isolated quasifree heavy
quarks as the initial state. The final states are heavy quarks and a
certain number of gluons and light quarks. It is the difference between the
actual hadronic and quark-gluon transitions that resides at low excitation
energies.

Therefore, in order to make the sum rules meaningful, we must cut off
the sums at some energy $\mu$ which then makes the expectation values
$\mu$-dependent. The simplest way is merely to extend the sum only up
to $\epsilon_k<\mu$, and this is the convention we normally use. 
Thus, in reality all the sums in the relations (\ref{bj})--(\ref{fourth})
must include the condition $\epsilon_k < \mu$, which we omitted there
for the sake of simplicity, and all the heavy quark parameters are 
normalized at the scale $\mu$. The exception are superconvergent
spin-nonsinglet sum rules (\ref{ur}) and (\ref{ur2}) where in the
perturbative domain $\mu \ll \Lam$ such $\mu$-dependence is power
suppressed by factors $\alpha_s\Lam/\mu$ and can be neglected.
For analytic computations it is often convenient to apply the exponential
cutoff factor ${\rm e}^{-\epsilon_k/\mu}$, which is essentially the Borel
transform of the related correlation functions. 
Since at large $\mu$ the cutoff factors differ only in the
perturbative domain, the difference between various renormalization
schemes can be calculated perturbatively.

The high-energy tail of the transitions to order $\as$ is given by the
quark diagrams in Figs.~5 with
$$
2\sum_m ... + \sum_n ... \; \ra \; \int \frac{{\rm d}^3\vec k}{2\omega}
$$
where $(\omega,\vec{k})$ is the momentum of the real gluon. The
spin-singlet 
amplitudes are just a constant proportional to $g_s$, and performing the
simple calculations we arrive at the first-order term in the evolution
of $\mu_\pi^2(\mu)$:
\beq
\frac{{\rm d}\mu_\pi^2(\mu)}{{\rm d}\mu^2}\; = \; \frac{4}{3}\,
\frac{\as}{\pi} + \ldots
\;.
\label{160}
\eeq
Purely perturbatively, the
continuum analogies of $\tau_{1/2}$ and $\tau_{3/2}$ are equal and a
similar additive renormalization of $\mu_G^2$ and
$\rho_{LS}^3$ is absent.

\thispagestyle{plain}
\begin{figure}[hhh]
 \begin{center}
 \mbox{\epsfig{file=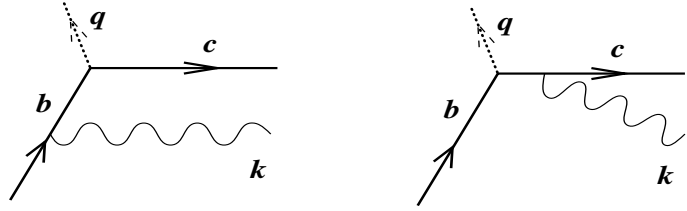,height=2.7cm,width=9cm}}
 \end{center}
\caption{ \small
Perturbative diagrams
determining the high-energy asymptotics of the heavy quark transition
amplitudes and renormalization of the local operators.}
\end{figure}

The perturbatively obtained evolution equations (\ref{160}),
(\ref{140}) allow one to determine the asymptotic values of
$\tau_{1/2}$ and $\tau_{3/2}$  at $\epsilon \gg \Lam$:
\bea
2\sum_{m} \epsilon_m^2|\tau_{3/2}^{(m)}|^2 +
\sum_{n} \epsilon_n^2 |\tau_{1/2}^{(m)}|^2 \!\!& \to & 
\frac{8}{9}\frac{\as(\epsilon)}{\pi}\: \epsilon\,  {\rm d}\epsilon\, 
\;,
\label{170}
\\
\sum_{m}
\epsilon_m^2|\tau_{3/2}^{(m)}|^2 -
\sum_{n}
\epsilon_n^2|\tau_{1/2}^{(n)}|^2  \!&\to & \!
-\frac{3\as(\epsilon)}{2\pi}\, \frac{{\rm d}\epsilon\,}{\epsilon}
\left\{
\sum_{\epsilon_m <\epsilon} \epsilon_m^2|\tau_{3/2}^{(m)}|^2 \!-\!
\sum_{\epsilon_n <\epsilon} \epsilon_n^2|\tau_{1/2}^{(n)}|^2
\!\right\}. \;\;\;\qquad
\label{171} 
\eea
The last bracket is simply $\frac{1}{6}\mu_G^2(\epsilon)$.
Eq.\,(\ref{170}) can be extended to higher orders in $\alpha_s$, this
simply reduces to using the dipole coupling $\alpha_s^{(d)}(\epsilon)$
discussed in Sect.~2.4.1:
\beq
\alpha_s^{(d)}(\epsilon) = 
\bar\alpha_s\left({\rm e}^{-5/3+\ln{2}} \epsilon\right)
- \left(\frac{\pi^2}{2}-\frac{13}{4} \right)
\frac{\as^2}{\pi} \,+\,{\cal O}(\as^3)\;.
\label{new96}
\eeq
($\bar\alpha_s$ is the standard $\overline{\rm MS}$ strong coupling).
Therefore we get a number of exact perturbative evolution equations
\cite{dipole}
\bea
\mu \frac{{\rm d}\rho^2(\mu)}{{\rm d}\mu} &=&
\frac{8}{9}\,\frac{\alpha_s^{(d)}(\mu)}{\pi} \\
\frac{{\rm d}\La(\mu)}{{\rm d}\mu} &=&
\frac{16}{9}\,\frac{\alpha_s^{(d)}(\mu)}{\pi} \\
\frac{{\rm d}\mu_\pi^2(\mu)}{{\rm d}\mu} &=&
\frac{8}{3}\,\frac{\alpha_s^{(d)}(\mu)}{\pi}\, \mu
\label{new100}
\eea
(the evolution for the Darwin operator is already more complicated
since includes the usual anomalous dimension \cite{vslog}).
The similar questions for the alternative combination of the oscillator
strengths entering $\mu_G^2$ and $\rho_{LS}^3$ has not been studied yet.

\subsection{On the saturation of the sum rules}

The question of the saturation of the heavy quark sum rules (in
particular, the lower ones Eqs.\,(\ref{bj})-(\ref{pig})) is of primary 
importance for phenomenology of the heavy quark expansion. The sum
rules state what is the asymptotic value in the r.h.s.\ at large
enough normalization point $\mu$, and perturbation theory tells us
their $\mu$ dependence. From which scale $\mu_0$ this
behavior applies, is a dynamical question. For superconvergent sum
rules (\ref{ur}) and (\ref{ur2}) this is the question at which scale
the sums approach the stated values with a reasonable accuracy. In
order to sensibly apply quantitative $1/m_Q$ expansion, one must have
$m_Q \!>\! \mu_0$, possibly, $m_Q\!\gg\!\mu_0$. 
While this is probably the case for
$b$ particles, such a hierarchy is not obvious {\it  a priori} in charm.

The existing numerical evaluations of $\La$ and $\mu_\pi^2$ at the
scale near $1\GeV$ suggest rather large values, approximately
$0.7\GeV$ and $0.6\GeV^2$, respectively, which impose rather tight
constraints. These facts are often neglected under various pretexts,
including challenging the accuracy of the QCD sum rules evaluations of
$\mu_\pi^2$. The significant uncertainties, in principle, cannot be
excluded. In a certain respect determination of $M_B-m_b$ from
$e^+e^- \!\to\! b\bar b$ is somewhat indirect as well including mild
theoretical assumptions. Indeed, even though QCD ensures that the
quark mass appearing in the analysis of $B$ decays and $b\bar b$
annihilation is the same, extrapolating from the decay (scattering) 
kinematics appropriate for $B$ decays with small negative $q^2$ for
the $b\!\to\! b$ transitions, to the timelike domain 
$q^2 \!\simeq \! 4m_b^2$ in 
the annihilation channel involves the long path. It brings in
significant perturbative effects which are to be accounted for. If 
actual dynamics of gluonic degrees of freedom sets in to the
perturbative regime late in energy, the associated uncertainties may,
in principle, increase.

Nevertheless, we point out that certain constraints following from the 
sum rules can be confined within the world with a single heavy quark,
and they still are tight. Namely, the value of $\mu_G^2\simeq
0.4\GeV^2$ has barely been challenged as extracted almost directly
from $B^{(*)}$ and  $D^{(*)}$ masses. By virtue of the sum rules, the
value of $\mu_\pi^2$ is at least as large. Thus, regardless of the
accuracy in evaluations of the kinetic expectation  value, the
question can be phrased in terms of the generally accepted value of
$\mu_G^2$. At which minimal scale $\mu_0$ the value of
$\mu_G^2(\mu_0)$ reaches $0.3$ or $0.4\GeV^2$? If this scale is below
$1\GeV$, large $\La$ and $\mu_\pi^2$ are almost inevitable. If,
however, $\mu_G^2(1\GeV)$ is significantly below $0.4\GeV^2$, the
chances for success in $1/m_Q$ expansion in charm are slim.

The estimates for $\tau_{3/2}$ and $\tau_{1/2}$ for the lowest $P$
wave states group around $0.4$, with  
$\epsilon^{(1)}_{3/2}\gsim \epsilon^{(1)}_{1/2}\simeq 400\;
\mbox{to}\; 500\MeV$ (for the review see \cite{fazio}). Similar values 
were reportedly extracted from the overall experimental yield of the
corresponding charmed $P$ wave states \cite{tauexp}. 
It is evident that such oscillator strengths fall short in saturating
the sum rules:
$$
\delta^{(1)}_{3/2}\,\mu_G^2 \simeq 0.2\GeV^2\,, \qquad
\delta^{(1)}_{3/2}\, \La \simeq 0.3\GeV\;.
$$
In principle, they alone should not necessarily,
even though the idea of the dominance of the lowest states contribution is 
very appealing. Let us mention that in the 't~Hooft model all the
heavy quark sum rules are saturated with amazing accuracy by the
first excitations \cite{lebur,burkur}. Is such a possibility 
excluded in QCD? Probably, not completely. The dominance of the first
excitation with $\epsilon\simeq 500\MeV$ (recall that one must use the
asymptotic $m_Q\!\to\infty$ values of the excitation energies and
amplitudes) is still possible if QCD sum
rules underestimate the value of $\tau^{(1)}_{3/2}$.\footnote{Let us
note that the technology of the QCD sum rules assumes the approximate
duality starting $\epsilon\!=\!1\GeV$ or even lower (the energy are
counted from the heavy quark mass there). Therefore accepting poor
saturation of the exact heavy quark sum rules at this scale and
still relying on the QCD sum rules predictions is not self-consistent.} 
Experimental determinations of $\tau$'s are also questionable since
$1/m_c$ corrections are not accounted for there. The estimates in the
't~Hooft model suggest that they can be very large. In the cases when
they are known explicitly in QCD, the $1/m_c$ terms generally turn out 
to be very significant as well \cite{vain}.

Another -- and, apparently, the most natural -- option is that there
are new states with the masses around $700\MeV$ with similar, or even
larger $\tau^{(2)}_{3/2}\simeq 0.4 \:\mbox{to}\: 0.5$; they 
can be broad and not 
identified with clear-cut resonances. The $\frac{1}{2}$-states must be yet
depleted up to this scale. All such states can be produced in
semileptonic $b$ decays and observed as populating the domain of
hadronic invariant mass below or around $3\GeV$. It will be important
to explore these questions in experiment.

The similar constraints and the pattern of saturation follow from
the sum rule (\ref{ur2}) and, in particular, (\ref{ur}). The most
natural ``favorable'' solution would imply existence of $\frac{3}{2}$ states
around $\epsilon\approx 700\MeV$ with significant $\tau_{3/2}\approx
0.5$, and suppressed transition amplitudes for the $\frac{1}{2}$ states in the
whole domain of $\epsilon$ below $1\GeV$.

\section{Heavy Flavor Sum Rules; Finite $m_Q$}

In practical applications we usually need to know the decay
amplitudes for actual $b$ particles; therefore, the mass of the initial 
heavy quark, although large compared to $\Lam$, is finite and the
corresponding corrections must be taken into account. The $c$ quark in
the final state is only marginally heavy, and these corrections are
typically very significant. In $b\to u$ transitions the final state
quark is light and the literal $1/m_Q$ expansion becomes inapplicable.
However, in certain cases -- for example, in the sum rules -- the
energy of the final hadronic state rather than $m_q$ itself is an
expansion parameter, and one still can use the similar expansion as 
long as
the transferred spacelike momentum is large \cite{optical}. 

The finite-$m_Q$ sum rules are less compact for a number of reasons.
First, one needs to account for the various $1/m_Q$-suppressed terms;
yet the corrections are expressed in terms of the
hadronic parameters we have encountered already analyzing the static SV sum
rules. This is true if one expresses everything in terms of the
quark-level kinematics. Since in practice we count all the
energies from the observable hadronic thresholds, the nonlocal
correlators which determine the expansion of the heavy hadron mass in
powers of $1/m_Q$ enter as well. This is the only way how, for example,
$\La$ enters the sum rules.
Another source of nonlocal correlators of 
heavy quark operators is the deviation of the expectation
values over the initial hadron ($B$ meson) from their asymptotic values
which would exist if $m_b \to \infty$.

Second, beyond the static approximation external weak currents
proliferate, and as much as five independent structure functions
describe the decays even for the standard $V\!-\!A$ current. Likewise,
we often have to go beyond the SV approximation and do not assume that
the final state velocity is small. Therefore, the structure functions
as in usual deep inelastic scattering (DIS) become dependent on two
kinematic variables, say $q_0$ and $q^2$. However, to keep the parallel
with the heavy quark and the SV limit it is often advantageous to choose
instead $q_0$ and $|\vec{q}\,|=\sqrt{q_0^2\!-\!q^2}$. (This 
choice is also convenient in a number of practical applications 
to $b\to c$ decays.)  

The sum rules are derived in QCD using the standard methods of the
short-distance expansion \cite{optical}. One starts the analysis from
the forward transition amplitude
\beq
T^{(12)}(q_0; \vec{q}\,)\;=\; \frac{1}{2M_B} \int\; {\rm d}^3\vec{x}\: 
{\rm d}x_0\; {\rm
e}\,^{i\vec{q}\vec x -iq_0x_0}\, \matel{B}{iT\{\bar c \Gamma^{(1)}
b(x), \,
\bar b \Gamma^{(2)} c(0)\}}{B}
\label{s2}
\eeq
where $\Gamma^{(1,2)}$ are some spin structures or, more generally,
local operators. The transition amplitude contains a lot of information
about the decay probabilities. As usual in QCD, one cannot calculate it
completely in the physical domain of $q$. The amplitude (\ref{s2}) has
several  cuts corresponding to different physical processes. The
discontinuity at the physical cut 
$q_0 < M_B\!-\! \sqrt{M_D^2\!+\!\vec{q}^{\:2}}$ 
describes the inclusive decay probabilities at a given energy released
into the final hadronic system. The cut continues further than the
domain accessible in the actual decays, see Fig.~6.

\thispagestyle{plain}
\begin{figure}[hhh]
 \begin{center}
 \mbox{\epsfig{file=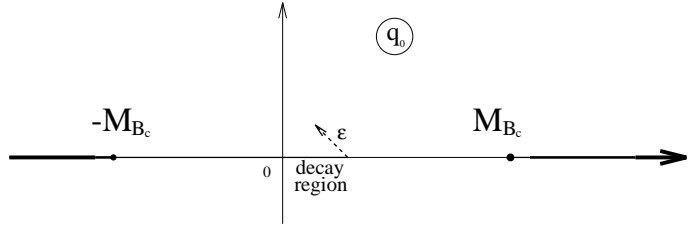,width=9.cm}}
 \end{center}
\caption{ \small
Cuts of the transition amplitude in the complex $q_0$ plane. The
physical cut
for the weak decay starts at $q_0\!=\!M_B \!-\!
(M_D^2\!+\!\vec{q}^{\:2})^{1/2} $ and continues towards
$q_0\!=\!-\infty$.
Other physical processes generate cuts starting near
$q_0\!=\!\pm
(M_{B_c}^2\!+\!\vec{q}^{\:2})^{1/2}
$ (one pair); another pair of cuts originates
at close values of $q_0$.
An additional channel opens at
$q_0 \!\gsim \!2m_b \!+\!m_c$. }
\end{figure}

For the practical case of left-handed currents $\Gamma^{(1)}\!=\!
\gamma_\mu(1\!-\!\gamma_5)$, $\,\Gamma^{(2)}\!=\!
\gamma_\nu(1\!-\!\gamma_5)$ and 
the hadronic tensor $T^{(12)}(q_0;\,\vec{q}\,)$ can be decomposed into
five covariants \cite{koyrakh}
\beq
T_{\mu\nu} = -h_1 g_{\mu\nu}
+ h_2 v_\mu v_\nu - ih_3 \epsilon_{\mu\nu\alpha\beta} v_\alpha q_\beta
+ h_4 q_\mu q_\nu + h_5 (q_\mu v_\nu\!+\! q_\nu v_\mu)
\;,
\label{98opt}
\eeq
each having the same analytic and unitarity properties. In
particular, they obey the dispersion relation
\beq
h_i(q_0) =\frac{1}{2\pi}\int\;
\frac{w_i ({\tilde q}_0)d{\tilde q}_0}{{\tilde q}_0 \!-\! q_0}
\; + \mbox{ polynomial}
\;,
\label{99opt}
\eeq
where $v_\mu\!=\!(1,0,0,0)$ is the four-velocity of the decaying meson and 
$w_i$ are observable structure functions:
\beq
w_i =2\,{\rm Im}\, h_i \;.
\eeq
For decays into massless leptons the structure functions $w_4$ and $w_5$
do not contribute due to conservation of the lepton current; they are
proportional to $m_{\ell}^2$ and are relevant only for the decays into
$\tau$ lepton \cite{koytau}.

In QCD we  calculate the amplitude (\ref{s2}) away from all its cuts.
Essentially, it can be expanded  in the inverse powers of the distance
from the physical cut, $\epsilon\,$:
\beq
\epsilon =  M_B\!-\!\sqrt{M_D^2\!+\!\vec{q}^{\:2}} \!-\! q_0 =
m_b \!-\! \sqrt{m_c^2\!+\!\vec{q}^{\:2}}+\delta(\vec q^{\:2}) 
\!-\! q_0\;,
\label{s3}
\eeq
where
$$
\delta(\vec q^{\,2})=
(M_B\!-\!m_b)-\left(\sqrt{M_D^2\!+\!\vec{q}^{\:2}}-
\sqrt{m_c^2\!+\!\vec{q}^{\:2}}\right)
\sim
{\cal O}\left(\Lam\frac{\vec{q}^{\:2}}{m_Q^2}\right) +
{\cal O}\left(\frac{\Lam^2}{m_Q}\right)\;.
$$
The ``off-shellness" $\epsilon$ must be chosen in such a way that that
$|\epsilon|$ and $|\epsilon \cdot \arg{\epsilon}| \gg \Lam$ but,
simultaneously, $|\epsilon| \!\ll \!\sqrt{m_c^2\!+\!\vec{q}^{\,2}}\, ,\,m_b$. 
The second requirement allows us to ``resolve" the contributions of the
separate cuts.

How does the deep Euclidean expansion of $T^{(12)}$ help to constrain the
amplitude in the physical domain of real $\epsilon$ lying just on the
cut? The dispersion relations  Eq.\,(\ref{99opt}) tells us that the 
coefficients
in the expansion of the amplitude in powers of $1/\epsilon$ are given by
the corresponding moments of the spectral density in $T^{(12)}$,
\beq
T^{(12)}(\epsilon; \vec{q}\,)\;=\; \frac{1}{\pi} \,\int\;
\frac{\Im T^{(12)}(\epsilon'; \vec{q}\,)}{\epsilon' \!-\!\epsilon}\;
d\epsilon'
\;=\;
-\frac{1}{\pi} \, \sum_{k=0}^\infty \:\frac{1}{\epsilon^{k+1}}\,\int\;
\Im T^{(12)}(\epsilon'; \vec{q}\,)\,\epsilon'^k\; d\epsilon'
\;;
\label{s4}
\eeq
the similar expansions hold for separate invariant structures.

On the other hand, we can build the large-$\epsilon$ expansion of the
transition amplitude {\em per se}, treating it as the propagation of the
virtual heavy quark submerged into a  soft medium. The expansion takes
the general form 
$$
T^{(12)}(q_0;\vec{q}\,)\;=\;
\frac{1}{2M_B} \int\; {\rm d}^3\vec{x}\: {\rm d}x_0 \;{\rm
e}\,^{i\vec{q}\vec x -iq_0x_0} \;\times
$$
$$
\matel{B}{\bar b(x) \Gamma^{(1)}
\: \left[(m_c\!+\! i\!\not\!\!{D})
\frac{1}{m_c^2 \!-\!(iD)^2 \!-\! \frac{i}{2}\sigma G}\right]_{x0}
\:\Gamma^{(2)} b(0)}{B}\;=
$$
$$
\aver{\Gamma^{(1)}  [m_c \!+\! m_b\gamma_0 -\!\!\not\!{q} +\!\not\!\pi ]
\frac{1}{m_c^2\!-\!(m_b\!-\!q_0)^2 \!+\!\vec q^{\,2} \!-\! 2m_b 
\pi_0\!+\!2q\pi \!-\! \pi^2\!-\!\frac{i}{2}\sigma G}
\:\Gamma^{(2)}}\;=
$$
\vspace{0.1cm}
$$
\aver{\Gamma^{(1)}
[m_c\!+\!(E_c\!+\!\epsilon\!-\!\delta(\vec q^{\,2}))\gamma_0\!+\!
\vec q\vec\gamma+\!\not\!\pi] \times
$$
\vspace{0.1cm}
\beq
\sum_{n=0}^\infty
\frac{\left[2(E_c\!+\!\epsilon\!-\!\delta(\vec q^{\,2}))\pi_0 
\!+\! 2\vec q \vec \pi
\!+\! \pi^2\!+\!\frac{i}{2}\sigma G\right]^n}
{\left[-(\epsilon\!-\!\delta(\vec q^{\,2}) )(2E_c \!+\! \epsilon
\!-\!\delta(\vec q^{\,2}))\right]^{(n+1)}}
\:\Gamma^{(2)}}\;.
\label{s7}
\eeq
\vspace{0.1cm}
Here $E_c\!=\!\sqrt{m_c^2\!+\!\vec{q}^{\:2}}\,$, the symbol $[...]_{x0}$
denotes the $0x$ matrix element of the corresponding operator
(it can be understood as the integral operator in the coordinate
representation), and we use the short-hand
notation
$$
\aver{...} = \frac{1}{2M_B}\, \int\; {\rm d}^3\vec{x}\: {\rm d}x_0
\,\matel{B}{\bar b(x) [...]_{x0} b(0)}{B}\;.
$$

In the first equation we used the full QCD fields and then passed  to
the low-energy fields according to Eqs.~(\ref{25})--(\ref{26}).
Picking up the corresponding term $1/\epsilon^{k+1}$ in the expansion
over $1/\epsilon$ at $\epsilon \gg \Lam $ and evaluating the expectation
value of the resulting {\em local} operators (e.g., $\bar b(0)\pi_\mu
b(x) \equiv \delta^4(x)\, \bar b\pi_\mu b(0)$), one gets the  sum rules
sought for. Taking $k\!=\!0$ yields the sum rule for the equal-time
commutator of the currents $\bar c \Gamma^{(1)} b $ and $\bar b
\Gamma^{(2)} c$; $\,k\!=\!1$ selects the commutator for the time 
derivative, {\it etc.} 

Applying this general procedure to the heavy quark transition amplitude,
we face a subtlety which is specific to the OPE in heavy quarks.
Namely, speaking formally the dispersion relation represents the
amplitude as a Cauchy integral over {\em all} cuts it has, while we try
to reconstruct the structure functions (or their moments) referring to a 
particular cut
describing the heavy quark decay process. It was first discussed in
Ref.\,\cite{optical}. Performing the OPE we identify the integrals over
the separate cuts in the quark Green functions with the corresponding
cuts in the physical amplitudes. This identification is established
order by order in $1/m_Q$ expansion since the contributions of the physical
cut behave as powers in $1/\epsilon$ while the contributions of the
other, ``distant'' cuts behave like $1/(\epsilon \!+ \!M)^n$  where $M\sim
m_Q $ is the distance to the other cut, and therefore they are regular at 
$\epsilon \!\ll\! m_Q$. This property was given the name of {\it global
duality} \cite{optical}. Beyond the practical version of the OPE global
duality is, in general, violated and may play the role of
additional theoretical background together with local duality
violations.\footnote{It is probable, however, that global duality is
not too sensitive to the peculiarities of Minkowskian kinematics, and
its violations are truly exponentially suppressed at large $m_Q$.}

The problem of separating the extra cut contributions would become
serious, for example, in the $b\to u$ transitions if $\sqrt{q^2}$
approaches $m_b$. The $u$-channel cut then becomes too close to the
physical decay cut \cite{cgg}, and their contributions in the correlator
Eq.\,(\ref{s2}) cannot be  taken
apart. This complication originates from the presence of the 
reverse ordering of the two weak decay vertices in the correlator. 
Naively, it could have been avoided if one considered the retarded --
instead of the time-ordered -- product in the correlation function, or
simply defined 
the hadronic tensor as the dispersion integral only over the discontinuity
of the decay cut (continued towards larger $q_0$ accessible, in
principle, in the process of scattering of the weak current on the heavy
hadron). However, in order to apply the technique of the OPE, we must
have two necessary ingredients: first, the amplitude must have the proper
analytic properties, and, second, be represented as the correlator of
local field operators for which we can write the  functional
representation in the background gluon field. The combination of the
two requirements dictates using the $T$-ordered product in
Eq.\,(\ref{s2}), and thus makes the question of global duality
unavoidable.

We quote here the expressions for the first three moments of the
structure functions $w_{1,2,3}$ for the standard $V\!-\!A$ weak currents
\cite{optical}; the higher moments require accounting for the
nonperturbative effects beyond ${\cal O}(\Lambda_{\rm QCD}^2)$.
The variable $\epsilon$ describing the excess of the final state hadronic 
energy above the threshold is defined as 
\begin{equation}
\epsilon=M_B - \sqrt{M_{D^*}^2 \!+\! \vec{q}\,^2}-q_0
\;,
\label{q1}
\end{equation}
i.e., is counted from the energy of the $D^*$ meson (rather from the
lowest state $D$ contributing to the sum rules); this convention 
does not affect the zeroth moments. Moments of the structure 
functions are
\begin{equation}
I_n^{(i)}(\vec{q}\,^2)=  \frac{1}{2\pi}
\int \;{\rm d}\epsilon \,\epsilon^n\,w_i (\epsilon,\, \vec{q}\,^2)
\label{q2}
\end{equation}
where $i$ labels the structure function; they can be considered
separately for axial current ($AA$), vector current ($VV$) and for the
interference of the two ($AV$).

For the zeroth moments we have
\begin{eqnarray}
&&I_0^{(1)AA}  =
 \frac{E_c\!+\!m_c}{2E_c}-\frac{\mu_\pi^2\!-\!\mu_G^2}{4E_c^2}
\frac{m_c}{E_c}\left[\frac{m_c^2}{E_c^2}\!+\!\frac{E_c^2}{m_b^2}
\!+\!
\frac{2}{3}\frac{m_c}{m_b}\right]-\frac{\mu_G^2}{3E_c^2}
\frac{m_c}{E_c} \frac{E_c^2\!+\!3m_c^2}{4E_c^2} \nonumber \\
&& \label{m0a1}\\
&& I_0^{(2)AA} = \frac{m_b}{E_c} \left\{
1-\frac{\mu_\pi^2\!-\!\mu_G^2}{3E_c^2}
\left[2\!-\!\frac{5}{2}\frac{E_c^2}{m_b^2}+
\frac{3}{2}\frac{m_c^2}{E_c^2}\right]
- \frac{\mu_G^2}{3E_c^2}
\left[\frac{1}{2}\!+\!\frac{m_c}{m_b}\!+\!
\frac{3}{2}\frac{m_c^2}{E_c^2}\right]\right\}\nonumber \\
&&\label{m0a2}\\
&& I_0^{(3)AV} = -\frac{1}{2E_c} \left\{
1-\frac{\mu_\pi^2\!-\!\mu_G^2}{3E_c^2}
\left[1\!+\!\frac{3}{2}\frac{m_c^2}{E_c^2}\right]
- \frac{\mu_G^2}{2E_c^2} \left[1 \!+\!
\frac{m_c^2}{E_c^2}\right]\right\}\;\;.
\label{m0av3}
\end{eqnarray}
Expressions for the $VV$ functions are obtained from the axial ones by
replacing $m_c \to -m_c$; the structure functions $w^{(1,2)AV}$
and $w^{(3)AA,VV}$ vanish. The above equations are analogies of the
Bjorken sum rule; however they incorporate nonperturbative effects which
appear at $1/m_Q^2$ level; the corrections are not universal and differ
explicitly for different currents and structure functions.

The first moments look as follows:
\begin{eqnarray}
&&I_1^{(1)AA} =  \frac{E_c\!+\!m_c}{2E_c}
\left\{ \frac{\mu_\pi^2\!-\!\mu_G^2}{2E_c}
\left[1\!-\!\frac{E_c}{m_b} \!-\! 
\frac{1}{3}\left(1\!-\!\frac{m_c}{E_c}\right)
\left(1\!+\!3\frac{m_c}{E_c}
\!+\! 2\frac{E_c}{m_b}\right)\right]+
 \right. \nonumber \\
&&~~~~~~~~~~~~~~~~~~~~~
  + \, \left. \frac{\mu_G^2}{2E_c}\left[1 \!-\!
\frac{2}{3}\frac{m_c}{E_c} \!+\! \frac{m_c^2}{E_c^2}\right]+
\left[(M_B\!-\!m_b) \!-\! (E_{D^*}\!-\!E_c)\right] \right\}
\label{m1a1} \\
&& I_1^{(2)AA} = \frac{m_b}{E_c} \left\{\frac{\mu_\pi^2\!-\!\mu_G^2}{3E_c}
\left[2 \!-\!\frac{7}{2}\frac{E_c}{m_b} \!+\!
\frac{3}{2}\frac{m_c^2}{E_c^2}\right]
+
\frac{\mu_G^2}{3E_c}
\left[\frac{1}{2}\!-\!\frac{E_c\!-\!m_c}{m_b}\!+\!
\frac{3}{2}\frac{m_c^2}{E_c^2}\right]+
\right. \nonumber \\
&&~~~~~~~~~~~~~~~~~~~~~~~
  +  \, \left.
\left[(M_B\!-\!m_b)\!-\!(E_{D^*}\!-\!E_c)\right] \right\}
\label{m1a2}  \\
&& I_1^{(3)AV} = -\frac{1}{2E_c} \left\{\frac{\mu_\pi^2\!-\!\mu_G^2}{3E_c}
\left[1\!-\!\frac{5}{2}\frac{E_c}{m_b} \!+\!
\frac{3}{2}\frac{m_c^2}{E_c^2}\right]
+
\frac{\mu_G^2}{2E_c}
\left[1\!+\!\frac{m_c^2}{E_c^2}\right]+ \right. \nonumber \\
&&~~~~~~~~~~~~~~~~~~~~~~~
\left.
+ \, \left[(M_B\!-\!m_b)\!-\!(E_{D^*}\!-\!E_c)\right] \right\}
\label{m1av3}
\end{eqnarray}
At $\vec q \!=\!0$ these relations determine $1/m_Q$ terms in the masses of
heavy mesons. Their derivatives with respect to $\vec q\,^2$ near zero
recoil give the Voloshin's ``optical'' sum rule. Here they are obtained
with better accuracy for arbitrary, not necessary small, velocity and
incorporate $1/m_Q$ relative corrections. The latter appear to be quite
sizable when $\vec q$ is not particularly large.

The third sum rules, which are relations for the second moments of the
structure functions, are calculated only in the leading non-trivial
approximation. They look rather simple and manifestly satisfy the heavy
quark symmetry relation \cite{third}: 
$$
\frac{2E_c}{E_c\!+\!m_c}I_2^{(1)AA}= \frac{2E_c}{E_c\!-\!m_c}I_2^{(1)VV}=
\frac{E_c}{m_b} I_2^{(2)AA} = \frac{E_c}{m_b} I_2^{(2)VV} = -2E_c
I_2^{(3)AV} = 
$$ 
\begin{equation}
=\;
\frac{\mu_\pi^2}{3}\:\frac{E_c^2\!-\!m_c^2}{E_c^2}
\,+\,\overline\Lambda^2 \!\left(1\!-\!\frac{m_c}{E_c}\right)^2 
\;\;.
\label{second} 
\end{equation} 
All higher moments vanish in our approximation. We used above the
notation $E_{D^*}$ for the energy of $D^*$ and $E_c$ for the energy of
the $c$ quark in the free quark decay:
\begin{equation}
E_{D^*}=\sqrt{M_{D^*}^2\!+\!\vec{q}\,^2}\;\;,
\;\;\;\;E_c=\sqrt{m_c^2\!+\!\vec{q}\,^2}\;\;.
\label{Ec}
\end{equation}
The quantity $(M_B\!-\!m_b) \!-\!(E_{D^*}\!-\!E_c)$ which enters 
first and second moments determines the difference in
kinematics between the free quark and the quasielastic $B\to D^*$
transitions. At zero recoil $\vec{q}\!=\!0$ it is of order $\Lam^2$,
however in the general situation scales like  $\Lam^1$:
$$
(M_B\!-\!m_b)-(E_{D^*}\!-\!E_c)\;\simeq
$$
\begin{equation}
\overline\Lambda\left(1\!-\!\frac{m_c}{E_c}\right) - 
(\mu_\pi^2\!-\!\mu_G^2)
\left(\frac{1}{2E_c}\!-\!\frac{1}{2m_b}\right)
- \frac{2\mu_G^2}{3E_c} -
\frac{\overline\Lambda^2}{2E_c}\left(1\!-\!\frac{m_c^2}{E_c^2}\right)
\;+\;{\cal O}\!\left(\frac{1}{m^2}\right)\,.
\label{Delta}
\end{equation}

The explicit expressions for the structure functions themselves
including these nonperturbative corrections, and the semileptonic decay
distributions are given in Ref.\,\cite{koyrakh}. Pure perturbative
corrections to the heavy quark structure functions were computed in
Ref.\,\cite{czarsf}.

\subsection{Zero recoil sum rules; $|V_{cb}|$ from $B \ra D^* \ell \nu $
at zero recoil.}

\subsubsection{Sum rules for $\mu _{\pi}^2$ and $\mu _G^2$}

Considering the semileptonic transitions driven by the pseudoscalar weak
current $J_5= \int {\rm d}^3 \vec{x}\, \bar c i\gamma_5 b(x)$ 
(i.e., at zero recoil $\vec q \!=\! 0$) one obtains the sum rule 
\beq 
\frac{1}{2\pi} \:\int_{0}^{\mu} w^{(5)}(\epsilon)\;  d\epsilon 
\; = \; 
\sum_{\tilde \epsilon_k<\mu} |\tilde F_k|^2 \; = \; 
\left(\frac{1}{2m_c}\!-\!\frac{1}{2m_b}\right)^2 
\left( \mu_\pi^2(\mu) \!-\!\mu_G^2(\mu)\right)  
\label{z5} 
\eeq 
yielding the inequality $\mu_\pi ^2(\mu) \!\ge\!  \mu_G^2(\mu)$; this
is the field-theoretic analogue of the relation Eq.\,(\ref{p31a}). This
correspondence is transparent keeping in mind the nonrelativistic 
expansion of the pseudoscalar current $\bar c i\gamma_5 b \simeq
(1/2m_c\!-\!1/2m_b)\, \varphi_c ^+ \vec\sigma \vec\pi \varphi_b$ at zero
momentum transfer. Only the transitions to the $P$-wave states survive
here in the leading in $1/m_Q$ approximation.  

The sum rule (\ref{3.7a}) for $\mu_G^2$ can also be readily obtained in
this way. For example, one can consider the antisymmetric (with
respect to $i$ and $j$)  part of the correlator of the vector currents
($\Gamma^{(1)}\!=\! \gamma_i$, $\Gamma^{(2)}\!=\! \gamma_j$); 
such a 
sum rule must be considered for $B^*$ mesons (in $B$ the expectation value
of $\vec{B}_{\rm chr}$ vanishes). 
Alternatively, the sum rule for the correlator
of the nonrelativistic currents $\bar c \pi_j b$ and  $\bar b
\vec\sigma \vec \pi c$ directly can be considered. The OPE guarantees
that all such relations are equivalent.

\subsubsection{$F_{D^*}$ at zero recoil}

The concept of the heavy quark symmetry was very important for the
evolution of studies of heavy flavor hadrons. The fact of the fixed
normalization of the $B\!\to\! D$ and $B\!\to\! D^*$ formfactors at 
zero recoil
in the limit $m_{b,c}\!\to\! \infty$ was of special significance in
applications since suggested the method for determination of $|V_{cb}|$
from the $B\ra D^*$ semileptonic decay channel near zero recoil.
To this end one measures its differential rate, extrapolates to the 
point of zero recoil and gets the quantity $|V_{cb}\,F_{D^*}(0)|$, where
$F_{D^*}$ is the axial $B \to D^*\, \ell \nu $ formfactor. Since the charm
quark is only marginally heavy, it is very important to estimate the
corrections, in particular, nonperturbative. The exclusive transition
amplitudes are not genuinely short-distance, such transitions proceed in
time intervals $\sim \!1/\Lam$. Nevertheless, it turns out that the
large-distance effects appear in these kinematics only suppressed by
$1/m_{c,b}^2 $:
\beq
F_{D^*}(0) = 1 + {\cal O}\left(\frac{\as}{\pi}\right) +
\delta^A_{1/m^2} + \delta^A_{1/m^3}\,+\; ...
\label{8.10}
\eeq

The absence of $1/m_Q$ corrections was first noted in
Ref.~\cite{vshqs}.\footnote{The first preprint version of the paper
\cite{shifheid} discussing the heavy quark symmetry in QCD was given to
me by M.~Shifman in July 1986.} It can be readily understood. Let us
consider, for example, the vector $B\!\to\! D$ transition at zero recoil:
\beq
\matel{D}{\bar c
\gamma_0 b}{B}\;=\; 2\sqrt{M_B M_D} \left\{1+\frac{a}{m_c}-
\frac{a}{m_b}+\,...\right\}
\label{190}
\eeq 
(short-distance effects are neglected). The relative magnitude of
$1/m_b$ and $1/m_c$ terms is fixed since at $m_b=m_c$ all corrections
must vanish identically. $T$-invariance, however, says that the
coefficients for $1/m_c$ and $1/m_b$ terms must be equal since $B$
differs from $D$ by only the value of the heavy quark mass. Thus, both
terms must vanish. Actually, this statement is the heavy-quark analogue of
the Ademollo-Gatto theorem for the $SU(3)_{\rm fl}$ breaking effects
which is routinely exploited in determinations of $|V_{us}|$ from
$K\ra\pi\ell\nu$ and semileptonic hyperon decays. This observation 
was later studied in more detail in Ref.\,\cite{luke} and is usually
called Luke's theorem. It improves the credibility of this method of
determination of $|V_{cb}|$ in spite of certain experimental
difficulties.

Since the corrections to the heavy quark limit are governed by the 
mass of the charm quark, even the $1/m_Q^2$ effects {\it a priori} can
be  significant. The need in evaluation of the $1/m_Q^2$ corrections in 
Eq.~(\ref{8.10}) for practical purposes was realized quite early
\cite{fn}. In these days the theory of the power corrections in heavy
quarks was immature, so that it was hard to decide even the  sign of
$\delta^A_{1/m^2}$. Quantitative application of the heavy quark 
symmetry to charm was generally viewed overly optimistic; it was believed
that the deviations from the symmetry limit in $F_{D^*}$ must be very
small, at the scale of $2\%$, and suggestions that they could be as
large as $10\%$ were categorically refuted \cite{fn}. At present, with
the application of dynamic methods in the heavy quark expansion, the
actual estimates of $F_{D^*}$ rather fall close to $0.9$ \cite{vcb}. 

The existing estimates of the power nonperturbative corrections in
$F_{D^*}$ are based on the sum rules for heavy flavor transition
amplitude discussed in the previous section. Studying the zero-recoil
transition we fix the spacelike momentum $\vec q\!=\!0$. The axial current
$\bar c \gamma_{i}\gamma _5 b$ produces $D^*$, $D\pi$ and higher
excitations in semileptonic $B$ decays at zero recoil. A straightforward
derivation yields the following sum rule for this current (cf.\
Eq.\,(\ref{m0a1}) for $3h_1^{AA}$ at $\vec q\!=\!0$):  
\beq
|F_{D^*}|^2 +
\sum_{0<\epsilon _i < \mu}|F_i|^2
\; = \;
\xi_A(\mu) \;-\; \Delta^A_{1/m^2} \;-\;\Delta^A_{1/m^3}\;+
{\cal O}\left(\frac{1}{m_Q^4}\right) \,,
\label{8.16}
\eeq
where
\beq
\Delta^A_{1/m^2} = \frac{\mu_G^2(\mu)}{3m_c^2} +
\frac{\mu_\pi^2(\mu)\!-\!\mu_G^2(\mu)}{4}
\left(\frac{1}{m_c^2}+\frac{1}{m_b^2}+\frac{2}{3m_cm_b}
\right)
\label{8.17}
\eeq
(the $1/m_Q^3$ corrections to the sum rule
$\Delta^A_{1/m^3}$ are also known, Eq.\,(8.21) of Ref.~\cite{rev}),
and the elastic formfactor $F_{D^*}$ is defined as  
$\matel{D^*(\vec q \!=\! 0)}{\bar c \gamma_i \gamma_5 b} {B} = 
\sqrt{2M_BM_{D^*}} F_{D^*} \varepsilon_i^*$. 
$F_i$ denote the axial-current transition formfactors to
excited charm states $i$ with the mass $M_i=M_{D^*}\!+\!\epsilon_i$, and
$\xi_A$ is a short-distance renormalization factor. Contributions from
excitations with $\epsilon$ higher than $\mu$ are dual to perturbative
contributions and get lumped into the coefficient $\xi_A(\mu )$ of the
unit operator, the first term in the right-hand side of 
Eq.\,(\ref{8.16}). 

The role of $\mu$ is thus two-fold: in the left-hand side it acts as an
ultraviolet cutoff in the effective low-energy theory, and by the same
token determines the normalization point for the local operators;
simultaneously, it defines the infrared cutoff in the Wilson
coefficients.

The sum rule Eq.\,\,(\ref{8.16}) leads to the upper bound:
$$
|F_{D^*}|^2 \simeq \xi_A(\mu)- \Delta^A_{1/m^2} - \Delta^A_{1/m^3}-
\, \sum _{0<\epsilon _i < \mu}|F_i|^2 \;,
$$
\beq
-\delta^A_{1/m^2}\;>\; \frac{1}{2}\Delta^A_{1/m^2} \,\ge\,
\frac{M_{B^*}^2 \!- \!M_B^2}{8m_c^2} \;\simeq\; 0.035  \;\,.
\label{200}
\eeq
The last relation is a model-independent lower bound for the $1/m^2$
corrections to $F_{D^*}$ at zero recoil \cite{vcb}.

To obtain an actual estimate of the nonperturbative corrections rather
than a bound, we need to know something about the contribution of the
excited states in the sum rule Eq.\,(\ref{8.16}). Unfortunately, no
model-independent answer to this question exists at present. The best we
can do is to assume that the sum over the excited states is a
fraction $\chi$ of the local term given by $\mu_\pi^2$ and $\mu_G^2$,
\beq
\sum _{\epsilon _i < \mu}|F_i|^2 \; = \; \chi\: \Delta^A_{1/m^2}\;,
\label{z8}
\eeq
where  on general grounds $\chi \!\sim\! 1$.
The contribution of the continuum $D\pi$ state can be calculated
\cite{vcb}, however theoretically it is expected to constitute only a
small fraction of the sum over resonant states.
Trying to be optimistic, we rather arbitrarily limit $\chi$ by
unity on the upper side, that is, put $\chi\!=\!0.5\!\pm \!0.5$ \cite{vcb}; 
the larger is $\chi$, the smaller is $F_{D^*}$.

Arguments in favor of the assumption $\chi \!\le\! 1$ are rather soft. In 
perturbation theory $\chi\!=\!1$ holds in the first order (see 
\cite{optical}, Sect.~VII), but this relation changes in higher orders.
The fraction  $\chi$ was recently computed analytically \cite{burkur} in
the exactly solvable 't~Hooft model (QCD in (1+1) dimensions in the
limit of large number of colors) and turned out to be close to $0.55$;
in this model it is almost saturated by the first radial excitation with
$\epsilon \approx 700 \MeV$.

The short-distance renormalization factor $\xi _A(\mu )$ is calculated 
perturbatively. To the first order it was computed in \cite{optical},
all-order BLM resummation performed in \cite{blmope}. The most
technically involved part of genuine ${\cal O}(\as^2)$ corrections to
$\xi _A(\mu )$ was computed in \cite{czareta}, and the complete result
given in \cite{xi}; the non-BLM corrections turn out to be small
slightly decreasing $\xi_A(\mu)$, with the overall estimate 
$\xi_A(\mu)\simeq (0.99)^2$ at $\mu$ around $0.7\GeV$. These
computations,  however were accomplished in expansion in $\mu/m_Q$ and
the terms to  order $\mu^2/m_Q^2$ were retained, which is consistent if 
only $1/m_Q^2$ correction to the formfactor $F_{D^*}$ are addressed. In
view of the presence of terms suppressed by powers of $1/m_c$, the
higher-order terms are expected to be significant. This was suggested by
the perturbative BLM resummation  itself where, in particular, $1/m_c^3$
corrections in $\xi_A^{1/2}$ were found to be at least at the scale of
$2\%$. To estimate such effects, it is possible to evaluate $\xi_A(\mu)$
completely as a function of $\mu/m_Q$, in the BLM approximation which
presumably yields the dominant contribution. The typical numerical 
outcome of such computations is shown in Fig.~7. It is thus reasonable
to accept the value $\xi_A^{1/2}(\mu) \simeq 0.96\pm 0.015$ at $\mu
\!\approx\! 0.8\GeV$ for the short-distance renormalization of the axial
current.\footnote{A similar number, with much smaller uncertainty is
often cited in the literature for the so-called $\eta_A$ factor
introduced in HQET to denote `purely perturbative' scale-independent 
renormalization of the axial current at zero recoil. Technically $\eta_A$
coincides with $\xi_A^{1/2}(\mu)$ at $\mu\to 0$ to any order in
the perturbative expansion. Regardless of intrinsic deficiencies of
such a notion, the numerical value of $\eta_A$ has the irreducible
infrared renormalon uncertainty of at least $3$-$5\%$ which by itself 
is much 
larger than the overall uncertainty in $\eta_A$ quoted in the
literature, see, {\it e.g.}\ \cite{babar}. The numerical close 
coincidence of the value quoted for $\eta_A$ and 
$\xi_A^{1/2}(0.8\GeV)$ is accidental.} The uncertainty here 
in the short-distance renormalization 
can hardly be significantly reduced 
further since the inherent momentum scale is relatively low.
The ${\cal O}(\alpha_s)$ corrections to the Wilson coefficient 
of the kinetic operator in the sum rule Eq.\,(\ref{8.17}) was also
computed to the next-to-leading order in \cite{xi}; the correction turns
out to be insignificant.

\thispagestyle{plain}
\begin{figure}[hhh]
 \begin{center}
 \mbox{\epsfig{file=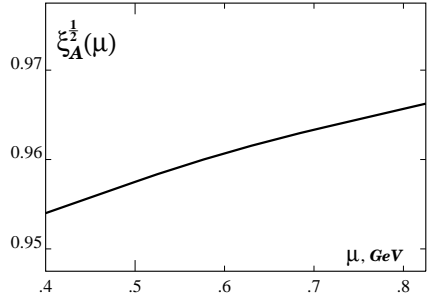,width=5.5cm}}
 \end{center}
 \caption{ \small
Short-distance renormalization $\xi_A^{\frac{1}{2}}(\mu)$ of the zero 
recoil $b\!\to\! c$ axial current as a function of $\mu$. $~\Lam\!=
250\MeV$.
}
\end{figure}

The $1/m_Q^3$ term in the zero recoil sum rule Eq.\,(\ref{8.16})
$-\Delta^A_{1/m^3}$ is estimated to constitute about $-3\%$. Although
it further suppresses the apparent value of $F_{D^*}$, we discard it in
our estimates. It is expected to be only a smaller fraction of the
overall $1/m_Q^3$ effects in the $B\!\to\! D^*$ formfactor, and keeping
$\Delta^A_{1/m^3}$ is hardly legitimate since the leading correction 
$\delta^A_{1/m^2}$ in the formfactor is not known completely. 

Assembling all pieces together we get for $\chi=0.5\pm 0.5$ 
\beq
F_{D^*}\;\simeq \; 0.89 \:-\:0.015\,
\frac{\mu_\pi^2\!-\!0.5\GeV^2}{0.1\GeV^2}\;\pm\;
0.025_{\rm excit}\;\pm\;0.015_{\rm pert}\;\pm\;0.025_{1/m^3}\;\,.
\label{z11}
\eeq
Estimates of the uncertainties in the contributions from higher 
excitations and of the magnitude of the ${\cal O}(1/m_Q^3)$
corrections are not very firm and are rather on the optimistic side; 
they can be larger. With $1/m_Q^2$ corrections amounting to $\sim 8\%$
one expects $|\delta_{1/m_Q^3}| \gsim 0.025$ simply on the dimensional
grounds.

\subsubsection{Quantum-mechanical interpretation}

We already illustrated the quantum-mechanical meaning of the inequality
between $\mu_\pi^2$ and $\mu_G^2$ in Sect.~3.2. Let us look at the sum
rule Eq.\,(\ref{8.16}) from the similar perspective \cite{optical}. 
From the point of view of light cloud in $B$ meson the semileptonic
decay of the $b$ quark is an instantaneous replacement of $b$ by
$c$ quark.
In ordinary quantum mechanics the overall probability of 
the produced state to
hadronize to some final state is exactly unity,
which is the first, leading term in the r.h.s.\ of Eq.\,(\ref{8.16}).
Why then are there any
nonperturbative corrections in the sum rule? The answer is that
the `normalization' of the weak current $\bar c \gamma_\mu \gamma_5 b$
is not exactly unity and depends, in particular, on the external gluon
field. This appears as presence of local higher-dimension
operators in the current. Indeed, expressing the QCD current in terms
of the nonrelativistic fields used in QM one has
\beq
\bar c \gamma_k \gamma_5 b \leftrightarrow \varphi_c^+\left\{
\sigma_k -
\left(\frac{(\vec\sigma
i\vec D)^2\sigma_k}{8m_c^2}+\frac{\sigma_k(\vec\sigma
i\vec D)^2}{8m_c^2}-
\frac{(\vec\sigma i\vec D )\sigma_k(\vec\sigma i\vec D)}{4m_cm_b}
\right) +
{\cal O}\left(\frac{1}{m^3}\right)
\right\} \varphi_b\;\;.
\label{8.26}
\eeq
The weak current $\bar c \gamma_5\gamma_k b$, according to
Eq.~(\ref{8.26}) converts the initial wavefunction $\Psi_b$ into
$\tilde\Psi$:
\beq
\Psi_B \;
\mbox{\Large 
$\stackrel{\mbox{\tiny $\bar c \gamma_5\gamma_k b\;\;\;$}}
{\longrightarrow}$} \;
\tilde\Psi = \sigma_k \Psi_B -
\left(
\frac{(\vec\sigma
i\vec D)^2\sigma_k}{8m_c^2}+\frac{\sigma_k(\vec\sigma
i\vec D)^2}{8m_c^2}-
\frac{(\vec\sigma i\vec D )\sigma_k(\vec\sigma i\vec D)}{4m_cm_b}
+ ...\right) \Psi_B\,.
\label{193}
\eeq
Then it is easy to calculate the normalization of $\tilde\Psi\,$:
\beq
\|\tilde\Psi\|^2 = \left\|\Psi_B\right\|^2 -
\frac{\mu_G^2}{3m_c^2} -
\frac{\mu_\pi^2\!-\!\mu_G^2}{4}
\left(\frac{1}{m_c^2}\!+\!\frac{1}{m_b^2}\!+\!\frac{2}{3m_cm_b}
\right)\:-\:...\;, \qquad \left\|\Psi_B\right\|^2=1\;.
\label{194}
\eeq
The additional terms are just the nonperturbative
correction in the right-hand side of the sum rule.
It is worth noting that the first two $1/m_Q^2$ terms in the bracket in 
Eq.~(\ref{8.26}) are the result of the
Foldy-Wouthuysen transformation Eq.~(\ref{28}). Additional 
discussion, including the pedagogical illustration of the role of the
latter transformation in the $1/m_Q$ expansion, can be found in
Ref.~\cite{varenna}.

\section{OPE for Inclusive Weak Decays}

Inclusive widths of the heavy flavor hadrons are examples of the genuine
short-distance processes. The decays proceed at the space-time intervals
$\sim \!1/m_b$ (more precisely, inverse energy release), and the widths
are affected by the soft strong dynamics to the minimal extent. The
decay widths are, perhaps, the best studied example of applying OPE to
quantify dynamical effects in heavy quarks. Heavy quark symmetry {\it
per se} is of little help here, in particular in $b\to u$ or nonleptonic
transitions. 

The general consideration of inclusive weak decays runs parallel to the 
treatment of  $\sigma (e^+e^- \!\!\to\!\! \mbox{ hadrons})$. One describes
the decay rate into an inclusive final state $f$ in terms of the
imaginary part of a forward scattering operator (the so-called
transition operator) evaluated to second order in the  weak
interactions \cite{SV}:
\begin{equation}
\Im \hat T(Q\rightarrow f\rightarrow Q)\;= \;
\, \Im \int {\rm d}^4x\ i\,T \left\{{\cal L}_w(x), {\cal
L}_w^{\dagger}(0)\right\}\
\label{OPTICAL}
\end{equation}
where $T$ denotes the time ordered product and ${\cal L}_w$ is the
relevant weak Lagrangian at the normalization point higher or about
$m_Q$.  The space-time separation $x$ in Eq.\,(\ref{OPTICAL}) is fixed by
the inverse energy release. If the latter is sufficiently large in the
decay, one can express the nonlocal operator product in  
Eq.\,(\ref{OPTICAL}) as an infinite sum of local heavy quark operators 
$O_i$ of
increasing  dimensions. The width for $H_Q\rightarrow f$ is then 
obtained by averaging $\Im \hat T$ over the heavy flavor hadron $H_Q\,$,
\beq
\frac{\matel{H_Q}{2\,\Im \hat T (Q\!\to \!f\!\to\!
Q)}{H_Q}}{2M_{H_Q}} =
\Gamma (H_Q\!\to\! f) =
\frac{G_F^2 m_Q^5 |V_{\rm KM}|^{2\!}}{192\pi^3} \sum _i  
\tilde c_i^{(f)}(\mu )
\frac{\matel{H_Q}{O_i}{H_Q}_{\mu }}{2M_{H_Q}}
\label{OPE}
\end{equation}
with $V_{\rm KM}$ denoting the appropriate combination of the CKM
parameters. A few comments are in order to elucidate the content of 
Eq.\,(\ref{OPE}).

(i) The parameter $\mu$ in Eq.\,(\ref{OPE}) is the normalization point,
indicating that we explicitly evolved from $m_Q$ down to $\mu$. The
effects of momenta below  $\mu$  are lumped into the matrix
elements of the operators $O_i$.

(ii) The coefficients $\tilde c_i^{(f)}(\mu )$ are dimensionful, they
contain powers of $1/m_Q$ that go up with the dimension of the operator
$O_i$. Using the normalization introduced in Eq.\,(\ref{OPE}), one
obtains on dimensional grounds
$$
\tilde c_i^{(f)}(\mu ) \, 
\frac{1}{2M_{H_Q}} \matel{H_Q}{O_i}{H_Q}_{(\mu )}
\sim
{\cal O} \left( \frac{\Lam^{d_i-3}}{m_Q^{d_i-3}},\;
\frac{\as\mu^{d_i-3}}{m_Q^{d_i-3}} \right)
$$
with $d_i$ denoting the dimension of operator $O_i$. The contribution
from the operators with the lowest dimension obviously dominates in 
the limit $m_Q \!\to \!\infty$.

(iii) 
It seems natural then that the expansion of total rates can be
given in powers of $1/m_Q$. The master formula (\ref{OPE}) holds for a
host of different integrated heavy-flavor decays: semileptonic,
nonleptonic and radiative transitions, CKM-favored or suppressed, etc.
For semileptonic and nonleptonic decays, treated through order
$1/m_Q^3$, it takes the following form:
$$
\Gamma (H_Q\ra f)=\frac{G_F^2m_Q^5}{192\pi ^3}|V_{\rm KM}|^2\times
$$
$$
\left[ c_3^{(f)}(\mu )\frac{\matel{H_Q}{\bar QQ}{H_Q}_{(\mu)}}{2M_{H_Q}}
+ c_5^{(f)}(\mu ) \,\frac{1}{m_Q^2}
\frac{\matel{H_Q}{\bar Q\frac{i}{2}\sigma G Q}{H_Q}_{(\mu )}}
{2M_{H_Q}\;\;}+ \right.
$$
\begin{equation}
\left. +\sum _i c_{6,i}^{(f)}(\mu ) \, \frac{1}{m_Q^3} 
\frac{\matel{H_Q}
{(\bar Q\Gamma_i q)(\bar q\Gamma _iQ)}{H_Q}_{(\mu )}}
{2M_{H_Q}\;} + {\cal O}(1/m_Q^4)\right]  \, ,
\label{WIDTH}
\end{equation}
where Wilson coefficients $c^{(f)}$ are of order unity.

We pause here to make a few explanatory remarks on this particular
expression. First, the main statement of the OPE is that there is no
correction of order $1/m_Q$ \cite{buv}.  This is particularly noteworthy
because the hadron masses, which control the phase space, do contain
such a correction: $M_{H_Q}\!=\!m_Q \left( 1 \!+\! \bar \Lambda /m_Q 
\!+\! {\cal O}(1/m_Q^2)\right)$; the parameter $\La$, different for different
hadrons, does not enter the width. The reason for the absence of the 
$1/m_Q$ correction in the total widths is twofold: the corrections to
the expectation value of the leading QCD operator $\bar Q Q_{(\mu)}$ is
only $\sim \mu^2/m_Q^2$, and there is no independent QCD operator of
dimension $4$ for forward matrix elements. Since the coefficients
functions are purely short-distance, infrared effects neither can
penetrate into them. 

A physically more illuminating way to think of the absence of corrections
of order $1/m_Q$ is to realize that the bound-state effects in the 
initial state (mass shifts, etc.) do generate corrections of order
$1/m_Q$ to the total width -- as does hadronization in the final state.
Yet local color symmetry demands that they cancel against each other, as
can explicitly be demonstrated in simple models. It is worth realizing
that this is a peculiar feature of QCD interactions -- other dynamical
realizations of strong confining forces would, generally, destroy the
exact cancellation. A detailed pedagogical discussion of physics behind
this cancellation can be found in Ref.\,\cite{varenna}, Sect.~3.1.

Second, the  leading nonperturbative  corrections are  $\sim {\cal
O}(1/m_Q^2)$,  i.e. small in  the total decay rates for beauty hadrons.
The first calculation of the leading nonperturbative corrections in the
decays of heavy flavors was done in \cite{buv,bs,dpf,prl}.

Third, the four-quark operators  $(\bar Q\Gamma q)(\bar q\Gamma Q)$
with different Lorentz and color structure depend explicitly on 
the light-quark flavors denoted by $q$. They,
therefore, generate differences in the weak transition rates for the 
different hadrons of a given heavy flavor.\footnote{Expanding $\langle
H_Q|\bar Q i \sigma G Q|H_Q\rangle /m_Q^2$ also yields contributions of
order $1/m_Q^3$; those are, however, practically insensitive to the
light quark flavors.} They describe the effects of Weak Annihilation
(WA) and Pauli Interference in $B$ mesons, and Pauli Interference and
Weak Scattering (WS) in heavy baryons. Their effects were calculated 
already in mid-eighties \cite{vslog}.

A note should be made regarding the subtlety in understanding the above 
four-quark expectation values \cite{WA,D2WA}. They must be taken in 
the effective low energy (nonrelativistic) theory with respect to the
heavy quark $Q$. In particular, the heavy quark field $Q(x)$ must
contain only the heavy quark annihilation operator, while $\bar Q(x)$
only the creation of the heavy quark. This differs from the full QCD
fields which contain both annihilation of the heavy quark and creation
of the antiquark in the $Q(x)$ field (and likewise for $\bar Q(x)$).
The full-QCD four quark expectation values over, say, $B$ meson 
$\matel{B}{\bar{b}\Gamma_1 q \, \bar{q} \Gamma_2 b}{B}$
include, strictly speaking, the intermediate states with the $b\bar b$ 
pair (not related to the gluon conversion into $b\bar b$), 
and not only light flavor hadrons. Even though the
energy gap between the two classes of states is large, about $2m_Q$, the
latter contributions to the expectation values are not necessarily 
short-distance and rather governed by strong coupling dynamics 
when their energy is
close to $2m_Q$. This additional nonperturbative component is absent
from the expectation values in the effective low-energy theory in the 
sector with a single heavy quark. Simultaneously, the analysis shows
that such contributions are absent from the inclusive widths as well.
Therefore, strictly speaking one cannot ``match'' in the usual
sense the full-QCD expectation values onto those in the effective
theory. The renormalization scale evolution of the four-fermion
operators in the effective theory is governed by the so-called
``hybrid'' anomalous dimensions introduced by Shifman and Voloshin in
the mid 80s \cite{vslog}.

Fourth, the  short-distance coefficients $c_i^{(f)}(\mu )$ in practice
are calculated in perturbation theory. However it is quite conceivable
that certain nonperturbative effects arise also in the short-distance
regime. They are believed to be rather small in beauty decays
\cite{inst}.

Fifth, a new matrix element appearing in  OPE, not discussed so far,
is the scalar heavy quark density. Its nonrelativistic expansion 
originally established in Ref.~\cite{buv} using the QCD equations of
motion for the heavy quark field, follows from the identity
(\ref{ident}) and takes the form
\beq
\matel{H_Q}{\bar QQ}{H_Q} = \matel{H_Q}{\bar Q\gamma_0 Q}{H_Q} -
\frac{\matel{H_Q}{\bar Q\left(\vec{\pi}^{\:2}\!-\!\frac{i}{2}\sigma G \right)
Q}{H_Q}}{2m_Q^2}+
{\cal O}(1/m_Q^4)\;\;.
\label{QQBAR}
\eeq
Since $\matel{H_Q}{\bar Q\gamma _0Q}{H_Q}=2M_{H_Q}$, the spectator
ansatz indeed emerges  as the asymptotic scenario universal for all
types of hadrons, and holds up to $1/m_Q^2$ corrections. In addition to
$\bar Q \vec\pi^{\,2} Q$, the second dimension-five operator is  the
chromomagnetic operator $\bar Q i\sigma G Q$. Since $\bar Q \vec D^{\,2}
Q$ is not a Lorentz scalar, it does not appear independently in  
Eq.~(\ref{WIDTH}).
For the pseudoscalar mesons, for example, we have
\begin{equation}
\frac{1}{2M_{P_Q}}\matel{P_Q}{\bar QQ}{P_Q}=1 -
\frac{\mu _{\pi}^2}{2m_Q^2}+
\frac{3}{8} \frac{M_{V_Q}^2-M_{P_Q}^2}{m_Q^2}+
{\cal O}(1/m_Q^3)\;.
\label{QQBARPQ}
\end{equation}

The reason for the kinetic operator term to appear is quite transparent.
The  first two quantities on the right-hand side of the equation
represent the mean value of the factor  $\sqrt{1\!-\!\vec v^{\,2}}$ 
reflecting the time dilation which slows  down the  decay of the quark
$Q$ moving inside $H_Q$ \cite{prl,WA,Ds}.

Equation (\ref{QQBAR}) is readily obtained in the heavy quark expansion
if one uses proper nonrelativistic heavy quark spinors incorporating
the Foldy-Wouthuysen transformation; in the context of HQET this
procedure was advocated in Ref.~\cite{korner}, but was largely ignored
for a few years.

Finally,  Eqs.\,(\ref{WIDTH})--( \ref{QQBAR}) show that the two 
dimension-five operators do produce differences in $B$ versus
$\Lambda_b/\Xi_b$ versus $\Omega_b$ decays of order  $1/m_Q^2$. To a
small extent they can also differentiate $B$ and $B_s$ {\it via} 
the $SU(3)$
breaking in their expectation values. Differences in the transition
rates inside the  meson family  are generated at order $1/m_Q^3$ by 
dimension-six four-quark operators. They are usually estimated in the
vacuum saturation approximation which -- although cannot be exact --
represents a reasonable starting approximation. There is an intriguing
way to check factorization experimentally \cite{WA}: similar
four-fermion operators enter semileptonic $b\to u$ transition rates.
Moreover, in the heavy quark limit the four-fermion operators populate
mainly the transitions into the hadronic states with low energy, and
thus show up, for example, in the end-point domain of the lepton
spectrum where their relative effect is enhanced. More precisely, these
effects are present at large values of lepton invariant mass $q^2$ and
thus can be isolated in the cleanest way studying double differential
semileptonic distributions. Considering the
difference of the decay characteristics of the charged and neutral
$B$'s in this domain, one can measure these matrix elements and even
feel their scale dependence. Further details regarding the
``flavor-dependent" preasymptotic effects can be found in dedicated
papers \cite{BELLINI,four}.
 
\thispagestyle{plain}
\begin{figure}[hhh]
 \begin{center}
 \mbox{\epsfig{file=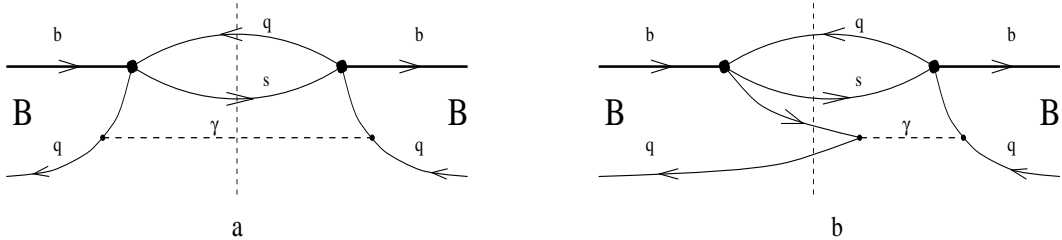,width=14cm,height=3.1cm}}
 \end{center}
\caption{ \small
{\bf
a)} \,Spectator effect leading to $1/m_b$ correction in the decay
width $b\rightarrow  s +\gamma\, $.
{\bf b)}~\,A~different cut of the same diagram leads to an
electromagnetic
correction in the hadronic decay width. Terms $1/m_b$ and  $1/m_b^2$
cancel out in the sum of two decay widths. The solid dot in the vertex
denotes the penguin-induced $ b\rightarrow s \bar q q$ interaction.  }
\end{figure}

It is important to keep in mind that the OPE approach discussed in this
section implies that {\em all} decay channels induced by a given term in
the short distance Lagrangian in Eq.~(\ref{OPTICAL}) are included. It is
not  enough to consider the states that  appear at the free quark level.
The final state interactions can annihilate, for example, the $\bar{c}
c$ quark pair into light hadrons; electromagnetic interaction, if
considered, can do the same. Disregarding the channels that can emerge
due to such final state interactions can violate the general theorems:
even $1/m_Q$ terms can appear in such incomplete ``inclusive" widths.
For example, the correction to the $b\ra s+\gamma$ width does have
nonperturbative corrections scaling like $1/m_b$ due to the effect of
weak annihilation (in mesons, or weak scattering in baryons) of the
light quarks with emission of a hard photon in the penguin-induced weak
decay $b\ra s\bar{q}q$, see Fig.~8a. This effect would only cancel
against the (virtual) electromagnetic correction to the hadronic
penguin-induced width, Fig.~8b.\ \cite{mirage,WA}. 

For the very same reason the spectator-independent nonperturbative
effects in the actual  $b\ra s+\gamma$ width can also have in practice
small corrections which scale like $1/m_b$ \cite{volbsg}. The effective
$b\ra s+\gamma$ interaction is generated by loop diagrams and,
including contributions from $c\bar{c}$ ($u\bar{u}$) pairs with momenta 
not large compared to $m_b$, is not genuinely pointlike. This part of the
decay interaction is properly treated considering the original $b\to
c\bar{c} s$ ($b\to u\bar{u} s$) operators as the weak decay Lagrangian
${\cal L}_w$, and
then truly inclusive widths for such an interaction have to include the
probability of the usual decay $b\to c\bar{c} s$. The latter is affected by
electromagnetic interaction, again by virtue of the same diagram as for
the $b\ra s+\gamma$ decay, but with a different cut. Of course,
in experiment -- in contrast to the OPE -- these processes are
completely different and are taken separately. 

The OPE predictions for the nonperturbative effects in the inclusive 
width differences are rather nontrivial. Consequently, a number of
questions are often raised. First, is this analysis
operating in terms of quarks and gluons, applicable to actual $b$
hadrons, with confinement leading to the drastic change of the physical
spectrum which consists only of the colorless hadrons but not quarks?
Second, if this is correct, how to formally justify this OPE? 
Third, how to compute the leading nonperturbative corrections to the
decay rates? And, finally, it is advantageous to have a transparent 
physical interpretation of the OPE machinery.

The answer to the first question is definitely positive, although is not
trivial; the validity of the OPE approach has been challenged more than
once, even over the recent years. The raised criticism covered different
aspects, but basically reduced to one main point: due to `brutal
confinement' there is no
``duality'' between the OPE computations operating in terms of quarks
and gluons, and actual hadronic decay rates, either for nonleptonic, or
for any type of the underlying quark transitions. We do not dwell here
on discussing the reasons (usually poorly substantiated) behind such
suggestions. This question was recently discussed
in much detail in the context of exactly solvable two-dimensional 
't~Hooft model where, in principle, all hadron masses and decay 
amplitudes can be computed. In Refs.~\cite{D2,D2WA,D2plb} the analytic
summation of the rates for open decay channels was performed in the
expansion in powers of $1/m_Q$, and compared to the OPE predictions for
the model. It was shown that to all orders in $1/m_Q$ the OPE series 
could be computed, they reproduced exactly the asymptotic expansion of
the actual inclusive widths, for both semileptonic and nonleptonic
decays. 

The physical interpretation of the main OPE result, the absence of the
$1/m_Q$ corrections to the decay widths has been briefly mentioned
above. A dedicated discussion can be found in Ref.~\cite{varenna},
Sect.~3.1. In the following sections we briefly address the third and
the second questions, respectively.

\subsection{Sample computation}

Here we illustrate the computation of the leading nonperturbative
corrections in semileptonic, nonleptonic and $b\to s + \gamma$ -type
decays. For simplicity, we do not go beyond order $1/m_b^2$, and also
neglect masses of the final state quarks and leptons; this will make
expressions more compact. For semileptonic and nonleptonic decays we
still will refer to one of the final state quarks as the charmed one.
The corresponding decay Lagrangian is given (we omit the CKM factors)
\beq
{\cal L}_{\rm w} =
-\frac{G_F}{\sqrt{2}}\: \bar{c} \gamma_\mu(1\!-\!\gamma_5) b
\cdot \bar{d} \gamma_\mu(1\!-\!\gamma_5) u
\;;
\label{sample9}
\eeq
this will describe the semileptonic decays as well if we switch off
strong interaction of the $d\bar u$ pair. We do not introduce here the 
color contraction schemes explicitly; color indices can be easily taken 
care of in the end.
For $b\to s + \gamma$ we
take the decay Lagrangian in the form 
\beq
{\cal L}_{\rm w} = \frac{\lambda}{2} \:
\bar{s} i\sigma_{\mu\nu} F^{\mu\nu} (1\!+\!\gamma_5)b
\;.
\label{sample11}
\eeq

\thispagestyle{plain}
\begin{figure}[hhh]
 \begin{center}
 \mbox{\epsfig{file=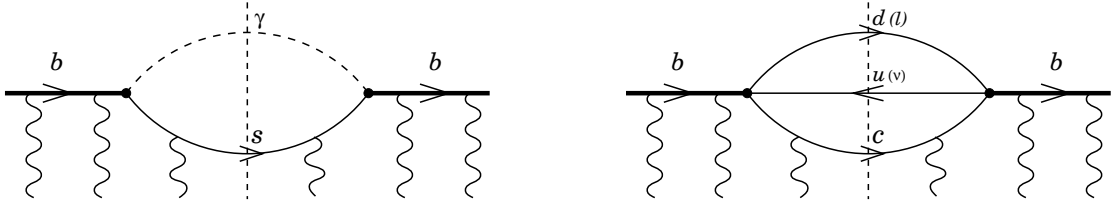,width=14.7cm}}
 \end{center}
 \caption{ \small
Diagrams describing the the transition amplitude in the external 
field for computing the $b\to s+\gamma\;\,$ (left) and nonleptonic or
semileptonic widths (right). Interaction with the gluon medium enters 
through Green functions for the final state quarks (intermediate quark 
lines), and through equations of motion for the initial $b$ fields.
}
\end{figure}

The computation of the transition operator $\hat T$ in Eq.\,(\ref{OPTICAL})
is most simply done in the Fock-Schwinger (fixed point) gauge 
$x_\mu A_\mu(x)\!=\!0$ (see Ref.~\cite{buv9} for details). The transition
operator in the coordinate space is given by the products of the fermion
and photon Green functions like 
$\bar{b}(x)\,G(x,0)\!\cdot\! D(x,0)\, b(0)$, 
or by the product of three fermion Green functions 
$\bar{b}(x)\,G(x,0)\!\cdot\!
G(x,0) \!\cdot\! G(0,x) \,b(0)$ for the two cases, respectively. For 
massless particles the Green functions are 
\beq
G(x,0)= \frac{i \!\!\not \! x}{2\pi^2 x^4} - {\frac{ix_\alpha \tilde
G_{\alpha\beta}\gamma_\beta \gamma_5}{8\pi^2x^2} + ...}
\;, \qquad 
D_{\mu\nu}(x,0)= \frac{\delta_{\mu\nu}}{4\pi^2 x^2} 
\label{sample13}
\eeq
(a note of caution: in the fixed point gauge one should distinguish
$G(x,0)$ from $G(0,x)$ for quark Green functions), and the expansion of
the gauge field is given by 
\beq
A_\mu(x) = \frac{1}{2\cdot 0!} x_\nu G_{\nu\mu}(0) + 
\frac{1}{3\cdot 1!} x_\nu x_\rho (D_\rho G_{\nu\mu}) + ...
\label{sample15}
\eeq
$\tilde G$ above is the dual field strength, 
$\tilde G_{\mu\nu}=
\frac{1}{2}\epsilon_{\mu\nu\alpha\beta}G^{\alpha\beta\!}$. 
Note that we work directly in Minkowski space. 
For $b\to s + \gamma$ we thus have 
$$
\hat{T}_{bs\gamma} =  4 i \lambda^2 \,
\frac{x^2\delta_{\mu\rho}\!-\! 4 x_\mu x_\rho}{2\pi^2 x^6}
\:
\bar{b}(x) \sigma_{\mu\nu}\left[
\frac{i \!\!\not\! x}{2\pi^2 x^4} -
\frac{ix_\alpha \tilde G_{\alpha\beta}\gamma_\beta \gamma_5}{8\pi^2x^2}
\right]
\sigma_{\rho\nu} (1\!+\!\gamma_5) \, b(0) \;=
$$
\beq
-\lambda^2 \: \bar{b}(x)\left[ \frac{6}{\pi^4} \frac{\not \!x}{x^8} 
+  \frac{1}{2\pi^4} \frac{x_\alpha}{x^6} \tilde
G_{\alpha\beta} \gamma_\beta + \ldots 
\right] (1\!+\!\gamma_5) \, b(0)
\;,
\label{sample17}
\eeq
where the summation over $\mu,\, \nu$ is straightforward. For the decays
mediated by the four-fermion interaction we consider both semileptonic and
nonleptonic cases simultaneously, simply introducing the factor $\xi$
which would switch off interaction of leptons with the gluon field,
$\xi\!=\!1$ for $b\to c\,\bar{u}d$ and $\xi\!=\!0$ for 
$b\to c\,\ell \bar{\nu}$:
$$
\hat T_{\rm 4\mbox{-ferm}} = 
4 i G_F^2 \:\bar{b}(x) \gamma_\mu \left( \frac{i \!\!\not \! x }{2\pi^2 x^4}
\!-\!
\frac{ix_\alpha \tilde G_{\alpha\beta}\gamma_\beta
\gamma_5}{8\pi^2x^2}
\right)
\gamma_\nu \left( \!- \frac{i \!\!\not \! x }{2\pi^2 x^4}
\!-\!
\xi \frac{ix \tilde G \gamma \gamma_5}{8\pi^2x^2}
\right)\cdot
$$
$$
\cdot \; \gamma_\mu \left( \frac{i \!\!\not \! x }{2\pi^2 x^4}
\!-\!
\xi \frac{ix \tilde G\gamma \gamma_5}{8\pi^2x^2}
\right)
\gamma_\nu (1\!-\!\gamma_5) \, b(0)  \;=
$$
\beq
4 G_F^2\: \bar{b}(x) \left[\frac{\not \! x }{\pi^6 x^{10}} 
{
+\,\xi\, \frac{x_\alpha}{4\pi^6 x^8} \tilde G_{\alpha\beta}\gamma_\beta }
\, + \ldots
\right] (1\!-\!\gamma_5) \, b(0)
\;,
\label{sample19}
\eeq
where we have used the Fiertz transformation for one of the $V\!-\!A$
vertices, and the short-hand notation $x\tilde G\gamma$ 
denotes the repeating structure $x_\alpha \tilde 
G_{\alpha\beta}\gamma_\beta$. 
We observe that only the interaction with the
antiquark is
present to this order  (using a variant of the Fiertz transformation, it
is easy to show that this holds for arbitrary fermion masses
\cite{buv}). 

Now we need to make the Fourier transformation to pass to the momentum
representation of the transition operator:
\bea
\int {\rm d}^4x \,{\rm e}\,^{-ipx} \, \frac{x_\mu}{x^6} & =&
\;\;-\frac{^{\,}\pi^2}{4}  \;\;\;\;\;\; p_\mu\,
{\ln{(-p^2)}}  \;+\; \mbox{polynomial in } p^2 \vspace*{-1mm}\\
\int {\rm d}^4x \,{\rm e}\,^{-ipx} \, \frac{x_\mu}{x^8}
&= &
\;\;\;\;\;\frac{\pi^2}{48} \;\;\; p^2 p_\mu\,
{\ln{(-p^2)}}
\;+\; \mbox{polynomial in } p^2 \vspace*{-1mm}\\
\int  {\rm d}^4x \,{\rm e}\,^{-ipx} \, \frac{x_\mu}{x^{10}} &=&
\!-\frac{\;\pi^2}{64\!\cdot \!24}\; p^4 p_\mu\,
{\ln{(-p^2)}}
\;+\;
\mbox{polynomial in } p^2
\eea
(the divergent pieces come from infinite momenta running inside loops
and cannot yield imaginary part). The polynomials do not have
discontinuity and can be discarded for computing decay widths; it 
comes only from the logarithm. 

Now we need to evaluate the expectation values 
$\bar{b}\, p^4 \!\!\not \!p\, b$,
$\,\bar{b}\, p^2 \!\!\not \!p\, b\,$ {\it etc.} which enter 
at $p^2\!\simeq \!m_b^2$.
To do this we recall that $p_\mu = iD_\mu\!-\!A_\mu$ and in the
Fock-Schwinger gauge one has 
\beq
A_\mu(0)=0\;, \qquad D_\mu A_\nu \left\vert _0 \right.=
\frac{1}{2} G_{\mu\nu}(0)
\label{sample23}
\eeq      
(this implies, for example, that in our approximation $[D_\mu, A_\mu]$
can be replaced by zero). Our strategy is to pull the gluon potential
$A_\mu$ to the left since $A(0)\!=\!0$ and the result does not vanish
only if a derivative acts on $A_\mu$. In this way we arrive at the 
very simple rule that in this gauge 
\beq
\bar{b}\, p^{2n}\!\!\not\! p\, b =
m_b^{2n+1} \,\bar{b}b \,-\, n \,m_b^{2n-1}\:
\bar{b} \,\frac{i}{2} \sigma G \,b \,+ \,
{\cal O}\left(m_b^{2n\!+\!1}/m_b^3\right)\;.
\label{sample25}
\eeq
We also have an additional structure
which, for the forward matrix elements reduces to the chromomagnetic 
operator using equations of motion for the $b$ field:
$$
\bar{b}\, p_\alpha \tilde G^{\alpha\beta} \gamma_\beta \gamma_5 \,b
=  m_b\, \bar{b}\, \frac{i}{2} \sigma G\, b
\;.
$$

Recalling that $\Gamma = 2\, \Im T$ and collecting all terms, we arrive at
\cite{buv,dpf}
$$
\Gamma_{\rm sl} = \frac{G_F^2 m_b^5}{192\pi^3} \:
\frac{\matel{B}{\bar{b} b}{B}}{2M_B}\,
\left\{1- 2\,\frac{\mu_G^2}{m_b^2} +
{\cal O}\left(\frac{1}{m_b^3}\right)\right\}\;, \qquad \qquad
$$
$$
\Gamma_{b\to s\!+\!\gamma} = \;\frac{\lambda^{\!2} m_b^3}{4\pi}\;\:
\frac{\matel{B}{\bar{b} b}{B}}{2M_B}\,
\left\{1- (1\!+\!1)\,\frac{\mu_G^2}{m_b^2} +
{\cal O}\left(\frac{1}{m_b^3}\right)\right\}\;,\qquad\;\;\;\;\;
\label{sample29}
$$
\beq
\Gamma_{\rm nl} \!=\! \frac{G_F^2 m_b^5 N_c}{192\pi^3}
\frac{\matel{B}{\bar{b} b}{B}}{2M_B}
\left\{\!\left(\!\frac{c_+^2 \!+\! c_-^2}{2} \!+\! 
\frac{c_+^2\!-\!c_-^2}{2N_c}\!
\right)\!\left(\!1\!-\!\frac{2\mu_G^2}{m_b^2}\!\right)-
\frac{c_{+\!}^2\!-\!c_-^2}{2N_c} \frac{8\mu_G^2}{m_b^2} +
{\cal O}\!\left(\!\frac{1}{m_b^3}\!\right)\!\right\}\! .
\label{sample31}
\eeq
Account for the charm quark mass is straightforward and yields the
factor
$(1\!-\!m_c^2/m_b^2)^3$ for the direct contribution of the operator
$\mu_G^2$, the second term in $\Gamma_{\rm nl}$. An additional piece
expressed in terms of the derivative of the
free quark phase space factor $z_0$ with respect to $m_c^2$ comes from
translating the result from the Fock-Schwinger gauge,
cf.\ Eq.~(\ref{sample25}); it introduces
the factor $(1\!-\!m_c^2/m_b^2)^4$ for $\mu_G^2$ in the first term, and
likewise in $\Gamma_{\rm sl}$.
The correction to the nonleptonic width $b\to c\bar{c} s$ with two
heavy quarks in the final state was computed in Ref.~\cite{bs}.

\subsection{How OPE can be justified for inclusive widths}

In the previous section we illustrated the calculation of the leading 
nonperturbative corrections to the inclusive decay
widths. The expected question is: what such a computation of the quark 
decay width
in the external gluon field has to do with the decay width of the actual 
heavy flavor hadron? The justification does not differ conceptually
from, say, the classical case of $e^+ e^-$ annihilation to hadrons and
makes use of the analytic properties of the transition amplitude and
the dispersion relations. This was discussed in 
Refs.~\cite{WA,inst,D2,D2WA} with varying degree of detalization.

To study the analytic properties of the forward transition amplitude 
we must introduce an auxiliary complex variable $\omega$:
\begin{equation}
{\cal A}(\omega)\;=\; \int\, {\rm d}^4x\; {\rm e}\,^{-i\omega(vx)}\:
\langle H_Q|\,i\,T\left\{ {\cal L}_{\rm w}(x),\, 
{\cal L}_{\rm w}^\dagger(0)\right\}|H_Q\rangle
\label{72}
\end{equation}
(once again, $v$ is the four-velocity of the decaying heavy hadron).
This $\omega$-dependent amplitude
can be visualized as the transition amplitude governing the total
(weak) cross section of the scattering of a fictitious spurion particle
$S$ on the heavy quark,
\begin{equation}
S(q) + H_Q (p)\; \to \mbox{ light hadrons } \;,
\label{73}
\end{equation}
or the weak decay width in the process
\begin{equation}
Q\; \to \mbox{ quarks (leptons) } + \:S \;.
\label{74}
\end{equation}
Such processes would appear if the weak decay Lagrangian is modified
from,
say the conventional four-fermion form to the ``four-fermion + spurion''
interaction,
\begin{equation}
{\cal L}_{\rm w}(x) \; \to \; S(x)\,{\cal L}_{\rm w}(x)\;.
\label{75}
\end{equation}
For simplicity it is convenient to assume, as in Eq.\,(\ref{72}) that
the spurion field does not carry spacelike momentum but only energy. 

The amplitude ${\cal A}(\omega)$ has the usual analytic properties: it
is analytic and has a number of cuts describing the physical processes
in different channels ($s$, $t$ or $u$). The physical cut corresponding
to the weak decay of the heavy quark we are interested in, starts near
$\omega \simeq E_r$ where the energy release $E_r$ denotes $m_Q$ minus 
the sum of
the masses of the final state quarks and/or leptons. Other cuts are
located far enough 
from this point and from the physical point $\omega\!=\!0$.
The discontinuity across the physical cut at which the point
$\omega\!=\!0$
is located, describes the total decay width we are interested in. The
OPE for the inclusive widths relies on the fact that the short-distance
expansion of ${\cal A}(\omega)$ runs in $1/(\omega\!-\!E_r)$ and can be
applied near the physical point $\omega\!=\!0$ exactly as in $\rm e^+e^-$
annihilation near a positive value of $s\!\gg \!\Lam^2$. 
To the same extent, in principle, a certain smearing
can be required if the hadronic probabilities still exhibit the
resonance structure.

Thus, there is no theoretical peculiarity in the asymptotic applications
of the OPE, say for nonleptonic widths compared to semileptonic. It does 
not make a conceptual difference to perform a short-distance expansion 
of a single quark 
Green function (like in semileptonic widths or $b\to s+\gamma$), of the
product of two Green functions ($\rm e^+e^-$ annihilation) or of the
product of three quark Green functions (the nonleptonic widths).

Smearing in $\omega$ can often be phrased as smearing over
the interval of $m_Q$. Indeed, in the heavy quark limit the decay 
amplitudes depend on just the combination $m_b\!-\!\omega$, therefore
\begin{equation}
{\cal A}(\omega, m_Q) \;\simeq {\cal A}(0, m_Q\!-\!\omega)
\label{77}
\end{equation}
(there are power corrections to this relation associated with
explicit mass effects in the initial state). 
Smearing in the heavy quark mass may look more transparent physically
when the effect of opening new threshold is effectively averaged.

The structure of the cuts in the amplitude ${\cal A}(\omega)$ is
particularly simple if all the masses of the final state quarks are
large compared to $\Lam$. What happens if some final state quarks are
light, for example, even in semileptonic $b\to u$ decays? The physical 
cut corresponding to the decay into $u$ quark can come very close to the
cut which would describe the scattering of $\bar{u}$ on the heavy quark
-- the distance between them is twice the energy of the quark, and it can
be small if the heavy quark line is soft (i.e., both energy and spacelike 
momentum are small). We would not be able to distinguish between the
contributions of the two processes if their respective cuts are too
close in the scale of strong interactions. The answer is that in the OPE
we remove explicitly all soft lines (including the quark ones) which
carry small momenta; their contributions are described by the
``condensates'' where the soft legs are treated as the external fields
acting on the initial state. The remaining pieces describe only
energetic particles for which the proximity of the cuts cannot take
place. From this perspective the OPE cutoff $\mu$ in computing the Wilson
coefficients acts similar to assigning a mass $\mu$ to each potentially
soft quark line.

It is important to emphasize in this respect that the OPE treatment of
these contributions in the ``corners of the phase space'' differs from,
say, the computation of the regular $1/m_Q^2$ terms given by the
chromomagnetic operator. In the latter case the effect (at least a part
of it) comes from the kinematics where the corresponding final state
quark is hard, and we really use the short-distance expansion of the
propagator to compute this term. In the case of soft quark legs we do
not compute them in the OPE, but, in a sense, parameterize their 
contribution. The large momentum flowing through the rest of the diagram
only ensures that it is given by the expectation value of a  {\it
local} heavy quark operator. However, even if we knew the full 
transition amplitude with some accuracy near $\omega\!=\!0$, we still would
not be able to resolve the contribution of the physical and the
``$u$-channel'' cuts. These subtleties used to cause some confusion in
the literature; the proximity of the cuts was viewed as the failure of the
OPE to describe such contributions \cite{cgg}.

\subsection{$|V_{cb}|$ from the total semileptonic $B$ width}

Including the leading corrections, the semileptonic width has the
following form \cite{buv,bs,dpf,prl}:
\beq
\Gamma_{\rm sl} = \frac{G_F^2 m_b^5}{192\pi^3}\, |V_{cb}|^2 \left\{z_0
\left(1\!-\!\frac{\mu_\pi^2\!-\!\mu_G^2}{2m_b^2} \right)
-2\left(1\!-\!\frac{m_c^2}{m_b^2} \right)^4\frac{\mu_G^2}{m_b^2}
-\frac{2}{3} \frac{\as}{\pi} a_1 + ...
\right\}
\label{crad6}
\eeq
where ellipses stand for higher order perturbative and/or power
corrections,  $z_0$ and $a_1$ depend on $m_c^2/m_b^2$.
Regardless the exact value of $\mu_\pi^2$, the direct 
$\, 1/m_b^2$ corrections to $\Gamma_{\rm sl}$ are rather small, about
$-5\%$ and lead to the increase in the extracted value of
$|V_{cb}|$ by $2.5\%$. The higher order power corrections are at a
percent level. There is one new operator, the Darwin term which appears
to order $1/(m_b\!-\!m_c)^3$; it was evaluated in Refs.~\cite{bds,grekap}
and may decrease the width by up to $4\%$. 
The remaining $1/m_b^3$ corrections enter only through the $1/m_b$ terms 
in the expectation values of the kinetic and chromomagnetic 
operators in actual $B$ mesons. They include the `spin-orbital'
expectation value $\rho_{LS}^3$ as a $1/m_b$ part of the full QCD operator
$\bar{b}{\small\frac{i}{2}}\sigma G b$, and the nonlocal 
correlators introduced in
Sect.~2.2. Their effect is minor and is included in the uncertainties of
the chromomagnetic and kinetic expectation values. 

The leading-order ${\cal O}(\as)$ perturbative corrections are known
from the QED calculations in muon decay \cite{muon}. The 
BLM corrections \cite{blm}, a part of the higher-order perturbative
series associated only with running of $\as$ in the first-order loop 
diagrams, were calculated to all orders \cite{bbbsl}. Their impact
appeared to be small if the width is expressed in terms of the properly
defined short-distance masses \cite{upset}. The remaining non-BLM 
(`genuine') 
two-loop $\as^2$ corrections were expected to be moderate if one uses
the heavy quark masses normalized at the scale around $1\GeV$
\cite{five}. They were evaluated in a series of papers by Czarnecki and
Melnikov \cite{czareta,czarm} and were confirmed to be small. At present, no
significant theoretical uncertainty remains in the perturbative
corrections to the semileptonic width.

The running mass of the $b$ quark is known with high precision,
Sect.~3.1.2. To determine $m_c$, we can rely on relation (\ref{3.16})
for the difference between $m_b$ and $m_c$ in terms of the hadron
masses. It turns out that in this way the direct dependence of the
semileptonic width on $\rho_D^3$ almost cancels against the one coming
indirectly from $m_b\!-\!m_c$, therefore the width becomes sensitive
only to the exact values of $\mu_\pi^2$ and $\bar\rho^3$. 

Evaluating the theoretical prediction we get 
$$
|V_{cb}|=0.0412\left(\frac{{\rm BR}(B\rightarrow
X_c\,\ell\nu)}{0.105}
\right)^{\frac{1}{2}}\left(\frac{1.55\,\rm
ps}{\tau_B}\right)^{\frac{1}{2}}
\cdot \left(1-0.012\frac{(\mu_\pi^2\!-\!0.5\GeV^2)}{0.1\GeV^2}\right)
\times
$$
\beq
\left(1-0.01\frac{\delta m_b(1\GeV)}{50\MeV}\right)
\left(1+0.007\frac{\bar\rho^3}{0.1\GeV^3}\right).
\label{w12}
\eeq
The main theoretical uncertainty at the moment resides
in the value of $\mu_\pi^2$, which comes from constraining $m_b\!-\!m_c$
in the heavy quark expansion. It is worth noting that this is the only
place where we relied on the expansion in $1/m_c$. It is vulnerable
to possible late onset of the $1/m_Q$ expansion which would show up as
the large expectation values of higher-dimension operators. Yet 
we note that the meson masses are expected to be the most robust
observables; it is well known in ordinary quantum mechanics that
eigenvalues of the Hamiltonian are more stable than wavefunctions and 
their overlaps. Nevertheless, it seems important to have an independent
direct determination of $m_c$ to isolate this potential problem.

Finally, we arrive at the model-independent evaluation 
$$
|V_{cb}|=0.0412
\left( \frac{{\rm BR}(B\rightarrow X_c\,\ell\nu)}{0.105}
\right) ^{\frac{1}{2}}
\left( \frac{1.55\,\rm ps}{\tau_B}\right) ^{\frac{1}{2}}
\times
$$
\beq
\left( 1-0.012\frac{(\mu_\pi^2\!-\!0.5\,\rm GeV^2)}{0.1\,\rm
GeV^2}\right)
\cdot \left( 1 \pm 0.012_{\rm pert} \pm 0.01_{m_b} \pm
0.02
\right) \, ,
\label{w20}
\end{equation}
where the last error reflects $m_Q^{-3}$ and higher
power corrections as well as possible deviations from local duality.

\subsection{$\Gamma_{\rm sl}(b\to u)$ and determination of $|V_{ub}|$}

Similar to the treatment of $\Gamma (B\ra X_c \, \ell \nu )$,  it is
straightforward to relate the value of $|V_{ub}|$ to the total
semileptonic width $\Gamma(B\ra X_u\, \ell\nu)$. The dedicated analysis
was performed in Ref.~\cite{vub}: 
\beq
|V_{ub}|\!=\!0.00445\left(\frac{{\rm BR}(B^0\!\rightarrow\!
X_u\ell\nu)}{0.002}
\right)^{\!\frac{1}{2}} \!\left(\frac{1.55\,\rm
ps}{\tau_B}\right)^{\!\frac{1}{2}} \!
\cdot \left(1 \pm 0.01_{\rm pert} \!\pm\! 0.03_{m_b} 
\!\pm\! 0.015_{\rm nonpert}
\right).
\label{22r}
\eeq
The dependence on $\mu_\pi^2$ is practically absent here. The complete 
perturbative corrections are known analytically two two orders
\cite{timo}, and again turn out to be small if one uses $m_b(1\GeV)$ as
the input. The contribution at the level of a few percent can be
expected from nonfactorizable expectation values of the four-quark
operators at order $1/m_b^3$. At this level it would be
advantageous to measure the $b\to u$ semileptonic decay width for the 
charged and neutral $B$ mesons separately \cite{WA}.

The most direct way to disentangle $b\ra u\,\ell\nu$ decays from 
hundred times
more abundant $b\ra
c\,\ell\nu$ without tagging the secondary charm decay would be to study
the invariant mass of hadrons in the final state, $M_X$:
\beq
\frac{{\rm d}}{{\rm d}M_{X}} \Gamma (B \ra X\, \ell \nu )\;, \qquad
M_X^2 = \left(\sum_i P_{\rm hadr}^{(i)}\right)^2\;.
\label{4mx}
\eeq
For the free quark decay one has $M_X^2\simeq 0$ in $b\ra u$ and
$M_X^2= m_c^2$ for the $b\ra c$ transitions. In actual decays the mass
can take values exceeding $m_\pi$ and $M_D$, respectively. The increase
in mass can originate both perturbatively if a hard
gluon is emitted in the decay, or through soft bound-state or
hadronization processes. 

The leading soft effects turn out to
significantly modify the $M_X$ distribution due to the effects of
primordial `motion' of the heavy quark caused by bound state dynamics in
the initial $b$ hadron. It was pointed out already in \cite{prl} 
that the QCD-based OPE automatically leads to an analogue of this
``Fermi motion'' which had been introduced phenomenologically long
ago, first in \cite{ap} and then elaborated further to the status of a
well-formulated model in \cite{acm}. A detailed
description of the Fermi motion in the framework of the $1/m_Q$
expansion was later given in \cite{motion,bsg} and in
\cite{randall}; it goes outside the scope of the present review. 
Here we only briefly mention the basic facts.

The ``Fermi motion'' in QCD has certain peculiarities which are absent
in the phenomenological models \cite{motion,roman}. The
distribution over the `primordial' Fermi momentum $F(\vec{p}\,)$ is
replaced by a certain distribution function $F(x)$, where $x\le 1$
measures the momentum of the $b$ quark in the units of $M_B\!-\!m_b$. 
It is
similar to the leading-twist distribution function of deep inelastic
scattering (DIS) on usual light hadrons.
First distinction is that $F(x)$ is one-dimensional;
one can define only the distribution over a certain projection of the
momentum.

Second, $F(x)$ {\em depends} essentially on the final state quark mass
(more exactly, on its velocity). While it is the same for $b\ra u\,\ell
\nu$ and $b\ra s+\gamma$ where the final quark is ultrarelativistic and
is given by the light-cone distribution function like in DIS, it
is rather different for $c$ quark in $b\ra c\,\ell \nu$ where it is
closer to a nonrelativistic particle; in this case the light-cone
distribution is replaced by the so-called temporal distribution
function.

Finally, the QCD $F(x)$ is normalization-point dependent. This property 
is well known already from usual DIS; the evolution of the heavy quark
distribution function, where normalization point is well below $m_Q$ is,
however, much stronger.

The leading soft effects generating the $M_X$ spectrum emerge
due to same physics which gives rise to
the Fermi motion. They are quite significant. For example, the
average invariant mass square of hadrons in $b\ra u$ gets the
nonperturbative correction $\sim \Lam \cdot m_b$ \cite{WA}:
\beq
\aver{M_X^2}_{\rm nonpert} \:=\: \frac{7}{10} m_b\left(M_B-m_b\right)
\,+\,{\cal O}\left(\Lam^2\right)\;\simeq\;
1.5\GeV^2\;.
\label{290}
\eeq
The perturbative
corrections lead to $\aver{M_X^2}_{\rm pert} \sim \frac{\as}{\pi}
m_b^2$, however in $B$ decays this increase is still smaller than
through the nonperturbative effects.

\thispagestyle{plain}
\begin{figure}[hhh]
 \begin{center}
 \mbox{\epsfig{file=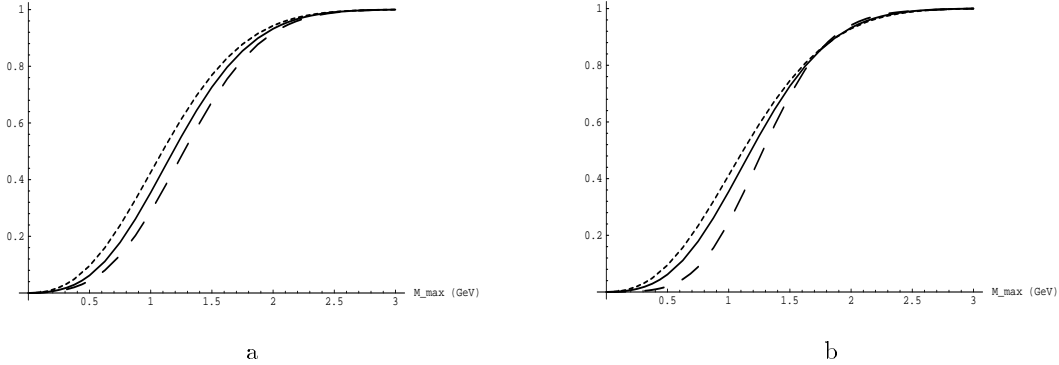,width=14cm}}
 \end{center}
\caption{ \small
The integrated fraction of the $b\ra u \,\ell \nu$ events $\Phi(M_X)$.
({\bf a}): Dependence on $m_b$ which is varied within $\pm 50\MeV$.
({\bf b}): Effect of variation of $\mu_\pi^2$ by $\pm 0.2\GeV^2$.
}
\end{figure}

To quantify the QCD effects on the $M_X$ distribution let us introduce,
following Refs.~\cite{dist,mx} the fraction of $b\!\ra\! u$ events
with $M_X$ below a certain cutoff mass $M_{\rm max}$:
\beq
\Phi(M_{\rm max}) = \frac{1}{\Gamma (b\ra u)}
\int _0^{M_{\rm max}} {\rm d}M_X \:\frac{{\rm d}\Gamma}{{\rm d}M_X}\;.
\label{18mx}
\eeq
$\Phi(0)\!=\!0$ and $\Phi(M_B)\!=\!1$ hold regardless of any dynamics.
The main question to theory is whether it can calculate
accurately enough $\Phi(M_{\rm max})$  with $M_{\rm max}\lsim 1.6\GeV$.

The dedicated analysis was carried out in \cite{dist,mx}, and the
conclusion appeared to be quite optimistic: $b\ra u \,\ell\nu$ decays 
are not expected to populate the domain above $M_X=1.6\GeV$ too
significantly. The typical theoretical predictions for $\Phi(M_{\rm
max})$ are shown in Fig.~10.

The corresponding experimental analysis was first attempted by ALEPH
\cite{alephmx} and recently performed by DELPHI \cite{delphimx}. The
$M_X$ spectrum was found in a nice agreement with the above theoretical
predictions, and the value of $|V_{ub}/V_{cb}|$ was found about $0.10$,
with the uncertainty of approximately $20$ percentage points; the latter 
is still dominated by the experimental error bars. 

\subsection{Summary on $|V_{cb}|$}

Here we give a brief summary of the two methods of
extracting $|V_{cb}|$. The most precise at the moment is the value
obtained from $\Gamma_{\rm sl}(B)$:
Eqs.~(\ref{w12}) and (\ref{w20}) with the 
central theoretical input values shown there
lead to 
\beq
|V_{cb}|\;\simeq\; 0.0413
\cdot
\left(1-0.012\frac{(\mu_\pi^2\!-\!0.5\GeV^2)}{0.1\,\rm
GeV^2}\right)\;.
\label{210}
\eeq
The overall relative theoretical uncertainty in this result is
$\delta_{\rm th} \lsim 5\%$. With future refinements we can expect 
reducing the uncertainty down to $2\%$.

The $B\ra D^*\ell \nu$ zero-recoil rate also provides a good accuracy.
The exact experimental status of the measurements extrapolated to
$\vec{q}=0$ is not completely clear at the moment. 
Until summer 2000 the results from the LEP experiments and from CLEO
seemed to be in a good agreement, yielding $F_{D^*}|V_{cb}|\simeq
0.035$  with a typical 
uncertainty of $\pm 0.001_{\rm stat} \pm 0.002_{\rm
syst}$. Using this reported average value and the literal estimate
$F_{D^*}\simeq 0.89$ only a bit lower value of $|V_{cb}| \simeq 0.0395$
emerges. Adding only one (systematic) standard deviation 
to $F_{D^*}|V_{cb}|$
would yield just the central value for $|V_{cb}|$ extracted from the
total semileptonic width -- even without varying the numbers within 
theoretical uncertainties. 

It is interesting that the central theoretical value in this method,
according to Eq.~(\ref{z11}) exhibits the dependence on $\mu_\pi^2$
similar in magnitude but opposite in sign to Eq.~(\ref{210}):
\beq
|V_{cb}|\;\simeq\; 0.0395
\cdot
\left(1+0.015\frac{(\mu_\pi^2\!-\!0.5\GeV^2)}{0.1\,\rm GeV^2}\right)\;.
\label{211}
\eeq
The theoretical uncertainty $\delta_{\rm th}$  here constitutes
probably $\delta_{\rm th} \!\approx\! 6\%$, however this 
estimate relies on
additional theoretical assumptions. It is not clear how it can be 
decreased. 

Recently CLEO came up with the new value for $F_{D^*}|V_{cb}|=0.0424\pm
0.0018_{\rm stat} \pm 0.0019_{\rm syst}$ \cite{cleoexcl} which is
significantly higher than both their old number and the current LEP average 
value  
$0.0345\pm .0007_{\rm stat} \pm 0.0015_{\rm syst}$
\cite{lepexcl}, 
and now has larger error bars. Taking it literally, one even
observes `overshooting' -- the central value for $|V_{cb}|$ under the
same theoretical assumptions would be even higher than the result from
$\Gamma_{\rm sl}(B)$. Since the shift exceeds the previously quoted
error bars, it may be premature to draw a definite conclusion at the
moment. 

In any case, it is remarkable that the values of $|V_{cb}|$ that emerged
from exploiting two theoretically complementary approaches are very 
close. The progress was not for free: it became possible only due to
essential refinements of the theoretical tools in the last several
years, which prompted us, in  particular, that the zero-recoil $B\!\to 
\! D^*$ formfactor $F_{D^*}$ is probably close to $0.9$, significantly
lower than previous expectations. The decrease in $F_{D^*}$ and more
accurate experimental data which became available shortly after, in
summer 1994 reduced
the gap between the exclusive and inclusive determinations of
$|V_{cb}|$.

It must be noted, however, that most of the existing theoretical
analyses essentially assume that the approximate duality between the 
actual hadronic amplitudes and the quark-gluon ones sets in already at
the excitation energies $\sim 0.7$ to $1 \GeV$. In the formfactor
analyses it is most obvious in applying the perturbative computations to
charm quarks and relying on $1/m_c$ expansion. In computation of the
decay widths we, alternatively, invoke the local quark-hadron duality.
While there are no experimental indications so far that this is not the
case (at least, in the semileptonic physics), the proof is not known
either. If that is not true, and duality generally starts only above
$1\GeV$, most probably  one would have to abandon the idea of accurate
determination of $|V_{cb}|$ from the exclusive $B\!\to\! D^{(*)}$
transitions. The only  option still open will be the inclusive
semileptonic decays where the energy release is large, $\sim 3.5\GeV$.
Of course, in such a pessimistic scenario the theoretical precision in
$|V_{cb}|$ will hardly be better than $5\%$.

\section{Challenges in the HQE}

Concluding the review it is appropriate to mention cases where the heavy
quark expansion apparently has problems.  It is probably premature to
speak of a direct contradiction to experiment; nevertheless, today's
question marks carry the seeds of  tomorrow's advances. Basically there
are two problems where our theoretical understanding is lagging behind.
Both are related to nonleptonic decays.

\subsection{Semileptonic branching ratio of $B$ and
$\Gamma (b \!\to \!c \bar c s )$ }
 
The theoretical attitude to this problem changes with time.
Twenty years ago the parton model gave a prediction ${\rm
BR_{sl}}(B)\!\simeq\! 13$--$15\%$, which was accurate enough according to
the existed standards. The variation reflected mostly the choice of
quark masses. While the rates of $c\ra c\,\ell\nu$ and $b\ra c \bar{u}d$
were affected by the choice of $m_c$ in more or less the same way, the
rate of the $b\ra c \bar{c}s$ channel decreased much faster when
larger masses (closer to the `constituent' rather than short-distance
`current' ones) were used. Therefore, ${\rm BR_{sl}}(B)$ increases for
larger masses. It should be remembered that the difference $m_b\!-\!m_c$ 
is fixed, so the choices of $m_c$ and $m_b$ are always
correlated. Even though the heavy quark symmetry had not been
formulated, the latter fact was clearly realized at least in the early
80's \cite{rueckl}.

The experimental situation was not very definite and indicated a rather
large ${\rm BR_{sl}}(B)$ which reasonably fitted the `larger' mass
option \cite{altp}. Since then it became standard to use the larger
quark masses, and the parton model prediction for ${\rm BR_{sl}}(B)$ was
accepted to be $13\,$--$\,14\%$, even though a smaller value could be
obtained as well. Since the impact of the nonperturbative corrections
was completely unknown and presumably significant, ${\rm BR_{sl}}(B)$ 
was not under intense
scrutiny even when the better data became available.

The situation changed when in 1992 Bigi {\it et al.} showed that there
are no $1/m_Q$ nonperturbative corrections to the inclusive widths, both
semileptonic and nonleptonic. The leading $1/m_b^2$ 
nonperturbative effects were readily calculated \cite{buv,bs,koytau}. 
They appeared to be suppressed, in particular, as a result of certain
cancellations. The overall effect $\Delta_{\rm npt}{\rm BR_{sl}}(B)$
was found to be about $-0.5\%$; the nonperturbative effects could not be
blamed for a discrepancy any more \cite{baffling}.

This prompted a more careful analysis of the perturbative corrections to
the widths. In particular, the ${\cal O}(\as)$ corrections to the
nonleptonic $b\ra c\bar{u}d$ width were calculated accounting for the
nonzero $m_c$ \cite{bagan}; this additionally enhanced the nonleptonic
width. Later the account for the charm mass was also completed for $b\ra
c\bar{c}s$ \cite{godz,volcc,lss}. Altogether, these ${\cal O}(\as)$
corrections further decreased ${\rm BR_{sl}}(B)$ down to
$11\,$--$\,12\%$. Although this shift naively seems very significant and
may raise concerns about the convergence of the perturbative
corrections, it actually is not dramatic if one starts with more
appropriate short-distance masses, the choice forgotten for historical
rather than rational reasons. The relative increase in the nonleptonic
width was particularly significant in the $b\ra c\bar{c}s$ channel.

The issue of the semileptonic branching ratio must be considered in
conjunction with the charm yield $n_c$, the number of charm states
emerging from $B$ decays. To measure $n_c$ one assigns charm
multiplicity {\em one} to $D$, $D_s$, $\Lambda _c$ and $\Xi _c$ and {\em
two} to charmonia. Zero is assigned to the charmless hadronic final
state. It is obvious that
\beq
n_c \simeq 1 +\mbox{BR}(B \ra c \bar c s \,\bar q)
-\mbox{BR}(B \ra \mbox{no charm})\, .
\label{8.32}
\eeq
Such a joint analysis was motivated already in \cite{buv}: the energy
release in $b\ra c\bar{c}s$  is rather small, and this can lead to
significant duality-violating and higher-order effects. The stability
of the perturbative expansion also downgrades. Measuring $n_c$ allows
one to effectively exclude this channel from the theoretical
calculations.

At present the semileptonic fraction coming both from CLEO and LEP seems 
to be $1\%$ to $1.5\%$ lower than the `preferred' theoretical
expectation. The latter can be decreased playing 
with the ratio of the quark
masses and boosting the effect of perturbative corrections -- for
example, taking strong coupling at a lower scale. This, however,
increases $n_c$ beyond the experimental limits. In principle, allowing
$n_c$ to lie in the upper corner of the experimental interval this would
almost accommodate the LEP value, but still is somewhat off the
CLEO interval for ${\rm BR_{sl}}(B)$ and $n_c$. 

It is quite possible that this discrepancy should be taken seriously,
and we would have to admit certain limitations in our ability to
compute the nonleptonic $B$ decay width. 
However, a more thorough analysis is badly needed which would be free
from prejudices. The existing compilations rely on the old analysis where,
among other things, pole masses of $b$ and $c$ are used as the input.
This leads to a number of problems: the perturbative corrections look too
significant, and simply rewriting the first order result in different
-- but equivalent to order $\alpha_s$ -- forms yields noticeably
differing numbers. This, somewhat artificial, ambiguity is sometimes
used to stretch the limits in a desired direction. Second, using the
pole masses for quarks does not allow to immediately constrain their 
values which are nowadays well known in the proper scheme. 

Another place where refinement is needed, is a more accurate
determination of the Wilson coefficient for $\mu_G^2$ in the
semileptonic fraction. So far all the analyses used the expressions 
(\ref{sample31}) from 
the original papers \cite{buv,bs} where it was obtained in the leading
logarithmic approximation only (barring the $b\to c\bar{c} s$ channel,
${\rm BR_{sl}}(B)$ is modified only via nonfactorizable diagrams
appearing to order $\alpha_s$), being proportional to a small
product of two different color coefficients $a_1 a_2$ of the weak
nonleptonic Lagrangian. This may be a poor approximation, and the
non-$\log$ piece coming from the extra gluon with $k \!\sim \!m_b$ can
dominate the result, as often happens with the LLA expressions in
practice. Here there are special reasons to expect significance of such
corrections: First, the corresponding diagrams can include 
gluon-gluon interaction, and such coupling is enhanced by color factors.
Second, the LLA term suffers accidental cancellation between the
different channels, the fact external to QCD itself. It is improbable
that the chromomagnetic interaction itself can lower semileptonic
fraction by, say, two percent. However, the existing apparent
discrepancy between theory and experiment is lower now than it 
appeared eight years ago, and this certainly prompts a more careful 
analysis, to draw the final conclusion.

\subsection{Lifetimes of beauty hadrons}

As stated before, differences between meson and baryon decay widths
arise already in order $1/m_Q^2$. The perturbative corrections to the
lifetime ratios are completely absent. The lifetimes of the various
mesons get differentiated effectively first in order $1/m_Q^3$. A
detailed review can be found in \cite{stone2,BELLINI,four}.

Because the charm quark mass is not much larger than typical
hadronic scale one can expect to make only semi-quantitative
predictions on the {\em charm} lifetimes, in particular for the
charm baryons. The agreement of the predictions with the
data is good. I would even say it is too good keeping in mind that the
$1/m_c$ expansion can hardly be justified. It is difficult to avoid mentioning
one particular example. It is generally understood that the large
lifetime ratio between charged and neutral $D$ mesons, 
$\tau_{D^+}/\tau_{D^0}\simeq 2.5$ 
is due to destructive Pauli
Interference in nonleptonic decays of $D^+$. However, it is the
semileptonic fraction of $D^+$ which is close to the `canonical' value
of $20\%$ obtained by simple-minded counting of the quark decay
channels, while ${\rm BR_{sl}}(D^0)$ is much smaller. Yet, if one
adds the $1/m_c^2$ shift in the semileptonic fraction due to
the chromomagnetic term computed in Ref.~\cite{buv}, the experimental 
values
of the semileptonic fractions are reproduced. Clearly, one should not
take such a coincidence too literally since the corrections of order
unity discussed here cannot be rigorously justified in the framework of
the standard $1/m_Q$ expansion itself. 

As far as the {\em beauty} lifetimes are concerned the $1/m_b$ expansion
is to be applicable. Table \ref{TABLE20} contains the recent 
data together with the predictions. The latter were actually
made before data (or data of comparable sensitivity) became available.

\begin{table}[t]
\caption{Predictions for Beauty Lifetimes (heavy quark
expansion) 
\label{TABLE20}}
\begin{center}
\begin{tabular} {|l|l|l|l|}
\hline
Observable & QCD Expectations ($1/m_b$ expansion)& Ref. &
Data from \cite{lepsg}\\
\hline
\hline
$\tau (B^-)/\tau (B_d)$ & $1+
0.05(f_B/200\, {\rm MeV} )^2 $ & \cite{mirage} & $1.070 \pm 0.02$ \\
\hline
$\bar \tau (B_s)/\tau (B_d)$ &$1\pm {\cal O}(0.01)$ &
\cite{stone2}
&  $ 0.945\pm 0.039$ \\
\hline
$\tau (\Lambda _b)/\tau (B_d)$&$\gsim 0.9 $ & \cite{stone2} &
$0.794\pm 0.053$ \\
\hline
\end{tabular}
\end{center}
\end{table}

Data and predictions on the meson lifetimes are 
nontrivially  consistent. Yet even so, a  comment is in order for
proper  orientation. The numerical predictions were based on the
assumption of factorization at a typical hadronic scale which is
commonly taken as the one where $\as(\mu_{\rm hadr})\!\simeq \!1$.
While there is no justification
for factorization at $\mu\sim m_b$, there exists ample
circumstantial evidence in favor of approximate factorization  at a
typical hadronic scale. More to  the point,  the
validity of factorization can be probed in semileptonic  decays of $B$
mesons in an independent way, as  was pointed out in \cite{WA}.

The possible effect of the nonfactorizable contribution has been
discussed in detail in \cite{WA,Ds}, and later in the dedicated paper
\cite{four}. They include the nonvalence
gluon mechanism discussed long ago in \cite{fritmink} in the simplified
language of the quark model. Significant nonfactorizable contributions
would in general lead to large effects of Weak Annihilation in $D_s$
mesons where experimentally such effects are quite small. Of course, we
cannot reliably treat $\Gamma_D$ in the $1/m_Q$ expansion, and
$\Gamma_{D_s}\!-\!\Gamma_{D^+}$ is sensitive to a particular combination of
the expectation values of a few four-fermion operators. Nevertheless, this 
is an indication that the effects due to nonfactorizable
expectation values must be suppressed. Later studies based on QCD sum
rules and preliminary lattice results reproduced within intrinsic
uncertainties the estimates one obtains relying on factorization 
at the low hadronic scale. The nonfactorizable effects may 
affect the difference between $\bar\tau (B_s)$ and $\tau (B_d)$.

The agreement of the data on $B$ meson lifetimes with experiment is
obscured by the apparent conflict for $\tau_{\Lambda_b}/\tau_{B_d}$.
To predict the $1/m_Q^3$ corrections to $\tau_{\Lambda_b}$ in the
$1/m_b$ expansion one needs to evaluate the baryonic expectation values
of two operators,
\beq
\matel{\Lambda_b}{\bar b b\,\bar{u}\gamma_0 u }
{\Lambda_b}\;,\qquad \matel{\Lambda_b}
{\bar b \mbox{\small $\frac{\lambda^a}{2}$}b\,
\bar{u}\gamma_0  \mbox{\small $\frac{\lambda^a}{2}$}u}{\Lambda_b}\;.
\label{305}
\eeq
They do not have the usual factorizable contribution, and their values
are rather uncertain. Nevertheless, it was shown that their
contributions cannot be too large \cite{boost} and the maximal
effect in $\Gamma_{\Lambda_b}$ does not exceed $10\,$--$\,12\%$.
To achieve larger corrections one would have to go beyond the usual
description of baryons where light quarks are ``soft''. The reason for
such a limitation is rather simple. The four-fermion expectation values,
in the language of ordinary quantum mechanics, are squares of the
wavefunction at the zero separation between the heavy and the 
corresponding light quark:
\beq
\Psi_{Qq}(0) = \int \frac {{\rm d}^3\vec{p}}{(2\pi)^3} 
\Psi_{Qq}(\vec p\,)\;,
\label{boost1}
\eeq
but, at the same time, the space integral of $|\Psi_{Qq}|^2$ is limited
by unity. A too large value of $\Psi_{Qq}(0)$ means that the wavefunction 
is sharply peaked at zero separation, which signifies presence of
high-momentum modes.
The above mentioned general bound agrees
with the fact that the constituent quark model estimates typically
yield only about $3$ to $5\%$ enhancement \cite{rosner}.  A similar
conclusion has been reached by the authors of Ref.~\cite{BARI} who
analyzed the relevant baryonic matrix elements through QCD sum rules.
A dedicated discussion of these questions can be found in 
Ref.~\cite{four}. 

Recently, the preliminary lattice results for the relevant four-fermion 
expectation values were reported from the UKQCD collaboration, with the
corresponding estimate $\tau_{\Lambda_b}/\tau_{B_d} \simeq 0.90$
\cite{ukqcd99}. We
note here that the $1/m_Q^4$ corrections to the lifetimes can well
constitute $30$ to $50\%$ of the $1/m_Q^3$ terms. Keeping in mind
that the dominant effect in the $\Lambda_b$ lifetime is Weak Scattering,
it is plausible that these effects further enhance the width. (This
depends on the corresponding dimension-$7$ four-fermion operators with
derivatives; taking the naive quark model picture one would simply
replace $m_b^2$ in the Wilson coefficient by the square of the average
diquark energy in $\Lambda_b$, to account for these effects.)

If future more accurate experimental data confirm the present deviation
of $\tau_{\!\Lambda_b}/\tau_{\!B_d}$ from unity, and refinements of the 
theoretical estimates do not reveal unexpected enhancement, the most
probable explanation of the discrepancy will be violation of local
quark-hadron duality in nonleptonic decays. As a matter of fact, its
significance in nonleptonic widths is theoretically expected {\it a
priori}. Indeed, the expansion parameter for the widths is not $m_b$
directly but rather the energy release which is noticeably smaller in
$b\!\ra\! c$. Moreover, the preasymptotic corrections depend on the concrete
form of the weak interaction involved. For the four-fermion interaction
they are enhanced: the arguments based on the high power of mass $m_Q^5$ 
in the decay rate \cite{five} suggest that the actual scale parameter is
smaller, $\propto  E_{\rm rel}/5$. In the semileptonic decays this does
not deteriorate the expansion since it is automatically protected by the
heavy quark symmetry when $m_c$ increases. 

For the nonleptonic decays the heavy quark symmetry does not generally
apply, and at insufficient energy release one expects significant
violations of duality. The real problem here is that the few leading
terms in the expansion have been evaluated and did not indicate 
the expansion blowing up at the level of $20\%$.

\section{Heavy Quark Expansion and Violations of Local Duality}

A large number of applications of the heavy quark theory
are based on quark-hadron duality, more exactly, its local
implementation. Although this notion becomes exact
at asymptotically high energies, at finite energy scale certain
deviations must be present. How fast duality sets in and how large are
these deviations are
important questions.
These and similar questions are among most difficult. Yet they are
important for phenomenology: determinations of $|V_{cb}|$ we discussed
rely on the assumption of local duality. 

The notion of duality in
general terms was first introduced in the
early days of QCD in Ref.~\cite{poggio} but not pursued for quite some
time. It was simply regarded as a problem of how the observables,
say $\sigma(e^+e^- \!\to\! \mbox{hadrons})$ as a function of energy computed 
through quarks (and
gluons) can be equated to the actual cross section. The former predict
small smooth deviations from the would be free quarks production, while
the latter has well-shaped resonant structure reflecting quarks
permanently confined in colorless hadrons. Since such hadronization
effects are nonperturbative in nature, the question of local duality is
tightly related to treatment of strong interactions beyond perturbation
theory. This was not really available in the early days of QCD;
confinement was simply assumed not to affect averaged cross sections
``significantly''.

Nowadays we have in our disposal methods which are applicable -- at
least in principle -- to quantify nonperturbative effects in a number of
observables, including those of the actual Minkowski world. These are
quantities amenable to study via the OPE. Local quark-hadron
duality was given a new consideration a few years ago by M.~Shifman
\cite{shiftasi}, who related its violations to the asymptotic nature of
the power expansion in inverse large energy scale provided by the
`practical' OPE.  These ideas were later reiterated and developed in a
number of papers, in particular in connection to heavy quark physics
(see, e.g., Refs.~\cite{inst,D2}).

The basic observation can be illustrated on the example of heavy quark
decay widths. The ``practical'' OPE yields the width in the power
expansion
\begin{equation}
\frac{\Gamma_{H_Q}}{\Gamma_Q}\;=\;
A_0 + \frac{A_1}{m_Q} + \frac{A_2}{m_Q^2}\:+\;... \;\;.
\label{1200}
\end{equation}
If the series in $1/m_Q$ were convergent (to the actual ratio),
$\Gamma_{H_Q}$ would have been an analytic function of $m_Q$
above a certain mass $m_0$ pointing to the onset of the {\em exact}
local parton-hadron duality. The actual $\Gamma_{H_Q}$ is definitely
non-analytic at any threshold (whether or not the amplitude vanishes at
the threshold). Thus, the `radius of convergence' cannot correspond to
the mass smaller than the threshold mass. Since in the actual QCD the
thresholds exist at arbitrary high energy, the power expansion in
Eq.\,(\ref{1200}) can be only asymptotic, with formally zero radius of
convergence in $1/m_Q$.

In practice, the true threshold singularities are expected to be
strongly suppressed at large energies, and the corresponding
uncertainties in the OPE series quite small. Eventually
they are expected to be exponentially suppressed, though, possibly,
starting at larger energies. In the intermediate domain they can
decrease as a
certain power and must {\em oscillate}. This reflects the peculiarity of
the Minkowski world. The terms left aside by the `practical' OPE are
exponentially suppressed in the Euclidean domain, $\sim
\exp{-\sqrt{Q^2}}$ where $Q^2$ generically denotes the square of the
large momentum scale. Continuing this expression to Minkowski domain
and taking discontinuity to determine transition probabilities, as
discussed in Sects.~5 and 6, we observe oscillating rather than 
exponentially suppressed effects.

As was illustrated in Ref.\,\cite{D2}, the power expansions like
Eq.\,(\ref{1200}) are meaningful even beyond the power suppression where
the duality-violating oscillations show up. In the case of the heavy
quark widths where mass $m_Q$ cannot be varied in experiment, the size
of the duality-violating component may set the practical bound for
calculating the widths. Thus, it is important to have an idea about its
size. At the same time one should always include the leading QCD
effects to the partonic expressions, rather than compare the
actual observable with the bare quark result. In the model considered in
Ref.\,\cite{D2}, incorporating the power corrections from the practical
OPE suppressed the apparent deviations by more than an order of
magnitude.

It is useful to keep in mind that violations of local duality,
although so far poorly understood dynamically, are not
arbitrary and must obey constraints following from the OPE. They cannot
be blamed, for instance, for the systematic excess or systematic
depletion of decay probabilities; the actual width can only oscillate 
around the OPE predictions as a function of energy or quark mass, or
the difference has to fade out exponentially.

Although the simplest illustration of the asymptotic nature of
the decay width $1/m_Q$ expansion and related violations of local
duality given above follows from the presence of hadronic
thresholds, violation of local duality is a
more universal phenomenon that is {\em not\/} directly related to
existence of hadronic resonances nor even confinement itself.  This
has been illustrated in Ref.~\cite{inst} by the example of soft
instanton effects that do not lead, at least at small instanton density, to
quark confinement -- but do indeed generate computable oscillating
duality-violating contributions to the total decay rates.

One can parallel the OPE with expanding the interaction between quarks
and gluons (and their related propagation) at small distances set up by
the inverse energy scale in the problem. This is 
transparent in toy quantum mechanical problems where the terms in the
OPE series can be traced back to the Taylor expansion of the potential
$V(x)$ at $x \!\to\! 0$. The behavior of $V(x)$ in the vicinity of $x\!=\!0$
does not exhaust the problem, however. The spectrum of final states
crucially depends on the asymptotics of the potential at large $x$; this
is analogous to the resonances affecting local duality. There are more
intriguing mechanisms associated with the finite-distance singularities
of QCD interactions -- the example is provided by instantons. The
interested reader can find the inspiring discussion in the dedicated paper by
Shifman in this volume \cite{shifmanioffe}.

While conceptual grounds for violation of local duality have become
more clear, its dynamical origin in actual QCD is still not well
understood. What type of physics lies behind the finite-$x$
singularities if they are relevant? How to quantify effects of
resonances at high energies? We do not have answers yet. The
instanton-based model \cite{inst}, for example, while capturing 
correctly the gross
features, clearly falls short in describing the size of the effects, 
at least under the standard assumptions. In all considered
cases, actually, the effects of local duality violation turn out very
small in the asymptotic domain of large energies or quark masses, much
below phenomenologically significant level. For practical purposes, we
would like to know the limitations imposed by duality in the domain of
intermediate energies where it {\it a priori} can be sizeable.
The use of analytic methods can be limited here.

Trying to get insights into the possible magnitude of violations of
local duality in semileptonic decays of heavy quarks, recent paper
\cite{lebur} studied their numerical significance in the exactly
solvable two-dimensional 't~Hooft model.  The model has built-in
``hard'' confinement related to linear Coulomb potential in 1+1
dimensions. Its spectrum consists of towers of  infinitely narrow
resonances. As was mentioned above, duality violation is not necessary
related to existence of resonances. Nevertheless, the intuition remains
that resonance dominance is not ``favorable'' for the OPE, and problems
might show up, for instance, through a delayed numerical onset of
duality, in that the approximate equality of the OPE predictions and the
actual decay widths may set in only after a significant number of
thresholds has been passed.  To address such issues, the 't~Hooft model
seems to represent the most certain testing ground for local duality in
the domain of decays of moderately heavy quarks.

Contrary to naive expectations, surprisingly accurate duality was found
between the (truncated) OPE series for  $\Gamma_{\rm sl}$ and the actual
decay widths. The deviations were suppressed to a very high degree
almost immediately after the threshold for the first excited final state
hadron is passed. No suspected delay in the onset of duality was found.
Remarkably, the `practical' OPE turns out to be the most efficient way 
to obtain numerical predictions for the total semileptonic width in
the model to a high accuracy hardly accessible to direct computations, 
down to rather low quark masses.

Some of the duality-violating features observed in those studies have
natural explanations.  At fixed energy release $m_b\!-\!m_c$ the
magnitude of the deviations is smaller if $m_b$, $m_c$ are both large
than if they are both small.  This is expected, since in the former case
the heavy quark symmetry for the elastic amplitude additionally enforces
approximate duality even when no expansion in large energy release can
be applied. However, at fixed $m_b$ the duality violation decreased
rapidly as $m_c$ decreases, in full accord with the OPE where the higher
order terms are generally suppressed by powers of $1/(m_b\!-\!m_c)$.
This is clearly a {\em dynamical\/} feature that goes beyond heavy quark
symmetry {\it per se}, the quality of which deteriorates as $m_c$
decreases.

To the extent the numerical findings of Ref.~\cite{lebur} can
be transferred to real QCD, violation of local duality in the total
semileptonic widths of $B$ mesons should not be an issue.  The scale of
duality violation lied far below the phenomenologically accessible
limits, and could not affect the credibility of extracting $|V_{cb}|$ or
$|V_{ub}|$.

In reality there are, of course, essential
differences between the two theories, including those aspects that are
expected to be important for local duality (for a discussion, see
Ref.~\cite{D2WA}).
Although many seem to optimistically suggest that
duality violation is more pronounced in the 't~Hooft model than for
actual heavy flavor hadrons, some differences may still work in the
opposite direction.  In $D\!=\!2$ there are no dynamical gluons, nor a
chromomagnetic field that in $D\!=\!4$ provides a significant scale of
nonperturbative effects in heavy flavor hadrons.  Likewise, there is
no spin in $D\!=\!2$, and no corresponding $P$-wave excitations of the
spin-$\frac{3}{2}$ light degrees of freedom which play an important role
in $D\!=\!4$.

Two-dimensional QCD neither has long perturbative ``tails'' of
actual strong interactions suppressed weakly (by only powers of
$\log$s of the energy scale).  In $D\!=\!2$ the perturbative
corrections are generally power-suppressed, as follows from the
dimension of the gauge coupling.  As discussed in Ref.~\cite{D2WA},
it is conceivable that the characteristic mass scale for freezing out
the transverse gluonic degrees of freedom is higher than in the
``valence'' quark channels.  This would imply a possibly higher scale
for onset of duality in $\alpha_s/\pi$ corrections to various
observables.

Regardless of these differences, it was demonstrated that presence of
resonance structure {\it per se} is not an obstacle for fine local
quark-hadron duality tested in the context of the OPE.  In
the 't~Hooft model resonances themselves do not demand a
larger duality interval.  As soon as the mass scale of the states
saturating the sum rules in a particular channel (quark or hybrid) has
been passed, the decay width can be well approximated numerically by
the expansion stemming from the OPE.

The same pattern is expected in QCD. To get a better idea regarding the
onset of local duality, at least in semileptonic decays, it is important
to study the saturation of the heavy quark sum rules, in particular the
static SV sum rules discussed in Sect.~4. Are the values of $\mu_\pi^2$
and $\mu_G^2$ we use reached at the energy scale of $0.7$ to
$1\GeV$ and right above we can use the perturbative description to
account for higher states, -- or we need to go higher in energy for that?
Experimental study of actual $b\to c$ transitions can give the answer.
There are ideas how this can be studied on the lattices. The question of
local duality in nonleptonic decays of beauty particles at present
remains largely unknown.

\section{Conclusions and Outlook}

Heavy quark theory is probably one of the youngest branches of the QCD 
tree, yet it has become mature. Many practically important problems
that laid dormant for years are now tractable. The development of the
heavy quark theory, often initiated in the quest for the most accurate 
extraction of electroweak properties of quarks, enriched the library
of available methods in QCD itself. At the same time, many natural
limitations of QCD show up in the heavy quark expansion as well. Quark 
confinement is not completely understood yet, the infrared part of dynamics is
more parameterized than solved. Therefore, every new result relying  
on the general properties of the QCD interactions brings in a
theoretical value. The success of the last decade was mainly related
to incorporating Wilson's approach to the heavy quark 
expansion. It allowed to quantify the deviations from the heavy quark
symmetry in many important cases, and led to a number of accurate
predictions not appealing to the symmetry itself. The theory of
preasymptotic corrections to the decay widths is one of such assets,
whether or not $b$ quark is heavy enough for precision applications of 
the heavy quark expansion.

A clear-cut recent manifestation of the power of the heavy 
quark theory is the framework it provides for determination of
$|V_{cb}|$. It is remarkable  that the values of $|V_{cb}|$ that 
emerged from exploiting two theoretically complementary  
approaches relying on quite different assumptions, are very close. The
progress was not for free: it became possible only
due to essential refinements of the available theoretical tools. 
They prompted us, in  particular, that the
zero recoil $B\!\to\! D^*$ formfactor $F_{D^*}$ is probably closer to $0.9$, 
with much larger deviations from the symmetry limit than previously
expected. The predicted decrease in $F_{D^*}$ brought more accurate
experimental data which appeared later in a good agreement with
$|V_{cb}|$ determined from $\Gamma_{\rm sl}(B)$. The confidence in the 
latter method, in turn required many improvements beyond theory of
nonperturbative effects as well, including accurate evaluation of
the perturbative corrections. The latter story, although eventually
successful, was far from simple (for the historical perspective see,
e.g.\ Ref.~\cite{varenna} and references therein). Sharpening the
numerical predictions was not possible without clarifying the nature
of the heavy quark mass entering theoretical expression, the questions 
which long caused confusion in the literature. It is encouraging that
now there is a consensus regarding also the numerical value of the running
$b$ quark mass -- it is extracted by different groups from the 
$\Upsilon$ sum rules using the state of the art NNLO resummation of
the perturbative effects. At the same time, a consistent theoretical
definition of such a low-scale mass required investigating the dipole
radiation in the non-Abelian gauge theory, a very general -- although
still somewhat theoretical -- physical phenomenon.

There are still problems to solve, ahead. Some of them are notoriously
difficult, like violations of local duality. Others may look at once
only technical, for example, developing the constructive Wilsonian OPE
including perturbation theory, where the soft parts have to 
be removed from all diagrams preserving such basic properties of QCD
interactions as gauge invariance. Like in the previous years, we may
expect new links in the chain of theoretical advances to emerge from
both types of studies.
\vspace*{.4cm}\\
{\bf Acknowledgments:}~\,I am indebted to my collaborators in the
heavy quark adventures I.~Bigi, M.~Shifman and A.~Vainshtein for too 
many -- to be mentioned separately -- crucial elements
which altogether made the studies possible. My special  
thanks to Misha Shifman for his encouragement and moral
support related to writing up this review. I often benefitted from
discussions with M.~Voloshin on various aspects of heavy quarks, and
acknowledge with gratitude useful discussions with M.~Eides on many
subjects including the new exact sum rules. 
This work was supported in part by NSF grant number PHY96-05080 and by
RFFI under grant No.\ 99-02-18355.

\end{document}